\documentclass[pre,showpacs,twocolumn,preprintnumbers,amsmath,amssymb]{revtex4}

\usepackage{graphicx}% Include figure files
\usepackage{dcolumn}% Align table columns on decimal point
\usepackage{bm}% bold math
\usepackage{parskip}
\usepackage{subfigure}
\usepackage{amssymb}
\usepackage{amsmath}
\usepackage{amssymb}
\usepackage{graphicx}
\newcommand{\bare}[1]{\mathaccent"7017{#1}}
\def\be{\begin{eqnarray}}
\def\ee{\end{eqnarray}}
\def\be{\begin{equation}}
\def\ee{\end{equation}}
\begin{document}
\title{Crossover from low-temperature to high-temperature fluctuations. \\ I.
Thermodynamic Casimir forces of isotropic systems}

\author{Volker Dohm}

\affiliation{Institute for Theoretical Physics, RWTH Aachen
University, D-52056 Aachen, Germany}

\date {17 August 2017}

\begin{abstract}
We study the crossover from low-temperature to high-temperature fluctuations including critical fluctuations in confined isotropic O$(n)$-symmetric systems on the basis of a finite-size renormalization-group approach at fixed dimension $d$ introduced previously [V. Dohm, Phys. Rev. Lett. {\bf 110}, 107207 (2013)].  Our theory is formulated within the  $\varphi^4$ lattice model in a $d$-dimensional
block geometry with periodic boundary conditions.
In contrast to the $\varepsilon = 4 - d$ expansion, the fixed-$d$ finite-size approach keeps the exponential form of the order-parameter distribution function unexpanded.
We derive the finite-size scaling functions $F^{\text ex}$ and $X$ of the excess free energy density and of the thermodynamic Casimir force,
respectively,
for $1\leq n \leq \infty$, $2<d<4$. Applications are given for $ L_\parallel^{d-1} \times L$ slab  geometries with a finite aspect ratio $\rho=L/L_\parallel$ as well as for the film limit $\rho \to 0$ at fixed $L$.
For $n=1$ and $\rho=0$ the low-temperature limits of  $F^{\text ex}$ and $X$ vanish whereas they are finite for  $n\geq 2$ and  $\rho = 0$ due to the effect of the Goldstone modes. For $n=1$ and $\rho>0$ we find a finite low-temperature limit of $F^{\text ex}$ which deviates from that of the the Ising model. We attribute this deviation to the nonuniversal difference between the $\varphi^4$ model with continuous variables $\varphi$ and the Ising model with discrete spin variables $s=\pm1$. For $n\geq 2$ and $\rho>0$, a logarithmic divergence of $F^{\text ex}$ in the low-temperature limit is predicted, in excellent agreement with Monte Carlo (MC) data for the $d=3$ $XY$ model. For $2\leq n \leq \infty$ and $0\leq \rho<\rho_0=0.8567$ the Goldstone modes generate a negative (attractive) low-temperature  Casimir force that vanishes for $\rho = \rho_0$ and becomes positive (repulsive) for $\rho > \rho_0$. At $T_c$ and for $1\leq n \leq \infty$ it vanishes for $\rho = 1$, $d=3$.
For $\rho \ll 1$ and $d=3$ a minimum of $X$ is found at $T_{\text min}\lesssim T_c$ for $n=1,2,3$ that is shifted to $T_{\text min}>T_c$ for $\rho \gtrsim 1/2$.
An exact description is given for the crossover from a $d$ dimensional transition to a $d-1$ dimensional transition in the large-$n$ limit for $d>3$. Our predictions are compared with MC data for Ising ($n=1$), $XY$ ($n=2$), and Heisenberg ($n=3$) models from far below to far above $T_c$ in slab geometries with $0.01\leq\rho\leq1$. Good overall agreement is found. In the subsequent paper [V. Dohm, Phys. Rev. E ...] our theory is extended to weakly anisotropic systems.
\end{abstract}
\maketitle

\renewcommand{\thesection}{\Roman{section}}
\renewcommand{\theequation}{1.\arabic{equation}}
\setcounter{equation}{0}
\section{ Introduction and summary}

Macroscopic forces arise from microscopic fluctuations in confined systems if the fluctuations have long-range correlations, i.e., if the ${\bf r}$-dependent correlation functions have a slow power-law decay rather than a fast exponential decay at large distances ${\bf r}$ in space. The most prominent example for such fluctuation-induced macroscopic forces is the Casimir force \cite{casimir,bordag} which is generated by vacuum fluctuations of the electromagnetic field, i.e., the quantum field of massless photons, confined between two neutral metallic plates. Analogous phenomena exist in various confined condensed matter systems at finite temperatures \cite{kardar,krech}
where classical thermal fluctuations rather than quantum fluctuations have long-range correlations  which then generate so called thermodynamic Casimir forces.

Among the systems with long-range correlations we consider two important examples which result from two fundamentally different sources: (i) from long-range classical fluctuations due to massless "Goldstone modes" \cite{goldstone,wagner} and (ii) from long-range critical fluctuations at a finite critical temperature $T_c$ \cite{fish-1}. Both types of fluctuations exist in $O(n)$-symmetric systems undergoing a second-order phase transition which is governed by the thermodynamic fluctuations of an $n$-component order parameter. (For bulk theories on systems with Goldstone modes near $T_c$ see, e.g., \cite{goldstone-crit,Burnett,str1999,str2003}). Examples of confined $O(n)$-symmetric systems where both types of thermodynamic Casimir forces have been found or predicted to exist are superfluids $(n=2)$ \cite{garcia,zandi2004}, superconductors $(n=2)$ \cite{wil-1}, $XY$ magnets $(n=2)$ \cite{dan-krech,vasilyev2009,hucht2007,hasenbusch2010}, isotropic Heisenberg magnets $(n=3)$ \cite{dan-krech}, spherical model systems  with $n=\infty$ \cite{danchev1996,krech1999,cd2004,dohm2009,dohm2011,diehl2012,DanRud} as well as $O(n)$-symmetric $\varphi^4$ models with $n \geq 2$ \cite{KrDi92a,KrDi92b,GrDi07,dohm2013,dohm2014}.

(i) In the low-temperature phase of such systems with $n>1$, the continuous symmetry (e.g. rotational symmetry of isotropic magnets) is spontaneously broken and transverse fluctuations of the finite order parameter exist, in addition to the longitudinal fluctuations occurring also in $n=1$ systems. The transverse fluctuations (e.g. rotations of fixed-length spin variables) have a vanishing restoring force in the long-wavelength limit and at vanishing external field ${\bf h}$. In infinite bulk systems, this implies the existence of transverse massless Goldstone modes in wave-vector space  and  a power-law decay of the correlation functions in real space which cause infinite transverse and longitudinal susceptibilities for all temperatures $T<T_c$. Likewise, in the presence of a confining geometry of size $L$, this implies a power-law dependence $\propto 1/L^x$ (rather than exponential $L$ dependence) for finite-size effects on thermodynamic quantities well below  $T_c$ \cite{hasen}. In particular, a finite thermodynamic Casimir force $\propto 1/L^3$ in a $^4$He film of thickness $L$ observed far below the superfluid transition \cite{garcia} as well as the low-temperature tails $\propto 1/L^3$ of Monte Carlo (MC) data for the Casimir forces of $XY$ and Heisenberg models in a three-dimensional slab geometry \cite{dan-krech,vasilyev2009,hucht2007,hasenbusch2010} have been attributed to Goldstone modes \cite{dan-krech,zandi2004,vasilyev2009,biswas2010,dohm2013}.

(ii) As the bulk critical temperature $T_c$ is approached, long-range {\it critical} correlations occur for $1\leq n \leq \infty$ due to the divergence of the bulk correlation length \cite{fish-1}
%$\xi_+$
which implies massless critical modes of the bulk system right at $T_c$. In a confining $d$-dimensional geometry of size $L$  this leads to a critical Casimir force $\propto 1/L^d$ at bulk $T_c$ \cite{cd2004,garcia,zandi2004,wil-1,vasilyev2009,hucht2007,hasenbusch2010,dan-krech,danchev1996,krech1999, dohm2009,dohm2011,diehl2012,DanRud,KrDi92a,KrDi92b,GrDi07,dohm2013,dohm2014,fisher78,night,krech,biswas2010,hertlein,rud2010,hucht2011,Jakub,toldin2013}.
This includes also systems with a one-component order parameter such as ordinary fluids \cite{fisher78}, binary fluid mixtures \cite{hertlein}, and Ising-like magnets \cite{rud2010,hucht2011}.

Casimir forces $F_\text{Cas}$ depend significantly on the boundary conditions (BC) and the geometry. In this work the focus is on the case of periodic BC in $d$-dimensional geometries with $2<d<4$ of some characteristic size $L$. We confine ourselves to systems with isotropic short-range interactions where, in the absence of noncubic lattice anisotropies  \cite{dohm2008,cd2004,dohm2006}, a unique second-moment bulk correlation length $\xi_+$ above $T_c$ can be defined, with the asymptotic critical behavior
\begin{equation}
\label{3dxi}
\xi_+(t) = \xi_{0+} t^{- \nu}, \;\;\; t=(T-T_c)/T_c.
\end{equation}
The Casimir force  $F_{{\text Cas}}=-\partial [Lf^{{\text ex}}]/\partial L$
can be derived from the excess free energy density (divided by $k_BT$) $f^{{\text ex}}=f-f_b$ where
$f$ and  $f_b$ are the free energy densities of the confined system and the bulk system, respectively. For large $L$ and small $|t|$, the finite-size scaling form
\begin{eqnarray}
\label{4a11x} F_\text{Cas}(t,L) \;& = &\; L^{-d} \;X(\tilde x),\\
\label{3jjx}
\tilde x\;&=&\;t(L/\xi_{0+})^{1/\nu},
\end{eqnarray}
can be inferred from the hypothesis of  two-scale-factor universality \cite{pri,priv} for the singular part of $f$ which implies that, for a given geometry, the scaling function $X(\tilde x)$ is universal within the subclass of isotropic systems \cite{cd2004,dohm2006,dohm2008} of a given $(d,n)$
bulk universality class.

For the case of isotropic systems in an $ \infty^{d-1} \times L$ film  geometry with periodic BC and for finite $n$, the structure of $(\ref{4a11x})$ has been confirmed and  analytic results have been derived that separately describe either (i) the amplitude  $X(-\infty)$ in the Goldstone-dominated regime deeply in the low-temperature phase for $n=2$ \cite{vasilyev2009,commentDantchev} or (ii) the scaling function $X(\tilde x)$ above bulk criticality for $n\geq 1$ \cite{KrDi92a,GrDi07,kastening-dohm}. An open problem remained, however, with regard to the crossover from $X(-\infty)$ to $X(0)$ in the region  $T \leq T_c$ where the scaling function $X(\tilde x)$ displays a characteristic minimum as detected by MC simulations for $n=1,2,3$ \cite{hucht2011,hasenbusch2010,dan-krech,vasilyev2009} for periodic BC. This lack of theoretical knowledge is related to the notorious difficulty of treating Goldstone modes in confined systems near $T_c$
which, for finite $n$, has been overcome only in a few cases (see, e.g., \cite{CDS1996}). For periodic BC, complete results for $X$ including the crossover between $X(-\infty)$ and $X(0)$ have been derived for the spherical model  and in the large-$n$ limit for a $d=3$ film geometry \cite{danchev1996,dohm2011}.

The results of \cite{KrDi92a,GrDi07,kastening-dohm} agreed well with the MC data well above $T_c$ but their scaling functions $X(\tilde x)$ contained
artificial cusp-like singularities at $T_c$ for general $n\geq 1$  (see Figs. 2-4 and 8) \cite{cusp}. This is due to the basic difficulty of describing the O$(n)$-symmetric systems in an $ \infty^{d-1} \times L$ film geometry in $d$ dimensions.  The problem is the existence of a film transition at a separate critical temperature $0< T_{cf}(L) < T_c$ for $n=1,d>2$, for $n=2,d\geq 3$, and for $n>2,d>3$
where the critical behavior is that of a $(d-1)$-dimensional bulk system.
A satisfactory analytic theory capturing the dimensional crossover from a $d$ dimensional transition at $T_c$ to a $d-1$ dimensional transition at $T_{cf}$ has not been developed so far, except for the case of the Gaussian model \cite{kastening-dohm}. In Sec. VI of this paper we present an exact description of this dimensional crossover in the large-$n$ limit for $d>3$.

The established theory of bulk critical phenomena \cite{fish-1} implies that, at the film transition for $d=3$, a logarithmically divergent slope of $X(\tilde x)$ should occur for $n=1$ \cite{hucht2011}, an  essential singularity for $n=2$ \cite{vasilyev2009}, and no film transition at all for $n>2$ at a finite temperature. These weak singularities for $n=1,2$ were not detected in MC simulations \cite{hucht2011,vasilyev2009,hasenbusch2010,smooth} which were carried out in finite  $L_\parallel^2\times L$ slab geometries with an aspect ratio
$\rho=L/L_\parallel$
where $\rho = 1/6$ \cite{vasilyev2009}, $\rho = 1/8, 1/16$ \cite{hucht2011}, and $\rho = 0.01$ \cite{hasenbusch2010}. The MC data demonstrate that the $\rho$-dependence of the Casimir force is quite weak for $\rho \ll 1$ which leads to the expectation that the shape of $X$ is not significantly changed when the film limit $\rho\to 0$ is taken.

The conclusion is that the problem of dimensional crossover in an idealized $\infty^2 \times L$ film geometry, though being an interesting theoretical topic in its own right, is not of primary relevance to the goal of explaining the shape of the scaling function $X$ observed in MC simulations and real systems where the film singularities of $X$ are not detectable. This conclusion was exploited in previous theoretical work for $n=1$ \cite{dohm2009,dohm2011} and more recently for general $n\geq1$ \cite{dohm2013} where the problems of an infinite film geometry were circumvented by considering a $L_\parallel^{d-1}\times L$ slab geometry.
The basic advantage of this finite geometry is the absence of both Goldstone and critical singularities
at finite temperatures and the existence of a {\it discrete} mode spectrum with a dominant lowest mode that is amenable to a simultaneous analytic treatment of the low-temperature and the critical regions.

Ordinary perturbation theory for finite systems in the sense of an expansion around bulk mean-field theory (see App. B) fails because of unphysical divergencies arising from the isolated lowest ({\bf k}={\bf0}) mode at bulk $T_c$ for $n\geq 1$ and from the massless Goldstone modes at the coexistence line below $T_c$ for $n>1$. A concept
of separating the lowest mode from the higher modes
was formulated within the framework of the $4-\varepsilon$ expansion for the $\varphi^4$ theory \cite{BZ,RGJ} and within the $2+\varepsilon$ expansion for the nonlinear $\sigma$ model \cite{BZ}. This method was further developed \cite{Esser,dohm2008,dohm2009,dohm2011} for $n=1$ within the framework of the minimal renormalization
at fixed dimension $d$ \cite{dohm1985}, without an $\varepsilon$ expansion,
and quantitative agreement with accurate MC data of the $d=3$ Ising model was found \cite{Esser,talapov}.
One of the advantages of the fixed-$d$ theory for finite systems is that it keeps the exponential form of the order-parameter distribution function unexpanded. This ensures that as much (perturbative) information as possible is taken into account whereas an expansion of this exponential form and a subsequent truncation (as is done in the $\varepsilon$ expansions) implies a partial and uncontrolled loss of information. This may lead to less reliable results as shown for the case $n=1$ \cite{Esser,dohm2008}.

An unresolved issue remained for $n=1$ with regard to the low-temperature behavior of
$f^{\text ex}$ in a slab geometry within  the lowest-mode separation approach of \cite{dohm2008,dohm2011}. It was noted in  \cite{dohm2010} that the result of \cite{dohm2008} disagreed with the exact analytic low-temperature result
$f^{\text ex} \approx - V^{-1} \ln 2$
\cite{Privman-Fisher} for the Ising model of volume $V$.  The latter result could be reproduced within the $\varphi^4$ theory by a one-loop expansion for $n=1$ \cite{dohm2010,dohm2011} around the two separate peaks of the order-parameter distribution function far below $T_c$. This result also agreed with earlier MC data of the $d=3$ Ising model
for $\rho=1$ \cite{hasenbusch2009}. Subsequent MC data  for $1/16 \leq \rho \leq 8$ \cite{hucht2011} agreed with
the analytic low-temperature prediction \cite{Privman-Fisher} but so far no comparison has been made between these MC data \cite{hucht2011} and the theory of \cite{dohm2010,dohm2011} in the low-temperature region.

More recently, the lowest-mode separation approach was extended to the general case $n\geq 1$ such that it was possible to treat simultaneously the critical and the Goldstone modes within a finite-size renormalization-group (RG) $\varphi^4$ theory at fixed $d$ \cite{dohm2013}. Unlike the $\varepsilon$ expansion, this approach is not based on a strict expansion in powers of the four-point coupling. A formula  was presented for the scaling function of $f^{\text ex}$ describing the crossover in a slab geometry for an arbitrary aspect ratio $\rho>0$ in $2<d<4$ dimensions for general $n$ in the entire temperature range from far below to far above $T_c$ including the critical and the Goldstone-dominated regions.

It is the purpose of this paper to present a detailed exposition of the RG approach of \cite{dohm2013} for the general case of a $d$-dimensional $ L_1 \times L_2 \cdot \cdot \cdot \times L_d$ block geometry with $L_d\equiv L$ and the aspect ratios
\be
\rho_\alpha=L/L_\alpha, \;\;\; \alpha= 1, 2, ..., d-1,\;\; \rho_d\equiv1,
\ee
and to further analyze the results.
This includes applications to slab geometry ($\rho_\alpha=L/L_\parallel=\rho$, $  \alpha= 1, 2, ..., d-1$) and a comparison with MC data  \cite{hucht2011,dan-krech,vasilyev2009,hasenbusch2010,hasenbusch2011} and with previous analytic work \cite{wil-1,KrDi92a,KrDi92b,GrDi07,dohm2011,Jakub}.
A summary of our main results is given below.

(a)  {\it Two-scale-factor universality for confined systems} \cite{pri}.
In the context of finite-size scaling, this hypothesis  is expressed by the
asymptotic (large $L$, small $t$) scaling form for the
singular part of the free energy density
\begin{equation}
\label{twoscalf}
f_s (t, \{L_\alpha\}) = L^{-d} \;  F (C_1 t L^{1/\nu},\{\rho_\alpha\}),
\end{equation}
with the universal scaling function $ F$  for given BC.
With the choice $C_1= \xi_{0+}^{-1/\nu}$, the structure of our result (\ref{scalfreeaniso}) for $f_s$ for isotropic systems
derived in this paper is in agreement with this hypothesis. The resulting scaling function $X(\tilde x,\{\rho_\alpha\})$  of isotropic systems describes the crossover  from far below to far above $T_c$ for $1\leq n \leq \infty$, $2<d<4$. It is not valid for weakly anisotropic systems as shown in  \cite{dohm2017II} (subsequent paper).

(b) {\it Minimum of the Casimir force scaling function.}
Quantitative predictions of the scaling
function $X(\tilde x, \rho)$ in slab geometry are made for $n = 1, 2, 3$ and $d=3$ from far below to far above $T_c$. For $\rho \ll 1$ a minimum is found at $T_{\text min}$ slightly below $T_c$ that is shifted to $T_{\text min}>T_c$ for $\rho \gtrsim 1/2$.
In Figs. 2-4 and 7 our predictions are compared with MC data for Ising ($n=1$) \cite{hucht2011}, $XY$ ($n=2$) \cite{vasilyev2009,hasenbusch2010}, and Heisenberg ($n=3$) \cite{dan-krech}  models in slab geometries with $0.01\leq\rho\leq1$. Good agreement for small $\rho$ is found for $n=1,2$.

(c) {\it Low-temperature behavior for $n>1$.}
The effect of the Goldstone modes at low-temperatures  in slab  geometry is analyzed. For $d=3$, $n>1$, and $\rho<\rho_0=0.8567$ our theory yields a finite negative (attractive) low-temperature  Casimir force that vanishes for $\rho = \rho_0$ and becomes positive (repulsive) for $\rho > \rho_0$.
This value of  $\rho_0$ is exact for $ n = \infty$ but the independence of $\rho_0$ on $n$ for finite $n$ is attributed to the approximations of our theory.
A logarithmic divergence of the scaling function $F^{\text ex}(\tilde x, \rho)$ of $f^{\text ex}$ for $\tilde x \to - \infty$, $\rho>0$, $n>1$ is predicted, in excellent agreement with MC data \cite{hasenbusch2011} for the $XY$ model (Fig. 3).

(d) {\it Low-temperature behavior for $n=1$.} The low-temperature amplitude $F^{\text ex}(-\infty, \rho)$ for $\rho>0$, (\ref{lowTemplx}), is finite, in contrast to the vanishing low-temperature limit of $F^{ex}$ for  film geometry for $n=1$ [see issue (f) below]. While  $X(\tilde x,\rho)$ agrees reasonably well with MC data \cite{hucht2011} of the $d=3$ Ising model for $\rho \lesssim 1/2$ (Fig. 2), there exist systematic differences of $F^{\text ex}(-\infty, \rho)$ with these data and with an exact analytic prediction \cite{Privman-Fisher,dohm2010} for the Ising model for $T \to 0$ (Fig. 2).
We do not attribute this deviation to a shortcoming of our approximation but rather to the nonuniversal difference between the $\varphi^4$ model with "soft-spin"  variables $\varphi$ and the Ising model with fixed-length spin variables $s=\pm1$. MC studies of  $F^{\text ex}$ for the $d=3$ $ \varphi^4$ model \cite{hasenbusch1999} rather than for the $d=3$ Ising model are desirable for resolving this issue. Our result for $X(\tilde x,\rho)$ and  $F^{\text ex}(\tilde x, \rho)$  constitutes an improvement over that of \cite{dohm2011} well below $T_c$ (see Fig. 11).

(e) {\it $\rho$ dependence at bulk $T_c$.}
An analytic expression is given in Sec. V. E for the monotonic $\rho$-dependence of $F^{\text ex}(0,\rho)$ and $X(0,\rho)$ at $T_c$  for $n\geq 1$. For $n=1$ good agreement with MC data \cite{hucht2011} is found (Fig. 7). Our result for $X(0, \rho)$ vanishes for $\rho = 1, d=3$, in agreement with the proof of \cite{hucht2011}. This proof does not apply to the Gaussian model (see Fig. 7 (b) and App. A) and not to weakly anisotropic systems \cite{dohm2017II}.

(f) {\it Film geometry $\rho \to 0$.} In the film limit (Sec. V. F), our finite-size theory correctly reproduces one-loop perturbation theory for film geometry at fixed $d$ above and below $T_c$ \cite{dohm2017II}. At $T_c$, a deterioration of the quality of our theory is expected \cite{dohm2013} since the separation between the lowest mode and the higher modes goes to zero for $\rho \to 0$. Indeed, if our result (\ref{scalfreeaniso}) derived for $\rho > 0$  is extrapolated to $\rho \to 0$, an artificial cusp at bulk $T_c$ occurs similar to that of previous approximate theories  \cite{wil-1,KrDi92a,GrDi07,kastening-dohm} for film geometry (Fig. 8). Well above $T_c$, the reasonable agreement with the $d=3$ MC data  for $n=1,2$ is remarkable in view of the fact that the computational effort in obtaining our fixed-$d$ RG result is considerably smaller than that for deriving the higher-order $\varepsilon$ expansion results \cite{KrDi92a,GrDi07}. The main achievements of our theory in film geometry are the scaling functions $F^{\text ex}(\tilde x, 0)$ and $X(\tilde x,0)$ for general $n$ {\it below} $T_c$ which are in good agreement with $d=3$ MC data for $n=1$ \cite{hucht2011} and $n=2$ \cite{hasenbusch2010} well away from $T_c$ (Fig. 8).
While the low-temperature amplitudes $F^{\text ex}(-\infty,0)$ and  $X(-\infty,0)$ for $n=1$ vanish, they are finite for $n\geq 2$ and, apart from a factor $(n-1)/n$, identical with the Gaussian critical amplitudes (\ref{FexfilmscalingGauss}) and (\ref{CasimirGaussAmp}) in film geometry, as anticipated for $d=3$ in Eq. (25) of \cite{vasilyev2009}.

(g) {\it Bulk limit.}  Our finite-size RG theory correctly reproduces bulk RG theory at fixed $d$ above and below $T_c$. The universal bulk amplitude ratios implied by (\ref{scalfreeaniso})  for $n = 1, 2, 3$ and $d=3$ are in good agreement with established numerical results \cite{pelissetto}, as shown in Sec. V. B.

(h) {\it Large-$n$ limit.} In Sec. V. G we test the quality of our approximate result for $X(\tilde x,\rho)$ by comparison with the exact result in the large-$n$ limit \cite{dohm2011}. We find that the large-$n$ limit of our approximate low-temperature amplitude $X(-\infty,\rho)/(n-1)$ agrees with the exact result for $n \to \infty$ (Figs. 5 and 10). Reasonable agreement is also found at $T_c$ up to $\rho \sim O(1)$ (Fig. 9) and away from $T_c$ for small  $0<\rho\ll 1$ (Fig. 10) but the agreement deteriorates for $\rho \gtrsim 1/2$ well away from $T_c$. In Sec. VI, an exact description is given for the crossover from a $d$ dimensional transition at $T_c$ to a $d-1$ dimensional transition at a shifted film transition temperature $T_{cf}(L)<T_c$ in the large-$n$ limit for $3<d<4$, with a universal amplitude (\ref{xfilmc}) of the finite fractional shift, in agreement with two-scale-factor universality.

(i) {\it Other finite-size scaling functions.} Our result for the finite-size scaling function of $f_s$ is derived from an order parameter distribution function $\propto \exp [- H^{eff}({{\bm\Phi}}^2) ]$, (\ref{distributionbare}) and (\ref{distribution}), with an exponential form whose exponent can be interpreted as an effective Hamiltonian. As shown in \cite{Esser} for the case $n=1$, the same distribution function determines the finite-size scaling functions of the susceptibility, the specific heat, the order parameter,  and the Binder cumulant. Thus these scaling functions can be calculated for general $n$ above and below $T_c$ parallel to the calculation of $f_s$ presented in this paper.

\renewcommand{\thesection}{\Roman{section}}
\renewcommand{\theequation}{2.\arabic{equation}}
\setcounter{equation}{0}
\section{ ${\bm \varphi^4}$ Model and basic definitions}
We start from the $\varphi^4$ lattice Hamiltonian divided by $k_B T$
\begin{eqnarray}
\label{hamiltonian} H  =   \tilde a^d \Big[\sum_{i=1}^N
\left(\frac{r_0}{2} {\bm \varphi}_i^2 + u_0 ({\bm \varphi}_i^2)^2  - {\bf h} \cdot {\bm \varphi}_i \right) \nonumber\\ +
\sum_{i, j=1}^N \frac{K_{i,j}} {2} ({\bm \varphi}_i - {\bm \varphi}_j)^2
\Big],
\end{eqnarray}
$r_0(T) = r_{0c} + a_0 t, \;\;t = (T - T_c) / T_c$,
with $a_0>0$,
$u_0>0$ where $T_c$ is the {\it bulk} critical temperature.  The
variables ${\bm \varphi}_i \equiv {\bm \varphi} ({\bf x}_i)$ are $n$-component
vectors on $N$ lattice points ${\bf x}_i$ of a $d$-dimensional
simple-cubic lattice with lattice constant $\tilde a$ and with periodic BC.
The couplings $K_{i,j} = K_{j,i}
\equiv K ({\bf x}_i - {\bf x}_j)$ and the temperature variable
$r_0 (T)$ have the dimension of $\tilde a^{- 2}$ whereas the variables
${\bm\varphi}_i$ have the dimension of $\tilde a^{(2-d)/2}$ such that $H$ is
dimensionless. The critical value $r_0 (T_c) = r_{0 c}$ depends on all couplings $K_{i,j}$. The components
$\varphi_i^{(\kappa)} \; , \kappa = 1, 2, \ldots, n$ of ${\bm \varphi}_i$ vary
in the continuous range $- \infty \leq \varphi_i^{(\kappa)} \leq
\infty$.
In the bulk limit (and for an appropriate class of couplings $K_{i,j}$ to be specified below), this model undergoes a second-order phase transition in $d$ dimensions at a finite $T_c$ at ${\bf h}={\bf 0}$ for $n=1, d>1$, for  $n=2, d\geq2$, and for  $n>2, d>2$. For $n=2, d=2$, this is a Kosterlitz-Thouless transition at a temperature $T_{KT}>0$. No finite $T_c$ exists for $n>2$, $d\leq2$. We assume the rectangular block geometry defined in Sec. I  with a volume $V = \prod^d_{\alpha = 1} L_\alpha = N \tilde a^d$.  Applications to a slab geometry with finite $0 < \rho \lesssim O(1)$ and to an infinite film geometry ($\rho \to 0$) with finite $L$ will be given. Our general results also contain the case of cylinder geometry ($\rho\gg 1$) which was treated in \cite{dohm2011}. This case will not be further discussed in this paper.
As far as critical phenomena are concerned, the "soft-spin" model (\ref{hamiltonian}) for $n=1,2,3$ belongs to the same bulk universality classes as the fixed-length spin models of the Ising, XY, and Heisenberg type. This does not imply, however, that these models have the same finite-size behavior at low temperatures far from $T_c$.

The homogeneous ordering field ${\bf h}$ breaks the  O$(n)$ symmetry of $H$. This field is coupled to the spatial average
\begin{eqnarray}
\label{Nullmodenamplitude}
{\bm \Phi}  =   N^{-1}\sum_{i=1}^N {\bm \varphi}_i
\end{eqnarray}
of the variables ${\bm \varphi}_i$. The $n$-component variable ${\bm \Phi}$ plays an important role in our theory in that it represents the amplitude of the lowest (homogeneous) mode.
The direction of  ${\bm \Phi}$ defines a {\it fluctuating} reference axis in ${\bm \varphi}$ space with respect to which the longitudinal and transverse parts of the fluctuating amplitudes ${\bm \sigma}_j$ of the higher-modes will be defined in Sec. IV. The variable ${\bm \Phi}$ has components both parallel and perpendicular to the external field ${\bf h}$. This is in contrast to the statistical average
\begin{equation}
\label{statistical average2} {\bf M}\equiv\langle{\bm \varphi}_j\rangle = \frac{\left[\prod_{i = 1}^N \int
d^n {\bm \varphi}_i \right]{\bm \varphi}_j \exp \left(- H \right)}{\left[\prod_{i = 1}^N \int
d^n {\bm \varphi}_i \right] \exp \left(- H \right)}\;= \langle{\bm \Phi}\rangle
\end{equation}
the orientation of which is parallel to the fixed orientation of ${\bf h}$. Because of the homogeneity of ${\bf h}$ and the periodic BC, $\langle{\bm \varphi}_j\rangle$
is independent of the position ${\bf x}_j$.
The dimensionless partition function and Gibbs free energy per unit volume divided by $k_BT$ are
\begin{eqnarray}
\label{partition} Z(t,  h, \{L_\alpha\}) &=& \Big[\prod_{i = 1}^N \frac{\int
d^n {\bm \varphi}_i}{\tilde a^{n (2-d) / 2}} \Big] \exp \left(- H \right),\\
\label{free energy} f(t,  h, \{L_\alpha\}) &=&  - V^{-1} \ln Z (t, h,\{L_\alpha\}),\;
\end{eqnarray}
where $h$ is the amplitude of the external field ${\bf h}= h\;{\bf e}_h$, with a $n$-component unit vector ${\bf e}_h$. The quantity (\ref{statistical average2}) is  identical with ${\bf M}= -\partial f/\partial {\bf h}$. The bulk Gibbs free
energy density, the order parameter ${\bf M}_b= M_b\;{\bf e}_h$, the excess free energy density, and the Casimir force per unit area in the $d$th (vertical) direction are defined as
\begin{eqnarray}
\label{bulk free}
f_b(t,h) &=&\lim_{ \{L_\alpha\} \to \infty}f(t, h,\{L_\alpha\}),
\\
\label{bulk order par}
M_b(t,h)&=& -\partial f_b(t,h)/\partial  h,\\
\label{2j}
f^{ex}(t, h, \{L_\alpha\}) &=&  f(t,  h, \{L_\alpha\}) - f_b(t,h),\\
\label{2k}
F_\text{Cas}(t,h, \{L_\alpha\}) &=&  - \partial[L
f^{ex}(t, h,\{L_\alpha\}) ]/\partial L
\end{eqnarray}
where the derivative is taken at fixed $ L_1, L_2, ..., L_{d-1}$.
A simplification of our model (\ref{hamiltonian}) is the assumption of a rigid lattice representing a system with a vanishing compressibility. The same assumption is made in models on which previous MC simulations of the Casimir force are based. Fluctuation-induced forces should exist also in solids with a finite compressibility \cite{dohm2011}.

For small $|t|$, $f_b(t, h)$ can be uniquely decomposed into singular and nonsingular parts
\begin{equation}
\label{3c}f_b(t,h) =  f_{b,s}(t,h) + f_{b,ns}(t).
\end{equation}
For large $L_\alpha/\tilde a$ and small $|t|$, a corresponding assumption is made \cite{priv} for
\begin{eqnarray}
\label{3a} f(t,h, \{L_\alpha\}) &=& f_{s}(t,h,\{L_\alpha\}) + f_{ns}(t, \{L_\alpha\})\\
\label{3a} f^{ex}(t, h,\{L_\alpha\}) &=& f^{ex}_{s}(t,h,\{L_\alpha\}) + f^{ex}_{ns}(t,\{L_\alpha\})\;\;\;\;\;\;\;\;
\end{eqnarray}
where $f_{ns}(t, \{L_\alpha\})$ and  $f^{ex}_{ns}(t, \{L_\alpha\})$ are regular functions of $t$  and where $f_{ns}(t, \{L_\alpha\})$ remains regular in the bulk limit, $f_{ns}(t,\{L_\alpha\}) \to f_{b,ns}(t)$, whereas  $f_s(t, h,\{L_\alpha\}) \to f_{b,s}(t,h)$ becomes singular in this limit. Following \cite{pri} we assume that, for periodic BC,  $f_{ns}$ is independent of $L_\alpha$  and  $h$ and is equal to the regular bulk part $f_{b,ns}(t)$ of $f_b(t)$ which implies $f^{ex}_{ns}=0$. This appears to be valid for the $\varphi^4$ theory with finite-range interactions $K_{i,j}$ (but not with long-range correlations \cite{dohm2008}) and is consistent with our results in Sec. IV.
The  critical behavior of  $F_\text{Cas}$ can then be calculated as
$F_\text{Cas} =  - \partial[L f^{ex}_{s}(t,h, \{L_\alpha\}) ]/\partial L$.

In terms of the Fourier components ${\bm \hat \varphi}({\bf k})= \tilde a^d\sum_{j=1}^N e^{-i {\bf k} \cdot {\bf x}_j}{\bm \varphi}_j $ the Hamiltonian reads
\begin{eqnarray}
\label{hamiltonian fourier} H  = V^{-1}  \sum_{{\mathbf k}} \frac{1}{2} [r_0 + \delta
\widehat K ({\mathbf k})] {\bm \hat \varphi}({\mathbf k}) \cdot {\bm \hat
\varphi}({-{\mathbf k}}))   -  {\bf h} \cdot {\bm \hat
\varphi}({{\mathbf 0}})\nonumber\\ + \frac{u_0}{V^3}
\sum_{{\mathbf{kp}}{\mathbf q}} [{\bm \hat \varphi}({\mathbf k})\cdot {\bm \hat
\varphi}({{\mathbf p}})] [{\bm \hat \varphi}({{\mathbf q}})\cdot {\bm \hat
\varphi}({-{\mathbf k}-{\mathbf p}-{\mathbf q}})] \qquad
\end{eqnarray}
with $\widehat K ({\bf k}) = N^{-1} \sum^N_{i,j = 1}K_{i,j}e^{-i {\bf k} \cdot ({\bf x}_i-{\bf x}_j)}$, $ \delta \widehat K({\bf k}) = 2 [\widehat K({\bf 0}) -
\widehat K ({\bf k})]$. The summations $\sum_{\bf k}$ run over $N $ discrete vectors
${\bf k} \equiv (k_1, k_2, \ldots, k_d)$ of the first Brillouin zone of the reciprocal
lattice including ${\bf k}={\bf 0}$ with ${\bm \hat \varphi}({\bf 0}) =\tilde a^d\sum_{j=1}^N {\bm \varphi}_j= V {\bm \Phi}$.
We assume finite-range interactions $K_{i,j}$ where
$\widehat K ({\bf 0}) = N^{-1} \sum^N_{i,j = 1}K_{i,j}$
has a finite limit for $N \to \infty$. In this paper we consider systems with an isotropic interaction at $O(k^2)$,
\begin{eqnarray}
\label{2hh} \delta \widehat K ({\bf k}) = c_0 {\bf k}^2 + \;\;O(k_\alpha k_\beta k_\gamma k_\delta)
\end{eqnarray}
with the standard choice $c_0=1$. (A constant $0<c_0\neq 1$ can be eliminated  by a simple transformation.) The bulk order-parameter correlation function  and the second-moment bulk correlation length at $h=0$ above $T_c$ are defined as
\begin{eqnarray}
\label{2l} G_b ({\bf x_i-x_j}, t) &=& \lim_{V \to \infty}
<{\bm \varphi}_i \cdot {\bm \varphi}_j>,\\
\label{3b} \xi_+(t) &=& \left(\frac{1} {2d} \frac{\sum_{\bf
x}\; {\bf x}^2 \;G_b ({\bf x}, t)}{\sum_{\bf x}\; G_b
({\bf x}, t)} \right)^{1/2} \;
\end{eqnarray}
with the asymptotic critical behavior (\ref{3dxi}).
The amplitude $\xi_{0 +}$ will
be needed as a reference length in the renormalized
theory. The assumption of isotropy at $O(k^2)$ is a significant restriction of the theory as shown in \cite{dohm2017II}.

The goal of the subsequent sections is to derive the universal finite-size scaling forms
\begin{eqnarray}
\label{4a1} f_s (t,0, \{L_\alpha\}) \; &=& \; L^{-d} \; F (\tilde x,\{\rho_\alpha\}),\\
\label{4a11} F_{Cas}(t,0,\{L_\alpha\}) \; &=& \; L^{-d} \;X (\tilde x,\{\rho_\alpha\}),
\end{eqnarray}
with the scaling argument (\ref{3jjx}).
The bulk part $F^\pm_b(\tilde x)$ of  $F(\tilde{x},\{\rho_\alpha\})$  is obtained in the limit of large $|\tilde x|$ as $L^d f_{s}(t,0, \{L_\alpha\}) \to F^\pm_b(\tilde{x})$ with
\begin{eqnarray}
\label{3jjbulk}
F^\pm_b(\tilde{x}) =\left\{
\begin{array}{r@{\quad \quad}l}
                         \; Q_1 \tilde x^{d \nu}\quad          & \mbox{for} \;T > T_c\;, \\
                         \; (A^-/A^+)Q_1\mid\tilde{x}\mid^{d\nu}& \mbox{for} \;T <
                 T_c \;
                \end{array} \right.
\end{eqnarray}
where $Q_1$   and the specific-heat ratio $A^-/A^+$ are universal quantities \cite{priv}. This implies
\begin{eqnarray}
\label{3k}
&&f^{ex}_{s}(t,\{L_\alpha\})= L^{-d}F^{ex}(\tilde{x},\{\rho_\alpha\}),\\
\label{3kkx}
&&F^{ex}(\tilde{x},\{\rho_\alpha\})= F(\tilde{x},\{\rho_\alpha\}) - F^\pm_b(\tilde{x}),\\
\label{3nn}
&&X(\tilde{x},\{\rho_\alpha\}) =(d-1)  F^{ex} (\tilde{x},\{\rho_\alpha\})\nonumber \\ && -
\frac{\tilde{x}}{\nu}
\frac{\partial F^{ex}(\tilde{x},\{\rho_\alpha\})}{\partial\tilde{x}}- \sum_{\beta=1}^{d-1}\rho_\beta
\frac{\partial F^{ex}(\tilde{x},\{\rho_\alpha\})}{\partial \rho_\beta}.\;\;\;
\end{eqnarray}
\renewcommand{\thesection}{\Roman{section}}
\renewcommand{\theequation}{3.\arabic{equation}}
\setcounter{equation}{0}
\section{Bulk critical behavior}
The purpose of this section is to determine the bulk Gibbs free energy density $f_b$,
to define the  RG framework that will be employed in the finite-size calculations,
and to discuss two-scale-factor universality.
\subsection{Unrenormalized bulk free energy density}
In one-loop order, the unrenormalized  perturbation expression of the bulk Helmholtz free energy density  is
\begin{eqnarray}
\label{Helmholtz one loop} g_b(t,M_b) = (1/2)r_0 M_b^2+u_0  M_b^4- [n/(2 \tilde a^d)]\ln (2 \pi) \nonumber\\+ (1/2)\widehat{\cal I}(\tilde r_{0{\rm L}})
+ [(n-1)/2]\widehat {\cal I}( \tilde r_{0{\rm T}}) + O(u_0)\;\;\;\;\;
\end{eqnarray}
with the longitudinal and transverse parameters
\begin{subequations}
\label{Longtrans}
\begin{align}
\label{Long}
\tilde r_{0{\rm L}}=r_0+12u_0 M_b^2 ,
\\
\label{Trans}
\tilde r_{0{\rm T}}=r_0+4u_0 M_b^2.
\end{align}
\end{subequations}
The bulk integral
\begin{eqnarray}
\label{bulklog}
\widehat{\cal I}(r)&=&\int_{{\bf k}} \ln \{[  r + \delta \widehat K (\mathbf k)]
\tilde a^2\},\\
\label{kintegral}
\int_{\bf k} \; &\equiv& \;
\prod^d_{\alpha = 1} \; \int^{\pi / \tilde a}_{- \pi /
\tilde a} \; \frac{d k_\alpha}{2\pi} \; ,
\end{eqnarray}
has finite lattice cutoffs $\pm \pi / \tilde a$ for each $k_\alpha$.
It can be decomposed as
\begin{subequations}
\label{bulkintegral}
\begin{align}
\label{bulkdecom}
 \widehat{\cal I}(r)=\widehat{\cal I}_1+\widehat{\cal I}_2(r)+\widehat{\cal I}_3(r),
\\
\label{bulk1}
 \widehat{\cal I}_1  =\int_{\bf k} \ln [ \delta \widehat K (\mathbf k)\tilde a^2],
 \\
\label{bulk2}
 \widehat{\cal I}_2(r)  =r \int_{\bf k} [\delta \widehat K (\mathbf k)]^{-1},
 \\
\label{bulk3}
  \widehat{\cal I}_3(r)=\int_{\bf k} \Big[ \ln\Big(1+\frac{r}{\delta \widehat K (\mathbf k)}\Big)-\frac{r}{\delta \widehat K (\mathbf k)}\Big]. \;\;
\end{align}
\end{subequations}
Unlike (\ref{bulk1}) and (\ref{bulk2}), the last integral is finite for $\tilde a \to 0$, $2<d<4$.
For $0\leq r \tilde a^2 \ll 1$, $2<d<4$  it is evaluated for the isotropic interaction (\ref{2hh}) as
\begin{eqnarray}
\label{bulkintsing}   \widehat{\cal I}_3(r)= -2 A_d\; r^{d / 2}/\;(d\;\varepsilon),\;\;
\end{eqnarray}
apart from additive corrections that vanish for $ r \tilde a^2 \rightarrow 0_+$.
The geometric factor $A_d$ is \cite{dohm1985}
\begin{eqnarray}
\label{geometric factor} A_d \;=\; \Gamma(3-d/2)\;[2^{d-2} \pi^{d/2}
(d-2)]^{-1}.
\end{eqnarray}
The integral (\ref{bulkintsing}) describes the unrenormalized bulk critical behavior for $r=\tilde r_{0{\rm L}} \to 0$ and the non-analytic (but non-divergent) behavior due to the Goldstone modes for $r=\tilde r_{0{\rm T}} \to 0$. We need to decompose $g_b$, (\ref{Helmholtz one loop}), into nonsingular and singular parts which are expressed as functions of $r_0 - r_{0c}$ rather than $r_0$ where $r_{0c}\sim O(u_0)$ is the critical value of  $r_0$.
In order to identify $r_{0c}$ we consider the contribution
\begin{eqnarray}
\label{termernullc}
&&\frac{1}{2}r_0M_b^2+ \frac{1}{2}\widehat{\cal I}_2(\tilde r_{0{\rm L}})
+ \frac{n-1}{2} \widehat{\cal I}_2( \tilde r_{0{\rm T}})=\frac{n}{2}r_0 \int_{\bf k} [\delta \widehat K (\mathbf k)]^{-1}\nonumber \\&&+ \frac{1}{2}\bigg\{r_0+4(n+2) u_0 \int_{\bf k} [\delta \widehat K (\mathbf k)]^{-1}\bigg\}\;M_b^2
\end{eqnarray}
where the term $\propto  M_b^2$ can be written as $\frac{1}{2}(r_0-r_{0c}) M_b^2$. The quantity
\begin{eqnarray}
\label{rnullc}
r_{0c}=-4(n+2) u_0 \int_{\bf k} [\delta \widehat K (\mathbf k)]^{-1}
\end{eqnarray}
is identical with the critical value of $r_0$ up to $O(u_0)$ at which
the inverse bulk susceptibility $\chi_b^{-1}$ \cite{dohm1985,cd1999}  vanishes as $r_0-r_{0c} \to 0_+$ for $h= 0$.
In the spirit of perturbation theory up to $O(1)$, it is justified to replace the arguments $\tilde r_{0{\rm L}}$ and $\tilde r_{0{\rm T}}$ of the integrals $\widehat {\cal I}$ in (\ref{Helmholtz one loop})  by
\begin{subequations}
\label{Longtransx}
\begin{align}
\label{Longx}
r_{0{\rm L}}=r_0-r_{0c}+12u_0M_b^2 ,
\\
\label{Transx}
r_{0{\rm T}}=r_0-r_{0c}+4u_0 M_b^2.
\end{align}
\end{subequations}
As a consequence, $g_b(t, M_b)$ can be expressed entirely in terms of $r_0 - r_{0c}$  as
\begin{eqnarray}
\label{helmholtz} g_b(t,M_b) = g^{(1)}_{b,ns}(t)+ g_{b,s}(t,M_b)
\end{eqnarray}
with the nonsingular bulk part
\begin{eqnarray}
\label{free bulk nonsing}
g^{(1)}_{b,ns}(t)= (n/2) \Big\{ -\tilde a^{-d}\ln (2 \pi) +\int_{\bf k} \ln \{[ \delta
\widehat K (\mathbf k)]  \tilde a^2\} \nonumber\\+ \;\; (r_0 - r_{0c}) \;
\int_{\bf k} [\delta \widehat K (\mathbf k)]^{-1}\Big\}+ O(u_0),\;\;\;\;
\end{eqnarray}
and the singular bulk part for isotropic interactions
\begin{eqnarray}
\label{helmholtz sing} g_{b,s}(t,M_b) = (1/2)(r_0-r_{0c})M_b^2+u_0  M_b^4 \nonumber\\-
[A_d/(\;d\;\varepsilon)] \;\big[ r_{0{\rm L}}^{d/2}+ (n-1) r_{0{\rm T}}^{d/2}\big]+ O(u_0)\;\;.\;\;
\end{eqnarray}
The nonanalytic dependence of $g_{b,s}$ on $ r_{0{\rm T}}$ through $ r_{0{\rm T}}^{d/2}$ leads to divergencies in the derivatives of the free energy arising from the Goldstone modes for $ r_{0{\rm T}}\to 0$ \cite{Burnett,str1999,str2003}.

In deriving the unrenormalized Gibbs free energy density
\begin{equation}
\label{Gibbs}
f_b(t,h)= g_b(t,M_b) - h M_b
\end{equation}
we need  $M_b$ as a function of $t$ and $h$.
For $h = 0$ above $T_c$ we have $M_b(t,0)=0$ and $ r_{0{\rm L}}=r_{0{\rm T}}=r_0-r_{0c}$ which yields the $n$ - dependent singular bulk part for $r_0 - r_{0c} > 0$
\begin{eqnarray}
\label{free bulk above sing}
f_{b,s}(t, 0)=  -n  A_d\; (r_0 - r_{0c})^{d/2}/\;(d\;\varepsilon)+ O(u_0),
\end{eqnarray}
up to corrections that  vanish for $ (r_0 - r_{0c})\tilde a^2 \rightarrow 0_+$.
For $h\neq 0$, $ M_b$ is determined via the bulk equation of state
\begin{eqnarray}
\label{orderparameter}
h= \partial g_b(t, M_b)/\partial M_b\;.
\end{eqnarray}
From  (\ref{Longtrans}), (\ref{helmholtz sing}), and  (\ref{orderparameter}) we obtain for $h\neq0$
\begin{eqnarray}
\label{implicit1}
&&M_b^2=\big[ -(r_0-r_{0c}) + \chi_{{\rm T}}^{-1}\big]/(4u_0) \nonumber \\
&&+(A_d/\varepsilon) \; \big[3  r_{0{\rm L}}^{(d-2) / 2}+ (n-1) r_{0{\rm T}}^{(d-2) / 2}\big]+ O(u_0)\;\;\;\;\;\;\;\;\;
\end{eqnarray}
which implicitly determines $M_b(t,h)$.
The dependence of $ M_b^2$ on $h$ enters only via the transverse bulk susceptibility
$\chi_{{\rm T}}= \; M_b/h$.
Equations (\ref{Longtrans}), (\ref{helmholtz sing}), and  (\ref{implicit1}) lead to
\begin{eqnarray}
\label{gibbsh}
f_b(t,h) = f^{(1)}_{b,ns} (t) +  f_{b,s} (t,h)
\end{eqnarray}
with the nonsingular bulk part
$f^{(1)}_{b,ns} (t)\equiv g^{(1)}_{b,ns}(t)$
and with the singular bulk part for $h\neq0$
\begin{eqnarray}
\label{free bulk below sing}
f_{b,s}(t, h)=- h M_b +   \big[\chi_{{\rm T}}^{-2}-(r_0 - r_{0c})^2\big]/(16 u_0) \nonumber \\ +\chi_{{\rm T}}^{-1}A_d\big[3  r_{0{\rm L}}^{(d-2) / 2}+ (n-1) r_{0{\rm T}}^{(d-2) / 2}\big]/(2\;\varepsilon)\;\nonumber\\-
A_d \;\big[ r_{0{\rm L}}^{d/2}+ (n-1) r_{0{\rm T}}^{d/2}\big]/(d\;\varepsilon)+ O(u_0).\;\;\;\;
\end{eqnarray}
We have presented here the leading dependence on $\chi_{{\rm T}}^{-1}$ in order to demonstrate that there are no divergencies due to Goldstone modes for the bulk free energy below $T_c$ for $h\to 0$ corresponding to $\chi_{{\rm T}}^{-1}\to 0$, in which limit (\ref{free bulk below sing}) yields the finite result for $r_0 - r_{0c} < 0$
\begin{eqnarray}
\label{free bulk below sing zero}
&& f_{b,s}(t,0) = - [-2(r_0 - r_{0c})]^2/(64 u_0) \;\nonumber\\ &&-
A_d \;[-2 (r_0 - r_{0c})]^{d/2}/(d\;\varepsilon) + O(u_0),\;\,
\end{eqnarray}
apart from corrections that vanish for $(r_0 - r_{0c})\tilde a^2  \rightarrow 0_-$.  The absence of Goldstone singularities is expected on general grounds \cite{david} for $O(n)$-symmetric quantities  (such as free energy, square of the order parameter, specific heat) in contrast to the true physical divergencies of the transverse and longitudinal susceptibilities for $h\to  0$ \cite{Burnett}. Spurious singularities at intermediate stages of perturbation theory for the Gibbs free energy below $T_c$ are known to appear at higher order which, however, have been shown to cancel among themselves \cite{Burnett,str1999,str2003}.

As expected, the unrenormalized expressions  given in Eqs. (\ref{free bulk above sing}) - (\ref{free bulk below sing zero}) do not yet correctly describe the leading bulk power laws of critical behavior. These shortcomings will be removed by turning to the renormalized theory.
Within unrenormalized perturbation theory, $ f_{b,s}(t,0)$ below $T_c$ is independent of $n$ up to $O(1)$, i.e.,  the transverse modes do not contribute to the {\it bulk} part $ f_{b,s}(t,0)$ below $T_c$ at $O(u_0^{-1})$ and $O(1)$. An $n$-dependence below $T_c$, however, enters through the subsequent renormalizations. Furthermore, we shall find that an explicit $n$-dependence appears in the {\it finite-size} part of the free energy at $O(1)$ of unrenormalized perturbation theory.
\subsection{Renormalization }
The singular part $ f_{b,s}$ of the bulk Gibbs free energy density at $ h= 0$ given in Eqs. (\ref{free bulk above sing}) and (\ref{free bulk below sing zero})  will be denoted by $\delta f_b(r_0 - r_{0c}, u_0)$. Within the minimal subtraction scheme at fixed dimension \cite{dohm1985} the renormalized quantities are defined as
\begin{eqnarray}
\label{5dx}   f_{R,b}(r, u,\mu) &=& \delta f_b(Z_rr,
       \mu^{\varepsilon}Z_{u}Z_{\varphi'}^{-2}A_d^{-1}u)\nonumber\\  &&- (1/8)\mu^{-\varepsilon} r^2 A_d
       A(u,\varepsilon)\; ,\\
\label{renormalized parameters}
\label{5a} u &=& \mu^{- \varepsilon} A_d Z_{u}(u, \varepsilon)^{-1} Z_{\varphi}(u, \varepsilon)^2 u_0,\;\;\;\;
\\
\label{rr}
r &=& Z_r(u, \varepsilon)^{-1} (r_0 - r_{0c}) ,
\\
{\bm \varphi}_R &=&
Z_{\varphi}(u, \varepsilon)^{- 1/2} {\bm \varphi},
\end{eqnarray}
with an arbitrary inverse reference length $\mu$. In view of the application to finite systems, we are using
$r$ rather than the bulk correlation lengths above and below $T_c$ $\xi_\pm$ \cite{str2003} as the appropriate
measure of the temperature variable. Using $r$ rather
than $\xi_\pm$ is advantageous in our finite-size theory where a
{\it single} analytic finite-size scaling function is derived for the whole temperature regime $-\infty \leq r \leq \infty$.
The $n$-dependent renormalization constants read up to one-loop order
\begin{subequations}
\label{renormalization constants}
\begin{align}
\label{Zr}
Z_r(u, \varepsilon) = 1 + 4(n+2) u / \varepsilon + O(u^2),
\\
\label{Zu}
Z_{u} (u,\varepsilon) = 1 + 4(n+8) u / \varepsilon + O(u^2),
\\
\label{Zphi}
Z_{\varphi} (u,\varepsilon) = 1 + O(u^2),
\\
\label{A}
A(u,\varepsilon)= - 2n/\varepsilon + O(u).
\end{align}
\end{subequations}
The same renormalization constants will be employed in our finite-size theory. Although they are primarily defined such that they  absorb the {\it ultraviolet} divergences at $d=4$ dimensions, they simultaneously govern the infrared (critical) singularities for $d\leq4$ \cite{dohm1985} via the field-theoretic functions $\zeta_r(u), \beta_u(u,\varepsilon),\zeta_\phi(u),  B(u)$ derived from $Z_i(u,\varepsilon)$ and $A(u,\varepsilon)$.
The resulting renormalized free energy density above and below $T_c$ is
\begin{eqnarray}
\label{bulk f+}
&& f_{R,b}(r, u,\mu)
=-n A_d\varepsilon^{-1}r^{d/2}
\big[d^{-1}-\big(r/\mu^2\big)^{\varepsilon/2}/4\big],\;\;\;\;\;\;\;\;\;\\
\label{bulk f-}
 &&f_{R,b}(r, u,\mu) = -\mu^d A_d\big(-2r/\mu^2\big)^2/(64 u)\nonumber \\&&-A_d\varepsilon^{-1}(-2r)^{d/2}
\big[d^{-1}-\big(-2r/\mu^2\big)^{\varepsilon/2}/4\big],\;\;\;
\end{eqnarray}
for $r>0$ and $ r<0$, respectively. $f_{R,b}$ has a finite
limit for $\varepsilon\rightarrow0_+$ at fixed $u>0$ and fixed $r$.
The dimensionless amplitude function
\begin{eqnarray}
\label{5i} F_{R,b}(r/\mu^2, u) =
 \mu^{-d} A_d^{-1} f_{R,b}(r,u,\mu)
\end{eqnarray}
satisfies the renormalization-group equation (RGE)
\begin{eqnarray}
\label{5j}
&&(\mu\,\partial_\mu+r\zeta_r\partial_r+\beta_{u}\partial_{u}+d)
   F_{R,b}(r/\mu^2,  u) \nonumber \\&&= -
  [r^2/(2 \mu^4)] B(u).\;\;\;\;\;\;\;
\end{eqnarray}
Integration of the RGE yields
\begin{eqnarray}
\label{5aaa}&& f_{R,b}(r, u,\mu)  = f_{R,b} \big(r(l),u(l),l\mu\big)  + \;\frac{A_d r(l)^2}{2(l\mu)^\varepsilon}
{\cal B}(l),\;\;\;\;\;\;\;\;\;\\
\label{r}
&&r(l) = r \exp\Big[\int_1^l\zeta_r(u(l'))\frac{dl'}{l'}\Big],
\\
\label{u}
&&l\big[du(l)/dl\big] = \beta_{u}(u(l),\varepsilon),\\
\label{calB}
&&{\cal B}(l) \equiv
\int_1^l
B(u(l'))\Big\{\exp\int_l^{l'}\Big[2\zeta_r(u(l'')) -
\varepsilon\Big]\frac{dl''}{l''}\Big\}\frac{dl'}{l'}, \nonumber\\
%\end{align}
%\end{subequations}
\end{eqnarray}
with $r(1)=r=at$, $a=Z_r(u,\varepsilon)^{-1}a_0$, $u(1)=u$. Asymptotically ($l \to 0$) we obtain the fixed point value $u^*= u(0)$ determined by $\beta_{u}(u^*,\varepsilon)=0$.
For the purpose of calculating scaling functions it is necessary to make an appropriate choice of the arbitrary reference length $\mu^{-1}$. Within the minimal subtraction approach at fixed dimension there exists the following exact relation between $\mu$ and the asymptotic amplitude $\xi_{0+}$ of the second-moment bulk correlation length (\ref{3dxi}) above $T_c$ \cite{dohm1985}
\begin{eqnarray}
\label{corrampl} \xi_{0 +} =  \frac{\mu^{2\nu-1}}{a^{\nu}}\Bigg[ Q^*
\exp \left(\int_{u}^{u^*} \frac{\zeta_r (u^*) - \zeta_r
(u')}{\beta_{u} (u', \varepsilon)}\, d u' \right)\Bigg]^\nu\;\;\;\;\;\;
\end{eqnarray}
where the dimensionless amplitude $ Q^* = Q (1, u^*, d)$ is the fixed point value of the $n$ dependent amplitude function $Q (1, u, d)$ related to $\xi_{+}$ \cite{dohm1985}.
As a natural choice for the reference length
$\mu^{-1}$ we take
\begin{eqnarray}
\label{my}
\mu^{-1} = \xi_{0 +}
\end{eqnarray}
which implies the exact representation \cite{dohm1985}
\begin{eqnarray}
\label{correlation amplitude}
\xi_{0 +}^2 =  a^{-1} Q^*
\exp \left(\int_{u}^{u^*} \frac{\zeta_r (u^*) - \zeta_r
(u')}{\beta_{u} (u', \varepsilon)}\, d u' \right).
\end{eqnarray}
Note that $\xi_{0 +}$ is a function of $a_0$ and $u_0$ through $a$ and $u$. In an exact theory the choice of the bulk flow parameter $l(t)$ is arbitrary. In a perturbative treatment of $f_{R,b}$, the right-hand side of (\ref{5aaa}) depends weakly on the choice of $l$. It can be chosen such that (\ref{5aaa}) provides a mapping from the critical region to the noncritical region where the perturbative treatment is valid. The function (\ref{5aaa}) contains the singular part of the Gibbs free energy density whose asymptotic ($l \to 0$) form is given in (\ref{asymfree}) - (\ref{6n}, after the flow parameter has been specified.

The asymptotic ($l \to 0$) form of $r(l)$, (\ref{r}) is, apart from Wegner corrections \cite{wegner1972},
\begin{eqnarray}
\label{rl}
r(l) \to l^{2-1/\nu}\;r \exp\left(\int_{u}^{u^*} \frac{\zeta_r
(u')-\zeta_r (u^*)}{\beta_{u} (u', \varepsilon)}\, d u' \right),
\end{eqnarray}
where $\zeta_r (u^*)= 2-1/\nu$ with the critical exponent $\nu$.
Eqs. (\ref{my})-(\ref{rl}) imply the asymptotic behavior
\begin{eqnarray}
\label{rasym}
r(l)/(\mu^2 l^2) \to Q^* t l^{-1/\nu}.
\end{eqnarray}
The most convenient choice of the bulk flow parameter $l_+(t)$ and $l_-(t)$ is made by requiring \cite{dohm1985}
\begin{eqnarray}
\label{bulk flow parameter}  \mu^2l^2 = \left\{
\begin{array}{r@{\quad \quad}l}
                         \mu^2 l_+^2 = r(l_+)& \mbox{for} \;T > T_c ,\\
                         \mu^2 l_-^2 = -2r(l_-) & \mbox{for} \;T <
                 T_c .
                \end{array} \right.
\end{eqnarray}
Equations (\ref{free bulk above sing}), (\ref{free bulk below sing zero}), and (\ref{5dx})-(\ref{renormalization constants}) then lead to $f_{R,b} \big(r(l),u(l),l\mu\big)\equiv f^\pm_{R,b} \big(l_\pm \mu,u(l_\pm)\big)$ with
\begin{eqnarray}
\label{5cc} &&f^+_{R,b} \big(l_+ \mu, u (l_+) \big) \; = \; - n\;
A_d (l_+ \mu)^d / (4 d), \\
\label{5dd}
&& f^-_{R,b} \big(l_- \mu, u (l_-)\big) \; =
%\nonumber\\
- A_d (l_- \mu)^d  \Big[ \frac{1}{64 u (l_-)} \; + \;
\frac{1} {4d}
\Big], \;\;\;\;\;\;\;\;\;\;
\end{eqnarray}
above and below $T_c$, respectively, apart from corrections of $O(u)$. Note that no explicit $n$-dependence appears  in  (\ref{5dd}).
Eqs. (\ref{my}) - (\ref{bulk flow parameter}) imply asymptotically
\be
\label{bulk flow} \mu l = \left\{
\begin{array}{r@{\quad\quad}l}
                \mu l_+ = Q^{* \nu} \; \xi_{0+}^{-1} t^\nu \;   & \mbox{for} \;\;\;  T > T_c\;, \\
\mu l_- = Q^{* \nu} \; \xi_{0+}^{-1} (2 | t |)^\nu & \mbox{for}
\;\;\;
                 T < T_c \;.
                \end{array} \right.
\ee
The integral (\ref{calB}) has the asymptotic ($l\rightarrow0$) behavior
${\cal B}(l) \rightarrow\; - \;(\nu/\alpha) \;B(u^*)$,
apart from a subleading term that contributes to the nonsingular bulk part
$f^{(2)}_{ns,b} \propto |t|^2$  \cite{dohm2008,str2003}.
We then obtain from (\ref{5aaa}), (\ref{5cc}), and (\ref{5dd}) the asymptotic form of the singular part $f_{s,b}(t,0)\equiv f_{s,b}^\pm(t)$ of the Gibbs free energy density of the isotropic system at $h =  0$
\begin{eqnarray}
\label{asymfree}
&&f_{R,b} (r, u, \mu ) \to f_{s,b}^\pm(t)=A^\pm|t|^{d \nu}\\
\label{6m}
&& A^+  =  - \; A_d \; Q^{*d\nu} \;
\left[n/(4d) + \nu B (u^*)/(2 \alpha)  \right]
\xi_{0+}^{-d} ,\;\;\;\;\;\;\;\;\;\;\\
\label{6n}
&& A^-  =  -  A_d \; (2Q^*)^{d\nu} \left[\frac{1}{64 u^*}  +
 \frac{1}{4d}  +  \frac{\nu B(u^*)}{8
\alpha}   \right]  \xi_{0+}^{-d}  \;\;\;\;\;\;\;\;\;\;
\end{eqnarray}
above and below $T_c$, respectively. In addition to the explicit  $n$-dependence in (\ref{6m}), an $n$-dependence of both  $A^+$ and  $A^-$ enters through the fixed point value $u^*$, through the critical exponents $\alpha$ and $  \nu$, and through the $n$ dependent functions $B(u^*)=n/2 + O(u^{*2})$ and $Q^*=Q(1, u^*, d)=1 + O(u^{*2})$.
In order to test our finite-size theory we shall compare it in Sec. VI with the exactly solvable case $n \to \infty$ at fixed $u_0 n$. The critical exponents and the universal bulk amplitude ratios
are well known in this limit \cite{priv}.
Within the minimal subtraction scheme at fixed $d$ \cite{dohm1985} we find
\begin{subequations}
\label{renormalization-constants-large-n}
\begin{align}
%label{Zr-large-n}
Z_r(u, \varepsilon) = Z_{u} (u,\varepsilon) = (1-4un/\varepsilon)^{-1},
\\
\label{A-large-n}
\lim_{n \to \infty} A(u,\varepsilon)/n = 0,
\end{align}
\end{subequations}
$Z_{\varphi} = 1$,  and the fixed-point values $\lim_{n \to \infty} u^*n=\varepsilon/4$ and $\lim_{n \to \infty} Q^* = 1$ \cite{dohm1985}.
\subsection{Two-scale-factor universality }
The singular bulk part $f_{b, s}$ has the asymptotic (small $t$, small $ h$) scaling form for general $n$ \cite{pri}
\begin{equation}
\label{3a} f_{b, s} (t, h) = A_1 |t|^{d \nu} \; W_\pm (A_2 h
|t|^{- \beta \delta})
\end{equation}
with  the universal scaling function $W_\pm (z)$. We use the normalization $W_+ (0) = 1$. The two amplitudes $A_1\equiv A^+$ and $A_2$ are nonuniversal. The hypothesis of two-scale-factor universality \cite{stau,hohenberg1976,weg-1} states that the nonuniversal amplitudes contained in the bulk correlation function $G_b$ near $T_c$ are fully determined once the amplitudes $A_1$ and $A_2$ have been chosen. In \cite{dohm2008} this issue was discussed for general $n$ above $T_c$ and $n=1$ below $T_c$, and results for the universal constants $Q_1$ and $A^+/A^-$ were given for $n=1$. Here we extend the discussion to $n>1$.

At $h=0$ the scaling form of $G_b$ near $T_c$ is for general $n$ above $T_c$ $(+)$ and for $n=1$ below $T_c$ $(-)$
\begin{eqnarray}
\label{3nbarAhNull} G_b ({\bf x}, t) = D_1|{\bf x}|^{2- d - \eta}\;
 \Phi_\pm \big(|{\bf x}|/ \xi_\pm(t)\big),\;\;\;\;\;
\end{eqnarray}
with $\xi_\pm(t)=\xi_{0\pm}|t|^{-\nu}$ and the universal scaling functions $\Phi_\pm$. This is valid for $\xi_\pm\gg \tilde a, |{\bf x}|\gg \tilde a$ at fixed $|{\bf x}|/ \xi_\pm$,  but not for large ${\bf x}$  at fixed $T\neq T_c$ (Fig. 2 of \cite{dohm2008}, shaded region) where the exponential decay of $G_b$ is governed by a "true" or "exponential" correlation length $\xi_{{\bf e}\pm}(t)\neq\xi_\pm(t)$
\cite{dohm2008,fish-2,cd2000-2,floeter,pelissetto}.
The ratio $\lim_{t \to 0}\xi_{{\bf e}\pm}(t)/\xi_\pm(t)$ is  universal \cite{floeter,pelissetto} and equal to $1$ for $n=\infty$ \cite{cd2000-2}.

The transverse correlation function $G_{b,{\rm T}}$ has an algebraic decay due to the Goldstone modes for large $|{\bf x}|$ \cite{hohenberg1976,pri}
\begin{eqnarray}
\label{trans correl large}
G_{b,{\rm T}} ({\bf x}, t) &\approx & {\cal C}_{\rm T}  M_b(t)^2 \big[\xi_{\rm T}(t)/|{\bf x}|\big]^{d-2},\\
%\end{eqnarray}
%
%
%\begin{eqnarray}
\label{constT}
{\cal C}_{\rm T}& = &\Gamma(d/2)/[2\pi^{d/2}(d-2)],
\end{eqnarray}
for $n>1$ below $T_c$  where $\xi_{\rm T}(t) = \xi_{0\rm T} |t|^{- \nu}$ is the transverse correlation length near $T_c$ and $ M_b(t)=B |t|^{\beta}$ is the order parameter. The same correlation length governs the algebraic decay of the longitudinal correlation function, see Eq. (2.48) of \cite{dohm2017II}. These long-ranged correlations are the origin of the finite Casimir force at low temperatures, as given in (\ref{lowXn2}) for a slab geometry.

Two-scale-factor universality
means that the nonuniversal amplitudes $\xi_{0+}$, $D_1$, $B$, $\xi_{0-}$, and $\xi_{0\rm T}$   are universally related to the two amplitudes $A_1$ and $A_2$
according to \cite{pri,dohm2008}
\begin{eqnarray}
\label{3f}
&&A_1  \xi_{0+}^d = Q_1 = \text {universal},\\\;\;\;\;\;\;\;
\label{3h}
&&D_1 A_2^{-2} A_1^{- 1 - \gamma / (d \nu)}  =
P_3  =  \text {universal},\;\;\;\;\;\;\;\;\;\;\\
\label{3n4}
&&B/(A_1A_2) = - W_1 = \text {universal},\;\;\;\;\;\;\;\\
\label{3ex}
&&\xi_{0 -} / \xi_{0 +} = \text {universal},\\
\label{3exT}
&&\xi_{0\text T} / \xi_{0 +} = \text {universal},
\end{eqnarray}
where $W_1\equiv \lim_{y\to 0}\partial W_-(y)/\partial y$ denotes the derivative of the universal scaling function $W_-(y)$, (\ref{3a}), for $  y \to  0$ \cite{correctA12}, as follows from $M_b =-\partial f_{b,s}(t, h)/\partial h$.
Eq. (\ref{3f}) is confirmed by  (\ref{6m}), with the universal constant
\begin{eqnarray}
\label{6oxx}  Q_1 = \; - \; A_d
Q^{* d \nu} \left[n/(4d) + \nu B (u^*)/(2 \alpha)  \right] \;.
\end{eqnarray}
Eqs.  (\ref{6m}) and (\ref{6n}) yield the universal amplitude ratio
\begin{eqnarray}
\label{6pxx} &&f^-_{b, s} (t)/ f^+_{b, s} (t)
 =A^-/A^+ \nonumber \\&&= \; 2^{d \nu} \; \frac {1 / (64
u^*) \; + 1/(4d) \; + \; \nu B (u^*) / (8 \alpha) }
{n/(4d) \; + \; \nu B (u^*) / (2 \alpha)}. \; \;\;\;\;\;\; \;\;\;\;\;\;
\end{eqnarray}
Eqs. (\ref{6oxx}) and (\ref{6pxx}) are the $n$-dependent generalizations of (6.19) and (6.20) of \cite{dohm2008}. For a comparison of these predictions with numerical results see Sec. V. B.
We introduce the volume
\begin{eqnarray}
\label{corrvoliso}
V^+_{corr}(t)=\xi_{0+}^d|t|^{-d\nu}
\end{eqnarray}
which is a measure of the spherical correlation volume above $T_c$. Then $f^\pm_{b, s}$ can be expressed in terms of the universal quantities $Q_1$ and $A^-/A^+$ as
\begin{eqnarray}
\label{3jjiso}
&&f^\pm_{b, s}(t) =\left\{
\begin{array}{r@{\quad \quad}l}
                          Q_1/V^+_{corr}(t) ,\;\;\;t > 0,\quad          &  \\
                         (A^-/A^+)Q_1/V^+_{corr}(t),\;\;\; t <                 0.&
                \end{array} \right.
\end{eqnarray}
In summary, only {\it two} nonuniversal bulk amplitudes are necessary to determine all other nonuniversal bulk amplitudes via universal relations. This is an important and unique feature for the subclass of isotropic systems near $T_c$ which does not hold for the subclass of weakly anisotropic systems \cite{cd2004,dohm2006,dohm2008,DG,kastening-dohm,dohm2017II,dohmphysik2009,dohm2005,chen-zhang}. %
\renewcommand{\thesection}{\Roman{section}}
\renewcommand{\theequation}{4.\arabic{equation}}
\setcounter{equation}{0}
\section{Finite-size RG approach}
\subsection {Unrenormalized free energy density}
The failure of ordinary perturbation theory for finite systems is due to the perturbative treatment of the dangerous ${\bf k} = {\bf 0}$ lowest mode (App. B). This is avoided by separating the lowest mode and performing perturbation theory only with respect to the higher modes \cite{BZ,RGJ,Esser,CDS1996,dohm2008}. Here we extend this approach such that it becomes applicable to general $n\geq1$ above and below $T_c$ at $ h =  0$. We decompose the variables of (\ref{hamiltonian})
\begin{eqnarray}
\label{phisigma}
{\bm\varphi}_j = {\bm\Phi} + {\bm\sigma}_j
\end{eqnarray}
into the lowest-mode amplitude ${\bm\Phi}$, (\ref{Nullmodenamplitude}),
and into higher-mode contributions
${\bm\sigma}_j = V^{-1} {\sum_{\bf k\neq0}} e^{i{\bf k} \cdot
{\bf x}_j}  \hat {\bm \sigma}({\bf k})$,
where $\hat {\bm \sigma}({\bf k}) \equiv \hat {\bm \varphi}({\bf k})$ for
${\bf k} \neq{\bf 0}$. We further decompose ${\bm\sigma}_j$  into "longitudinal" and "transverse" parts
${\bm\sigma}_j = {\bm\sigma}_{{\rm L}j} + {\bm\sigma} _{{\rm T}j}$
which are parallel and perpendicular with respect to ${\bm\Phi}$, i. e., ${\bm\sigma} _{{\rm T}j}\cdot {\bm\sigma} _{{\rm L}j}= 0$ and ${\bm\sigma} _{{\rm T}j}\cdot {\bm\Phi}= 0$ but ${\bm\sigma}_{{\rm L}j}\cdot {\bm\Phi}\neq 0$.
Correspondingly, the Hamiltonian $H$ is decomposed as
\begin{equation}
\label{decomp1}
H = H_0({{\bm\Phi}}^2) + \widetilde H ({\bm\Phi}, {\bm\sigma}),
\end{equation}
with the lowest-mode Hamiltonian
\begin{eqnarray}
\label{lowHamilton}
H_0 ({{\bm\Phi}}^2) = V[(r_0/2) {{\bm\Phi}}^2
 + u_0 ({{\bm\Phi}}^2)^2  ]
\end{eqnarray}
and the higher-mode Hamiltonian
\begin{eqnarray}
\label{higher Hamilton}
&&\widetilde H ({\bm\Phi}, {\bm\sigma})= H^{(2)} ({{\bm\Phi}}^2, {\bm\sigma}) +H^{(4)} ({\bm\Phi}, {\bm\sigma}),\\
\label{higher2}
&&H^{(2)}  = (\tilde a^d/2)\Big\{\sum_{j=1}^N
\left[( \bar r_{0{\rm L}}({{\bm\Phi}}^2) {{\bm\sigma}_{{\rm L}j}}^2 + \bar r_{0{\rm T}}({{\bm\Phi}}^2) {{\bm\sigma}_{{\rm T}j}}^2)\right] \nonumber\\&&\;\;\;\;\;\; +
\sum_{i,j=1}^N K_{i,j} \left[({\bm\sigma}_{{\rm L}i} - {\bm\sigma}_{{\rm L}j})^2 + ({\bm\sigma}_{{\rm T}i} - {\bm\sigma}_{{\rm T}j})^2\right]
\Big\},\\
\label{higher34}
&&H^{(4)} = u_0 \tilde a^d\sum_{j=1}^N
\big[
4{\bm\Phi}\cdot{\bm\sigma}_{{\rm L}j}({{\bm\sigma}_{{\rm L}j}}^2 +\; {{\bm\sigma}_{{\rm T}j}}^2)\nonumber\\&& \;\;\;\;\;\;\;\;\;\;\;\;\;\;\;\;\;\;\;\;\;\;\;\;\;\;\;\;\;\;+{{({\bm\sigma}_{{\rm L}j}}^2 + {{\bm\sigma}_{{\rm T}j}}^2)}^2 \big],\\
\label{rbarlong}
&&\bar r_{0{\rm L}}({{\bm\Phi}}^2) = r_0+12u_0{{\bm\Phi}}^2,
\\
\label{rbartrans}
&&\bar r_{0{\rm T}}({{\bm\Phi}}^2) = r_0+4u_0{{\bm\Phi}}^2.
%\end{subequations}
\end{eqnarray}
In (\ref{lowHamilton}) the lowest mode is stabilized for arbitrary $-\infty \leq r_0 \leq \infty $ through the $u_0({{\bm\Phi}}^2)^2$ term. The corresponding decomposition of the partition function is
\begin{eqnarray}
\label{2fx}
Z&=& V^{n/2}\tilde a^{-n} \int  d^n
{\bm\Phi} \exp \left\{- [H_0({{\bm\Phi}}^2)  + \bare{\Gamma}({{\bm\Phi}}^2) ] \right\},\;\;\;\;\;\;\;\\
\label{4h} {\bare{\Gamma}}({{\bm\Phi}}^2)  &=& -\; \ln
\Big[\prod_{{\bf k\neq 0}} \frac{ \int \;d^n
 {\bm \hat \sigma}({\bf k}) }{\tilde a^n V^{n/2}} \Big]  \exp [ -\widetilde H ({\bm\Phi}, {\bm\sigma}) ] \;\;\;\;
\end{eqnarray}
where ${\bare{\Gamma}}({{\bm\Phi}}^2)$ describes the higher-mode contribution. The integration measure in (\ref{4h}) is the $n$-component generalization of that defined for $n=1$ in  Appendix B of \cite{dohm2008}.
The quantity ${\bare{\Gamma}}({{\bm\Phi}}^2)$ can be interpreted as a constraint
free energy, with the constraint being that the zero-mode
amplitude ${\bm\Phi}$ is fixed. In the absence of an external field $ h$, $
{\bare{\Gamma}}({{\bm\Phi}}^2)$ is invariant against rotations in ${\bm\Phi}$ space, thus it is free of Goldstone singularities in an exact theory \cite{david}. Spurious singularities, however, may arise at intermediate stages of perturbation theory in higher order as is known from bulk theory below $T_c$ \cite{Burnett,str1999,str2003}.

The integrand of (\ref{2fx}) plays the role of an order-parameter distribution function of the finite system \cite{CDS1996},
\begin{eqnarray}
\label{distributionbare}
P({\bm\Phi}^2) \propto \exp \left\{- [H_0({{\bm\Phi}}^2)  + \bare{\Gamma}({{\bm\Phi}}^2) ] \right\},
\end{eqnarray}
which is a physical quantity in its own right. An important conceptual issue of our approach is that, although we shall make approximations for $\bare{\Gamma}({{\bm\Phi}}^2)$ within the exponent of (\ref{distributionbare}), we shall not make any expansion that would destroy the exponential form of $P({\bm\Phi}^2)$. This is in line with our minimal renormalization procedure at fixed dimension $2<d<4$ \cite{dohm1985} which avoids the $\varepsilon=4-d$ expansion where the fixed-point value $u^* \sim O(\varepsilon)$ of the coupling $u_0$ is treated as a smallness parameter \cite{BZ,RGJ}. Such an $\varepsilon$ expansion destroys the exponential form of $P({\bm\Phi}^2)$ and may lead to unreliable results below $T_c$, as shown previously \cite{Esser} in the context of the specific heat for a cubic system with a one-component order-parameter.

The main task of the theory is the calculation of ${\bare{\Gamma}}({{\bm\Phi}}^2)$. We shall make two approximations.  The first approximation is to neglect the effect of the higher-mode part $H^{(4)}({\bm\Phi}, {\bm\sigma})$, (\ref{higher34}), on the finite-size properties (but not on bulk quantities such as critical exponents and the fixed-point value $u^*$ which will be incorporated via the Borel-resummed results \cite{dohm1985,larin} within the renormalized theory at fixed $d$). Our approximation goes beyond ordinary one-loop perturbation theory (see App. B) not only because of the fourth-order term $\sim u_0{({{\bm \Phi}}^2)}^2$ in the lowest-mode Hamiltonian $H_0$ but also because $H^{(2)} ({{\bm\Phi}}^2,{\bm\sigma})$ contains the non-Gaussian couplings $\sim 12 u_0 {{\bm \Phi}}^2 {{\bm\sigma}_{{\rm L}j}}^2$ and $\sim  4 u_0 {{\bm \Phi}}^2 {{\bm\sigma}_{{\rm T}j}}^2$ between the lowest mode and the higher modes which arise from the terms $ \bar r_{0{\rm L}}({{\bm \Phi}}^2) {{\bm\sigma}_{{\rm L}j}}^2 $  and $\bar r_{0{\rm T}}({{\bm \Phi}}^2) {{\bm\sigma}_{{\rm T}j}}^2$ , respectively, in $H^{(2)}$. These couplings will produce the pole terms of $O\big(u_0{{\bm \Phi}}^2 \varepsilon^{-1}\big)$ and $O\big(u_0^2({{\bm \Phi}}^2)^2 \varepsilon^{-1}\big)$ given in (\ref{Gammalongbare}) and (\ref{Gammatransbare}) below.

Performing the integration over ${\bm \hat \sigma} ({\bf k})$  we obtain from (\ref{4h}), with $\widetilde H $  replaced by  $H^{(2)}$,
\begin{eqnarray}
\label{4hneu}
{\bare{\Gamma}} ({{\bm\Phi}}^2) &=& \big[ - n (N-1) \ln(2 \pi) +VS_0\big(\bar r_{0{\rm L}}({{\bm \Phi}}^2),\{L_\alpha\}\big) \nonumber\\ &+& (n-1)V S_0\big(\bar r_{0{\rm T}}({{\bm \Phi}}^2),\{L_\alpha\}\big) \big]/2 \;
\end{eqnarray}
with the sum over the higher modes
\begin{eqnarray}
\label{Snullx}
S_0(r,\{L_\alpha\})=\frac{1}{V} {\sum_{\bf k\neq0}} \ln \left\{\left[ r + \delta \widehat K (\mathbf k)\right] \tilde a^2\right\}. \;\;
\end{eqnarray}
The applicability of (\ref{4hneu}) and (\ref{Snullx}) is restricted to $\bar r_{0{\rm L}}({{\bm\Phi}}^2)\geq 0$, $\bar r_{0{\rm T}}({{\bm\Phi}}^2)\geq 0$. This restriction will be removed  by  our second approximation in Sec. IV.C.
Since the ${\bf k = 0}$ mode has been separated the sum $S_0(r,\{L_\alpha\})$ is finite for $r=0$ which is important at $T_c$  and below $T_c$ for the transverse part $\propto (n-1)$.
The sum $S_0$ can be expressed as
\begin{eqnarray}
\label{Snullxdelta}
S_0(r,\{L_\alpha\})&=&\Delta(r,\{L_\alpha\}) -V^{-1}  \ln( r  \tilde a^2)\nonumber\\ &+& \int_{\bf k} \ln \{[r + \delta \widehat K
(\mathbf k)] \tilde a^2\}, \;\;\\
\label{bb9xy}
\Delta(r, \{L_\alpha\}) &=& V^{-1}
{\sum_{\bf k}} \ln
\{[r + \delta \widehat K (\mathbf k)] \tilde a^2\}\nonumber\\
&&- \int_{\bf k} \ln \{[r + \delta \widehat K (\mathbf k)]
\tilde a^2\}. \;
\end{eqnarray}
Note that  $\Delta(r, \{L_\alpha\})$ depends on the lattice constant $\tilde a$. An asymptotically exact calculation of the function $\Delta(r,  \{L_\alpha\})$ for $L_\alpha/\tilde a\gg 1$, $0 < r\tilde a^2 \ll 1$,  $0 < r L_\alpha^2 \lesssim O(1)$,  and for finite $0 < \rho_\alpha < \infty$ can be carried out in a way similar to that in \cite{dohm2008,dohm2011}.
The result is independent of $\tilde a$ and reads for isotropic systems
\begin{eqnarray}
\label{F.155yy}
\Delta (r,\{L_\alpha\} ) &=& \;  L^{-d} {\cal G}_0
(r L^2, \{\rho_\alpha\}),\\
\label{calG0x}
{\cal G}_0(x, \{\rho_\alpha\}) \;& = &\;\int_0^\infty
 dy \; y^{-1} \exp \left(- \frac{x y}{4
\pi^2} \right) \nonumber\\
&\times& \Big\{\left(\pi/y\right)^{d/2} \; - \;\bar\rho\;^{d-1}\; K_d (y,
{\bf C}) \Big\},\;\;\;\;\;\\
\label{rhobarx}
\bar\rho &=& \Big[\prod^{d-1}_{\alpha = 1} \;\rho_\alpha\Big]^{1/(d-1)}.
\end{eqnarray}
where $\bar\rho$ is the geometric mean of the aspect ratios.
The function $K_d (y, {\bf C})$ is defined for $y>0$ by
\begin{eqnarray}
\label{Kd} K_d (y,{\bf C}) =\sum_{\bf n} \;\exp (- y
{\bf n} \cdot {\bf  C n})\;
\end{eqnarray}
where the $d \times d$ matrix ${\bf C}$ has the elements
\begin{eqnarray}
\label{matrixC}
C_{\alpha\beta}=\rho_\alpha \rho_\beta\;\delta_{\alpha \beta}.
\end{eqnarray}
The sum $\sum_{\bf n}$ runs over ${\bf n} = (n_1, n_2, ..., n_d) \; , n_\alpha = 0, \pm 1,
..., \pm \infty$.
By means of the Poisson identity \cite{morse} one can show that for $y>0$ this function satisfies
\begin{eqnarray}
\label{aneu} K_d (y, {\bf  C})  = (\det {\bf C})^{-1/2}
\left(\pi/y\right)^{ d/2} K_d \left(\pi^2/y , {\bf
C}^{-1} \right)  \;\; \;\; \;\; \;\;
\end{eqnarray}
which is useful for small $y>0$. The function ${\cal G}_0( x, \{\rho_\alpha\})$ decays exponentially for large $x$ and is logarithmically divergent for $x\to 0_+$ which comes from the large-$y$ behavior of $K_d (y,{\bf C}) \approx 1$ in the last term of (\ref{calG0x}).
This divergent part is separated in the exact decomposition
\begin{eqnarray}
\label{calG0decom}
&&{\cal G}_0(x, \{\rho_\alpha\}) \; = \;\bar\rho\;^{d-1}\big[\ln\big(x/(4\pi^2)\big)+ {\cal J}_0(x, \{\rho_\alpha\} )\big],\;\;\:\:\:\;\;\:\:\:\\
%\end{eqnarray}
%
%
%\begin{eqnarray}
\label{calJ3x}
&&{\cal J}_0(x, \{\rho_\alpha\} )=  \int_0^\infty
dy y^{-1}
  \Big\{\exp {\left[-xy/(4\pi^2)\right]}
  \nonumber\\&& \times \Big[\bar\rho\;^{1-d}
  (\pi/y)^{d/2}
  - \; K_d (y,{\bf C}) + 1 \Big]   - e^{- y}\Big\}\;\;\;
\end{eqnarray}
where the function ${\cal J}_0$  has a finite limit ${\cal J}_0(0, \{\rho_\alpha\})$ for $x \to 0_+$.
From (\ref{Snullxdelta}), (\ref{F.155yy}), and (\ref{calG0decom}) we obtain
\begin{eqnarray}
\label{c1111u}S_0(r,\{L_\alpha\}) = \int_{\bf k} \ln \{[r + \delta \widehat K
(\mathbf k)] \tilde a^2\} \nonumber\\
  + V^{-1}\Big\{ \ln[L^2/( 4\pi^2\tilde a^2)]+ {\cal
J}_0(r L^2, \{\rho_\alpha\})\Big\}.
\end{eqnarray}
For the case of a finite-slab geometry with aspect ratio $\rho$, the following substitutions are to be made,
\begin{eqnarray}
\label{substiso}
&&\bar\rho\;^{d-1} \to \rho^{d-1},
\\
\label{substK}
&&K_d (y,{\bf C})\to \big[ K(\rho^2y)\big]^{d-1}K(y),\\
\label{K}
&&K(y) = \sum^{\infty}_{m= -\infty} \;\exp (- y m^2),\\
\label{calG3spec}
&&{\cal G}_0( x, \{\rho_\alpha\}) \to{\cal G}_0(x,\rho )= \int_0^\infty
\frac{dy}{y}
  \Big\{e^{- xy/(4\pi^2)}
  \nonumber\\ &&\times \Big[
  (\pi/ y)^{d/2}
  - \;\big[\rho K(\rho^2y)\big]^{d-1}K(y)  \Big] \Big\},\\
\label{calJ3spec}
&&{\cal J}_0( x, \{\rho_\alpha\}) \to{\cal J}_0(x,\rho )= \int_0^\infty
\frac{dy}{y}
  \Big\{e^{- xy/(4\pi^2)}
  \nonumber\\ &&\times \Big[ \rho^{1-d}
  \Big(\frac{\pi}{ y}\Big)^{d/2}
  - \;\big[ K(\rho^2y)\big]^{d-1}K(y) + 1 \Big]   -  e^{- y}\Big\}. \;\;\;\;\; \;\;\;\;\;
\end{eqnarray}
The next step is to rewrite the exponential argument $H_0 + \bare{\Gamma}$ of $Z$, (\ref{2fx}), as a function of $r_0 - r_{0c}$ rather than of $r_0$. According to (\ref{Snullx}), (\ref{c1111u}), and (\ref{bulkdecom})-(\ref{bulk2}), the r.h.s. of (\ref{4hneu}) contains the bulk term
\begin{eqnarray}
\label{bulkterms}(V/2) \Big [ \bar r_{0{\rm L}}({{\bm\Phi}}^2)+ (n-1)\bar r_{0{\rm T}}({{\bm\Phi}}^2)\Big ] \int_{\bf k}
[\delta \widehat K (\mathbf k)]^{-1}
\end{eqnarray}
which we rewrite as
\begin{eqnarray}
\label{bulkterms1}(V/2) n r_0 \int_{\bf k}
[\delta \widehat K (\mathbf k)]^{-1} - (V/2) r_{0c}{{\bm\Phi}}^2
\end{eqnarray}
with $r_{0c}$ given by (\ref{rnullc}). The last term in (\ref{bulkterms1}) can be combined with the lowest-mode Hamiltonian (\ref{lowHamilton}) as
\begin{eqnarray}
\label{lowHamilton1}
&&H_0 ({{\bm\Phi}}^2) - (V/2)r_{0c}{{\bm\Phi}}^2  \nonumber\\&& = V
\left[ (r_0 - r_{0c}){{\bm\Phi}}^2/2 + u_0 {({{\bm\Phi}}^2)}^2  \right]  \equiv \widehat H_0({{\bm\Phi}}^2).\;\;\;
\end{eqnarray}
In all $r_0$ - dependent terms of the higher-mode contribution ${\bare{\Gamma}}({{\bm \Phi}^2})$, (\ref{4hneu}), we may replace $r_0$ by $r_0 - r_{0c}$ in the spirit of perturbation theory since $r_{0c} \sim O(u_0)$. Eqs. (\ref{bulklog}) - (\ref{geometric factor}) and (\ref{2fx}) - (\ref{lowHamilton1})  then lead to the unrenormalized free energy density in $2<d<4$ dimensions
\begin{eqnarray}
\label{free prime}
f(t,\{L_\alpha\})= f^{(1)}_{b,ns} (t) + \;\delta f(r_0 - r_{0c}, u_0,L,\{\rho_\alpha\})\;\;\;\;
\end{eqnarray}
where $f^{(1)}_{b,ns} (t)\equiv g^{(1)}_{b,ns} (t)$ , (\ref{free bulk nonsing}), and
\begin{eqnarray}
\label{deltafprime6}
&&\delta f(r_0 - r_{0c}, u_0,L,\{\rho_\alpha\})= \nonumber\\&&- V^{-1}\ln  \Big\{ \Big(2\pi V/{L}^2\Big)^{n/2}  \int  d^n {\bm\Phi}  \exp \Big[ -\widehat H_0({{\bm\Phi}}^2) \nonumber\\ &&-{\bare{\Gamma}}_{\rm L}({{\bm\Phi}}^2) - (n-1){\bare{\Gamma}}_{\rm T}({{\bm\Phi}}^2) \Big]\Big\},\\
\label{2fff}
&&\int  d^n {\bm\Phi} \equiv 2 \pi^{n/2}\; \Gamma (n/2)^{-1}\int_0^\infty  d |{\bm\Phi}||{\bm\Phi}|^{n-1},
\end{eqnarray}
with the longitudinal and transverse contributions
\begin{eqnarray}
\label{Gammalongbare}
{\bare{\Gamma}}_{\rm L}({{\bm\Phi}}^2)&=& -\frac{VA_d}{d\varepsilon}{r_{0{\rm L}}({{\bm\Phi}}^2)}^{d/2} + \frac{1}{2}{\cal J}_0( r_{0{\rm L}}({{\bm\Phi}}^2){ L}^2, \{\rho_\alpha\}),\nonumber\\\\
%\end{eqnarray}
%\begin{eqnarray}
\label{Gammatransbare}
{\bare{\Gamma}}_{\rm T}({{\bm\Phi}}^2)&=&-\frac{VA_d}{d\varepsilon}{r_{0{\rm T}}({{\bm\Phi}}^2)}^{d/2} + \frac{1}{2}{\cal J}_0( r_{0{\rm T}}({{\bm\Phi}}^2){ L}^2, \{\rho_\alpha\}),\nonumber\\\\
\label{rwidetilde}
\label{rlongprime}
 r_{0{\rm L}}({{\bm\Phi}}^2)&=&r_0 - r_{0c}+12u_0{{\bm\Phi}}^2,
\\
\label{rtransprime}
r_{0{\rm T}}({{\bm\Phi}}^2)&=&r_0 - r_{0c}+4u_0{{\bm\Phi}}^2.
\end{eqnarray}
The $\tilde a$ dependence in (\ref{2fx}) and (\ref{c1111u}) of the lowest-mode and higher-mode contributions to $\delta f$ have cancelled among themselves. Because of the exact treatment of $H^{(2)} ({{\bm\Phi}}^2,{\bm\sigma})$,  the effective action in (\ref{deltafprime6}) contains powers of $u_0{{\bm\Phi}}^2$ up to infinite order.
\subsection{Renormalized free energy density}

The multiplicative and additive renormalizations of $\delta f(r_0 - r_{0c}, u_0,L,\{\rho_\alpha\})$, (\ref{deltafprime6}), are the same as for the corresponding bulk quantity (\ref{5dx}) since $L$ and $\rho_\alpha$ are not renormalized. Thus we employ the minimal subtraction scheme at fixed dimension \cite{dohm1985} and define the renormalized counterpart of $\delta f$ in $2<d<4$ dimensions as
\begin{eqnarray}
\label{5ddfinite}
f_R(r, u,L,\{\rho_\alpha\},\mu) &=&\delta f(Z_rr, \mu^{\varepsilon}Z_{u}Z_{\varphi}^{-2}A_d^{-1}u,L,\{\rho_\alpha\}) \nonumber\\ && - (1/8)\mu^{-\varepsilon} r^2 A_d\; A(u,\varepsilon) ,
\end{eqnarray}
where the renormalized quantities $u,r$ and the renormalization constants $Z_r(u,\varepsilon),Z_{u}(u,\varepsilon), Z_{\varphi}(u,\varepsilon), A(u,\varepsilon)$ are the same as those for the bulk system in (\ref{renormalized parameters}) and (\ref{renormalization constants}). We also take the same choice $\mu^{-1}=\xi_{0+}$ of the inverse reference length as for the bulk system. Because of $Z_{\varphi}=1$ within our approximation, the renormalized lowest-mode amplitude of the transformed system is simply ${\bm \Phi}_R=Z_{\varphi}^{-1/2}{\bm \Phi}={\bm \Phi}$. In order to maintain the exponential form of the integrand of $\delta f$, (\ref{deltafprime6}), and to avoid any further approximation we rewrite the additive part of the renormalization in (\ref{5ddfinite}) in an exponential form
\begin{eqnarray}
\label{additive expo}  -(1/8)\mu^{-\varepsilon} r^2 A_d
       A(u,\varepsilon) =  n\mu^{-\varepsilon} r^2 A_d/(4\varepsilon)
       \nonumber\\= -V^{-1}\ln \big\{\exp\big[-V n\mu^{-\varepsilon} r^2 A_d/(4\varepsilon)
       \big]\big\}
\end{eqnarray}
which permits us to incorporate it in the exponent of the integral over ${\bm \Phi}$ in (\ref{deltafprime6}). In (\ref{additive expo}) we have already substituted $A(u,\varepsilon)= -2n/\varepsilon$ according to (\ref{A}). Consequently, after multiplicative renormalization of $r_0 - r_{0c}$ and $u_0$ in the lowest-mode Hamiltonian $\widehat H_0({{\bm\Phi}}^2)$ in (\ref{deltafprime6}), we also keep the corresponding pole terms $ 2(n+2)u r/\varepsilon$ and $ 4(n+8){u}^2/\varepsilon$ of $(1/2)Z_r(u,\varepsilon)r$ and $Z_{u}(u,\varepsilon)u$  in the exponent of the integral over ${\bm \Phi}$ in (\ref{deltafprime6}).

Now there exist two types of pole terms $\propto \varepsilon^{-1}$ in $f_R(r, u,L,\{\rho_\alpha\},\mu)$: (i) those arising from the lowest-mode Hamiltonian $\widehat H_0({{\bm\Phi}}^2)$ after multiplicative renormalization together with the additive pole term of (\ref{additive expo}), (ii) those of the higher-mode contributions ${\bare{\Gamma}}_{\rm L}$ and ${\bare{\Gamma}}_{\rm T}$ in (\ref{Gammalongbare}) and (\ref{Gammatransbare}). As far as the pole terms (i) are concerned, it is nontrivial that the corresponding terms in the exponential argument of (\ref{deltafprime6}) can be expressed as
\begin{eqnarray}
%\begin{align}
\label{umformung}
&&-\widehat H_0({{\bm\Phi}}^2)    -V n\mu^{-\varepsilon} r^2 A_d/(4\varepsilon) =-H_R({{\bm\Phi}}^2) \nonumber\\&&
-VA_d\mu^{-\varepsilon}\big[ r_{\rm L}{({{\bm\Phi}}^2)}^2+(n-1) r_{\rm T}{({{\bm\Phi}}^2)}^2\big]/(4\varepsilon),\;\;\;\;\\
\label{lowrenorm}
&&H_R({{\bm\Phi}}^2)=V{{\bm\Phi}}^2\big(r/2
+\mu^{\varepsilon}A_d^{-1}u{{\bm\Phi}}^2\big),\\
\label{rbarrenormlong}
&&r_{\rm L}({{\bm\Phi}}^2)=r+12\mu^{\varepsilon}A_d^{-1}u{{\bm\Phi}}^2,
\\
\label{rbarrenormtrans}
&&r_{\rm T}({{\bm\Phi}}^2)=r+4\mu^{\varepsilon}A_d^{-1}u{{\bm\Phi}}^2,
\end{eqnarray}
where $r_{\rm L}$ and $r_{\rm T}$ are the renormalized counterparts of
(\ref{rwidetilde}) and (\ref{rtransprime}).
As far as the pole terms (ii) are concerned, we may replace the unrenormalized parameters $ r_{0{\rm L}}({{\bm\Phi}}^2)$ and $r_{0{\rm T}}({{\bm\Phi}}^2)$  in  the higher-mode contributions ${\bare{\Gamma}}_{\rm L}$ and ${\bare{\Gamma}}_{\rm T}$ of (\ref{deltafprime6}) by their renormalized counterparts $ r_{\rm L}({{\bm\Phi}}^2)$ and $ r_{\rm T}({{\bm\Phi}}^2)$ in the spirit of perturbation theory. Then the sum of the pole terms $\propto V A_d/(d\varepsilon) $ in (\ref{Gammalongbare}) and (\ref{Gammatransbare})  reads
\begin{eqnarray}
\label{poleterms}
V A_d\big[{r_{\rm L}({{\bm\Phi}}^2)}^{d/2}+(n-1) {r_{\rm T}({{\bm\Phi}}^2)}^{d/2}\big]/(d\varepsilon).
\end{eqnarray}
Comparison between (\ref{poleterms}) and  (\ref{umformung}) shows that, for $d  \to 4$, indeed all pole terms (i) and (ii) of $f_R(r, u,L,\{\rho_\alpha\},\mu)$ cancel among themselves. We do not make an expansion of the expression (\ref{poleterms}) in powers of $u$ or in powers of $\varepsilon =4-d $.
We arrive at the renormalized free energy density
\begin{eqnarray}
\label{renorm fprimex}  && f_R(r, u,L,\{\rho_\alpha\},\mu) = - V^{-1}\ln  \Big\{  \big(2\pi V/{L}^2\big)^{n/2} \int  d^n {\bm\Phi}\nonumber\\&&\times   \exp \Big[ -H_R({{\bm\Phi}}^2)  - \Gamma_{\rm L}({{\bm\Phi}}^2)-(n-1)\Gamma_{\rm T}({{\bm\Phi}}^2) \Big]\Big\}\;\;\;\;\;\;\;\;\;\;
 \end{eqnarray}
with the longitudinal and transverse contributions
\begin{eqnarray}
\label{Gamma-longx}
&&\Gamma_{\rm L}({{\bm\Phi}}^2)=(1/2){\cal J}_0( r_{\rm L}({{\bm\Phi}}^2) {L}^2, \{\rho_\alpha\})\nonumber\\&&+\;\bar\rho\;^{1-d}\frac{A_d}{\varepsilon}\Bigg\{\frac{\big[{L}^2 r_{\rm L}({{\bm\Phi}}^2)\big]^2}{4(L\mu)^{\varepsilon}}-\frac{\big[{L}^2 r_{\rm L}({{\bm\Phi}}^2)\big]^{d/2}}{d}\Bigg\},\;\;\;\;\;\;\\
\label{Gamma-transx}
&&\Gamma_{\rm T}({{\bm\Phi}}^2)=(1/2){\cal J}_0( r_{\rm T}({{\bm\Phi}}^2) {L}^2, \{\rho_\alpha\})\nonumber\\&&+\;\bar\rho\;^{1-d}\frac{A_d}{\varepsilon}\Bigg\{\frac{\big[{L}^2 r_{\rm T}({{\bm\Phi}}^2)\big]^2}{4(L\mu)^{\varepsilon}}-\frac{\big[{L}^2 r_{\rm T}({{\bm\Phi}}^2)\big]^{d/2}}{d}\Bigg\},\;\;\;\;\;\;
\end{eqnarray}
where ${\cal J}_0$ is defined in (\ref{calJ3x}). Here we have written $V$ in the form
$V = {L}^d \bar\rho\;^{1-d}$.
We note that no approximation has been made in going from (\ref{deltafprime6}) to  (\ref{renorm fprimex}). In particular, the renormalized distribution function
\begin{eqnarray}
\label{distribution}
P_R({{\bm\Phi}}^2) \propto \;
\exp [ -H_R({{\bm\Phi}}^2) - \Gamma_{\rm L}({{\bm\Phi}}^2)-(n-1)\Gamma_{\rm T}({{\bm\Phi}}^2) ]\nonumber \\
\end{eqnarray}
contained in (\ref{renorm fprimex}) has maintained its exponential form  and is free of pole terms in $2<d\leq4$ dimensions. No expansion has been made with respect to the coupling $u$ contained in $H_R({{\bm\Phi}}^2)$, $r_{\rm L}({{\bm\Phi}}^2)$, and $r_{\rm T}({{\bm\Phi}}^2)$.
The dimensionless amplitude function
\begin{eqnarray}
\label{5iL} F_R(r/\mu^2,u,L\mu,\{\rho_\alpha\}) =
 \mu^{-d} A_d^{-1} f_R(r, u,L,\{\rho_\alpha\}, \mu)\;\;\;\;\;
\end{eqnarray}
satisfies the same RGE (\ref{5j}) as the bulk amplitude function $F_{R,b}$. Integration of the RGE yields
\begin{eqnarray}
\label{5aaaL} f_R(r, u,L,\{\rho_\alpha\}, \mu) &&  = f_R \big(r(l),u(l),L,\{\rho_\alpha\},l\mu \big) \nonumber\\ && + \;A_d r(l)^2(l\mu)^{-\varepsilon} {\cal B}(l)/2
\end{eqnarray}
[compare (\ref{5aaa})]. Equations (\ref{lowrenorm}) -(\ref{5aaaL}) provide an exact RG description of the contributions of the lowest mode and the higher-mode Hamiltonian $H^{(2)}$ to the free energy density. This description incorporates the exact bulk critical exponents (not only in one-loop order) via the effective parameters $r(l)$, (\ref{r}), and $u(l)$, (\ref{u}).

It is convenient to express (\ref{renorm fprimex}) in terms of the dimensionless variable
$s= (V\mu^\varepsilon u/A_d)^{1/4}|{\bm\Phi}|$
as
\begin{eqnarray}
\label{renorm fprime-s}   && f_R(r, u,L,\{\rho_\alpha\},\mu) \nonumber\\&&= \frac{n}{2V} \ln\Big\{\frac{{(L \mu)}^{\varepsilon/2} {[\Gamma(n/2)]}^{2/n}{u}^{1/2}  }{2 \pi^2 A_d^{1/2}}\bar\rho^{(d-1)/2}\Big\}\nonumber\\&&- \frac{1}{V}\ln  \Big\{  2 \int_0^\infty  ds s^{n-1} \exp \Big[ -\frac{1}{2} y s^2-s^4   \nonumber\\&&- \widehat\Gamma_{\rm L}(s^2)-(n-1)\widehat\Gamma_{\rm T}(s^2) \Big]\Big\}\;, \;\;\;\;\;\;\;\\
\label{ypsilon}
&&y  = r [V A_d/(\mu^{\varepsilon}u)]^{1/2} \; ,
%\ee
\end{eqnarray}
where $\widehat\Gamma_{\rm L}(s^2)$ and $\widehat\Gamma_{\rm T}(s^2)$ are given by (\ref{Gamma-longx}) and (\ref{Gamma-transx}) with ${{\bm\Phi}}^2={[A_d/(V\mu^\varepsilon u)]}^{1/2}s^2$.
The next step entails an appropriate choice of the flow parameter $l$ such that $F_R$ with the argument $r/\mu^2$
is  mapped to the noncritical region corresponding to $F_R$
with $l$-dependent arguments. An  appropriate choice will be made  after an approximation for the higher-mode contributions (\ref{Gamma-longx}), (\ref{Gamma-transx}).

\subsection{Approximation for the higher-mode contributions}

In its present form,  the renormalized order-parameter distribution function (\ref{distribution}) is applicable only to  $T \geq T_{c}$. This is a consequence of  neglecting the interaction $H^{(4)}$, (\ref{higher34}), which implies that the higher-mode contributions  $\Gamma_{\rm L}({{\bm\Phi}}^2)$ and $\Gamma_{\rm T}({{\bm\Phi}}^2)$  do not exist for negative $ r_{\rm L}({{\bm\Phi}}^2)$ and $ r_{\rm T}({{\bm\Phi}}^2)$. Both parameters indeed become negative in  the regime $r < 0$, i.e., $T < T_{c}$, for sufficiently small ${{\bm\Phi}}^2$ such that $0\leq {{\bm\Phi}}^2<-r/(12\mu^{\varepsilon}A_d^{-1}u)$. For this reason it is appropriate to make a second approximation which permits us to extend the theory to $T<T_c$.
The range of small ${{\bm\Phi}}^2$  where $ r_{\rm L}({{\bm\Phi}}^2)$ and $ r_{\rm T}({{\bm\Phi}}^2)$ become negative represents only an unimportant part of $P_R({{\bm\Phi}}^2)$.  The dominant part of $P_R({{\bm\Phi}}^2)$ is the region ${{\bm\Phi}}^2\approx M^2$ around the renormalized lowest-mode average
\begin{eqnarray}
\label{lowestmode}
\langle {{\bm \Phi}}^2 \rangle_{0,R}&\equiv&  {M}^2(r,u,V,\mu)  = \frac{
\int  d^n {\bm\Phi}{{\bm\Phi}}^2 \exp [-
H_R({{\bm\Phi}}^2)]}{\int  d^n {\bm\Phi} \;\exp [-  H_R({{\bm\Phi}}^2)]}
\nonumber\\&=& (V\mu^{\varepsilon}A_d^{-1}u)^{- 1/2} \; \vartheta_2 (y),\\
\label{thetax}
\vartheta_{2,n} (y) &=& \frac{\int_0^\infty d s \;s^{n+1} \exp (-
\frac{1}{2} y s^2 - s^4)} {\int_0^\infty ds s^{n-1} \; \exp (-
\frac{1}{2} y s^2 - s^4)} \;,
\end{eqnarray}
with $y$ given by (\ref{ypsilon}).
The distribution $ \exp [-H_R({{\bm\Phi}}^2)]$, with $H_R({{\bm\Phi}}^2)$ given by (\ref{lowrenorm}), is well defined  for $-\infty \leq r \leq \infty$. In the region ${{\bm\Phi}}^2\approx {M}^2$ both $ r_{\rm L}({{\bm\Phi}}^2)$ and $ r_{\rm T}({{\bm\Phi}}^2)$ are positive at finite $V$ for arbitrary $r$ and both quantities $\Gamma_{\rm L}({{\bm\Phi}}^2)$ and $\Gamma_{\rm T}({{\bm\Phi}}^2)$ are well behaved in this region.  This leads to our second approximation
\begin{eqnarray}
\label{rappr}
 r_{\rm L}({{\bm\Phi}}^2)&\approx&  r_{\rm L}({M}^2),\;\;\;\; r_{\rm T}({{\bm\Phi}}^2)\approx  r_{\rm T}({M}^2),\\
\label{distributionapprx}
P_R({{\bm\Phi}}^2)& \approx &
\exp [ -H_R({{\bm\Phi}}^2) )\nonumber\\&&- \Gamma_{\rm L}({M}^2-(n-1)\Gamma_{\rm T}({M}^2) ],\;\;\;\;\;\;\;\\
\label{Gamma-long-approx}
\Gamma_{\rm L}({M}^2)&=&(1/2){\cal J}_0( \bar r_{\rm L} {L}^2, \{\rho_\alpha\})\nonumber\\&&+\bar\rho\;^{1-d}\frac{A_d}{\varepsilon}\Big\{\frac{\big(\bar r_{\rm L}{L}^2 \big)^2}{4(L\mu)^{\varepsilon}}-\frac{\big(\bar r_{\rm L}{L}^2 \big)^{d/2}}{d}\Big\},\;\;\;\;\;\;\;\;\\
\label{Gamma-trans-approx}
\Gamma_{\rm T}({M}^2)&=&(1/2){\cal J}_0( \bar r_{\rm T} {L}^2, \{\rho_\alpha\})\nonumber\\&&+\bar\rho\;^{1-d}\frac{A_d}{\varepsilon}\Big\{\frac{\big(\bar r_{\rm T}{L}^2 \big)^2}{4(L\mu)^{\varepsilon}}-\frac{\big(\bar r_{\rm T}{L}^2 \big)^{d/2}}{d}\Big\},\;\;\;\;\;\;\;\;\;\\
\label{rbarrenormaverage}
\label{rbarrenormlongaverage}
\bar r_{\rm L}\equiv  r_{\rm L}({M}^2)&=&r+12\mu^{\varepsilon}A_d^{-1}u {M}^2 \nonumber\\&=&r+12(r/u)\;\vartheta_{2,n} (y),
\\
\label{rbarrenormtransaverage}
\bar r_{\rm T}\equiv  r_{\rm T}({M}^2)&=&r+4\mu^{\varepsilon}A_d^{-1}u {M}^2 \nonumber\\&=&\bar r_{\rm L}-8(r/u)\;\vartheta_{2,n} (y).
\end{eqnarray}
For a plot of the function $\vartheta_{2,n} (y)$ for $n=1,2,3$ see Fig. 1 of \cite{Esser}. The leading behavior for large $|y|$ is for $n\neq2$
\begin{subequations}
\label{theta2large}
\begin{align}
\label{theta2largeabove}
 \vartheta_{2,n} (y)  \longrightarrow  n/y \big[1+O(y^{-2})\big],
\\
\label{theta2largebelow}
 \vartheta_{2,n} (y)  \longrightarrow  -y/4-(n-2)y^{-1}\big[1+O(y^{-2})\big]
\end{align}
\end{subequations}
for $y \to \infty$ and  $y \to -\infty$, respectively. For $n=2$, $\vartheta_{2,2}(y)$ has an exponential approach to the bulk behavior $-y/4$ below $T_c$ according to the exact representation \cite{Esser}
\begin{eqnarray}
\label{theta22}
 \vartheta_{2,2} (y) =  -y/4 + \pi^{-1/2} \exp(-y^2/16)[\text {erfc}(y/4)]^{-1}.\;\;\;\;\;\;\;\;
\end{eqnarray}
The resulting expression for the renormalized free energy density reads for arbitrary $-\infty \leq r \leq \infty$
\begin{eqnarray}
\label{approx renorm fprime-s}   && f_R(r, u,L,\{\rho_\alpha\},\mu) \nonumber\\&&= \frac{n}{2V} \ln\Bigg\{\frac{{(L \mu)}^{\varepsilon/2} {[\Gamma(n/2)]}^{2/n}{u}^{1/2}  }{2 \pi^2 A_d^{1/2}}\bar\rho^{(d-1)/2}\Bigg\}\nonumber\\&&+V^{-1}\Big[{\cal W}_n(y) + \Gamma_{\rm L}({M}^2) +(n-1)\Gamma_{\rm T}({M}^2) \Big],\; \;\;\;\;\;\;\;\\
\label{calWx}
&&{\cal W}_n(y)= -\ln  \big[  2 \int_0^\infty  ds s^{n-1} \exp ( -y s^2/2-s^4    )\big].\;\;\;\;\;\;\;\;\;\;\;
\end{eqnarray}
The latter function is related to $\vartheta_{2,n} (y)$ by
$\partial\;{\cal W}_n(y)/ \partial y= \;\vartheta_{2,n} (y)/2$
which yields the integral representation
\begin{eqnarray}
\label{calWthetay}
{\cal W}_n(y)= {\cal W}_n(0)+\frac{1}{2}\int_0^{y}dy'\;\vartheta_{2,n} (y').\;\;\;\;\;\;
\end{eqnarray}
For further properties of the function  ${\cal W}_n(y)$ see App. C.
The quantities
${\cal W}_n(y)$, ${M}^2 $, $ \bar r_{\rm L}$, and $ \bar r_{\rm T}$ have the following limits for $V\to \infty$ at fixed $r$ above and below $T_c$ as obtained  from (\ref{theta2large}) and (\ref{theta22}) for large $|y|$,
\begin{eqnarray}
\label{calWbulk} \lim_{V \rightarrow \infty} \frac{{\cal W}_n(y)}{V}  = \left\{
\begin{array}{r@{\quad\quad}l}
                0 & \mbox{for} \;\;\; r\geq 0\;, \\ - r^2 A_d  /(16 \mu^\varepsilon u) & \mbox{for} \;\;\;
                 r\leq 0\;,
                \end{array} \right.\;\;\;\;\\
\label{4nn} \lim_{V \rightarrow \infty} M^2 = \left\{
\begin{array}{r@{\quad\quad}l}
                  0 & \mbox{for} \;\;\; r\geq 0\;, \\   -r A_d/( 4\mu^\varepsilon u) & \mbox{for} \;\;\;
                 r\leq 0\;,
                \end{array} \right.\;\;\;\;\\
\label{rprimelongx} \lim_{V \rightarrow \infty} \bar r_{\rm L}  = \left\{
\begin{array}{r@{\quad\quad}l}
                r & \mbox{for} \;\;\; r\geq 0\;, \\ -2r & \mbox{for} \;\;\;
                 r\leq 0\;,
                \end{array} \right.\;\;\;\;\\
\label{rprimetransx} \lim_{V \rightarrow \infty}\bar r_{\rm T}  = \left\{
\begin{array}{r@{\quad\quad}l}
                  r  & \mbox{for} \;\;\; r\geq 0\;, \\  0 & \mbox{for} \;\;\;
                 r\leq 0\;.
                \end{array} \right.\;\;\;\;
\end{eqnarray}
The large-volume limit $V \to \infty$ can be performed not only as a {\it bulk} limit where  $L_\alpha \to \infty$ for all $ \alpha = 1, ..., d$ but also as a {\it film} limit where $L_\alpha \to \infty$ for $ \alpha = 1, ..., d-1$ at fixed $L\equiv L_d$. For this reason, Eqs. (\ref{calWbulk})-(\ref{rprimetransx}) can be employed  in deriving the free energy density not only of the bulk system but also of the film system.
In the limit $V \to \infty$, the parameter $\bar r_{\rm L}$  goes to zero for  $T=T_c$ whereas $\bar r_{\rm T}$ vanishes  for all temperatures $T \leq T_c$. Thus the parameters $\bar r_{\rm L}$  and $\bar r_{\rm T}$ of the {\it finite} system play the role of effective distances  from bulk or film criticality and from the bulk or film coexistence line below $T_c$, respectively. It is a crucial advance that no spurious divergencies occur in $f_R$, (\ref{approx renorm fprime-s}),  as  $\bar r_{\rm L}$  and $\bar r_{\rm T}$ go to zero.

Since the limit of large $|y|$ can also be performed by letting $|r| \to \infty$ at fixed finite $V$,   the quantities ${\cal W}_n(y)/V$, ${M}^2 $, $ \bar r_{\rm L}$, and $ \bar r_{\rm T}$ interpolate smoothly between the limits (\ref{calWbulk})-(\ref{rprimetransx}) as the temperature variable $r$ is varied from far below to far above $T_c$ at finite $V$. Most important, consistency with bulk theory is guaranteed in that (\ref{approx renorm fprime-s}) yields the correct bulk result in one-loop order,
\begin{eqnarray}
\label{bulklimitx}
\lim_{\{L_\alpha \rightarrow \infty\}}f_R(r, u,L,\{\rho_\alpha\},\mu) = f_{R,b} (r,u,\mu)\;
\end{eqnarray}
where $f_{R,b}(r,u,\mu)$ is given by (\ref{bulk f+}) and (\ref{bulk f-}). The verification of (\ref{bulklimitx}) follows from the bulk limits
\begin{subequations}
\label{bulklimitj}
\begin{align}
\label{jlong}
\lim_{\{L_\alpha \rightarrow \infty\}}V^{-1}{\cal J}_0( \bar r_{\rm L} {L}^2, \{\rho_\alpha\})=0
,
\\
\label{jtrans}
\lim_{\{L_\alpha \rightarrow \infty\}}V^{-1}{\cal J}_0( \bar r_{\rm T} {L}^2, \{\rho_\alpha\})=0
,
\end{align}
\end{subequations}
\begin{subequations}
\label{bulklimit}
\begin{align}
\label{Gammalongbulkx}
&
\lim_{\{L_\alpha \rightarrow \infty\}}V^{-1}\Gamma_{\rm L}({M}^2)\nonumber\\&= -\frac{A_d}{\varepsilon}\left\{ \begin{array}{r@{\quad\quad}l}
                  r^{d/2}
[d^{-1}-(r/\mu^2)^{\varepsilon/2}/4]\hspace{1.0cm} & \mbox{for} \;\;\; r\geq 0\;, \\ (-2r)^{d/2}
[d^{-1}-(-2r/\mu^2)^{\varepsilon/2}/4] & \mbox{for} \;\;\;
                 r\leq 0\;,
                \end{array} \right.
\\
\label{Gammatransbulkx}
&
\lim_{\{L_\alpha \rightarrow \infty\}}V^{-1}\Gamma_{\rm T}({M}^2)\nonumber\\&= -\frac{A_d}{\varepsilon}\left\{ \begin{array}{r@{\quad\quad}l}
                  r^{d/2}
[d^{-1}-(r/\mu^2)^{\varepsilon/2}/4]\hspace{1.0cm} & \mbox{for} \;\;\; r\geq 0\;, \\ 0 & \mbox{for} \;\;\;
                 r\leq 0\;.
                \end{array} \right.
\end{align}
\end{subequations}
The result (\ref{approx renorm fprime-s}) is valid for $2<d<4$, with a finite limit for $d\to 4$. It is
the basis for describing the crossover from the low- to the high-temperature behavior of the finite system. For this purpose we consider (\ref{approx renorm fprime-s}) with (\ref{Gamma-long-approx})-(\ref{rbarrenormtransaverage}) in the $l$ dependent form given on the right-hand side of (\ref{5aaaL})  where the $l$ dependent parameters are given by
\begin{eqnarray}
\label{flowxx}
\bar r_{\rm L}(l)&=& r(l) +12\;\big[r(l)}/{y(l)\big]\;\vartheta_{2,n} (y(l)),\\
\label{flowxxtrans}
\bar r_{\rm T}(l)&=& \bar r_{\rm L}(l) - 8\;\big[r(l)}/{y(l)\big]\;\vartheta_{2,n} (y(l)),\\
\label{ypsilonflowx}
y(l)&=& \frac{r(l)\; {(l\mu L)}^{d/2}  A_d^{1/2}}{(l\mu)^2 \;u(l)^{1/2}}\;\;\bar\rho^{(1-d)/2} .
%\ee
\end{eqnarray}
So far the flow parameter $l$ is still undetermined. The most natural choice of $l$ is made by requiring
\begin{eqnarray}
\label{5v} \bar r_{{\rm L}} (l) = \mu^2l^2
\end{eqnarray}
which implies that $l=l(t,L,\{\rho_\alpha\})$ depends on $t$, $L$, $\rho_\alpha$, and $\xi_{0+}=\mu^{-1}$.
Because of (\ref{theta2large}) the choice (\ref{5v}) ensures the bulk choice (\ref{bulk flow parameter}) of $l$ in the large-volume limit
\begin{eqnarray}
\label{rprimelong} \lim_{V \rightarrow \infty} \bar r_{\rm L}(l)  = \left\{
\begin{array}{r@{\quad\quad}l}
                r(l_+) =\mu^2 l_+^2 & \mbox{for} \;\;\; T\geq T_c\;, \\ -2r(l_-)= \mu^2 l_-^2 & \mbox{for} \;\;\;
                 T\leq T_c\;,
                \end{array} \right.\;\;\;\;\\
\label{rprimetrans} \lim_{V \rightarrow \infty}\bar r_{\rm T}(l)  = \left\{
\begin{array}{r@{\quad\quad}l}
                  r(l_+) =\mu^2 l_+^2 & \mbox{for} \;\;\; T\geq T_c\;, \\  0 & \mbox{for} \;\;\;
                 T\leq T_c\;.
                \end{array} \right.\;\;\;\;
\end{eqnarray}
In summary, the renormalized free energy density in the regime $\-\infty \leq r \leq\infty$ is given by (\ref{5aaaL}) and (\ref{approx renorm fprime-s}), with $r(l)$, $u(l)$, ${\cal B}(l)$ defined by (\ref{r})-(\ref{calB}), with $l=l(t,L,\{\rho_\alpha\})$  determined by  (\ref{5v}), and with
\begin{eqnarray}
\label{approx renorm fprime-s-l}   && f_R(r(l), u(l),L,\{\rho_\alpha\},l\mu) \nonumber\\&&= \frac{n}{2V} \ln\Bigg\{\frac{{(l\mu L)}^{\varepsilon/2} {[\Gamma(n/2)]}^{2/n}{u(l)}^{1/2}  }{2 \pi^2 A_d^{1/2}}\bar\rho^{(d-1)/2}\Bigg\}\nonumber\\&&-\frac{A_d}{{L}^d }\Bigg\{\frac{(l\mu L)^d}{4d} -\frac{(n-1)}{\varepsilon}\Big[\frac{l_{\rm T}^2}{4(l\mu L)^\varepsilon}-\frac{l_{\rm T}^{d/2}}{d}\Big]\Bigg\}\nonumber\\&&+(2V)^{-1}\;\big[{\cal J}_0( l^2\mu^2 {L}^2, \{\rho_\alpha\})+ (n-1){\cal J}_0( l_{\rm T}, \{\rho_\alpha\})\big]\nonumber\\&&+ V^{-1}\;{\cal W}_n\Big(y(l)\Big)\;, \;\;\;\;\;\;\;\\
\label{lT}
&&l_{\rm T}(t,L,\bar \rho)\equiv \bar r_{\rm T}(l)\;{L}^2 = l^2\mu^2 {L}^2\nonumber\\&&-\;8\;\Big[(l\mu L)^\varepsilon\; u(l) A_d^{-1}\;\bar\rho\;^{d-1}\Big]^{1/2}\;\vartheta_{2,n} (y(l)),\;\;\;\;\;\;\;\;
\end{eqnarray}
where ${\cal J}_0$ is defined by  (\ref{calJ3x}). This result describes the crossover at finite volume $V$ for general $n$ from far below to far above $T_c$ including the critical regime near $T_c$. Eqs. (\ref{5aaaL}) and (\ref{approx renorm fprime-s-l}) still include nonasymptotic (Wegner) corrections to scaling \cite{wegner1972} via  $u(l)$ and  $r(l)$.
Taking into account (\ref{jlong})-(\ref{Gammatransbulkx}) one verifies that (\ref{approx renorm fprime-s-l}) yields
\begin{eqnarray}
\label{bulklimit}
\lim_{\{L_\alpha \rightarrow \infty\}}f_R(r(l), u(l),L,\{\rho_\alpha\},l\mu)=  f^\pm_{R,b} \big(l_\pm \mu,u(l_\pm)\big)\;\;\;\;\;
\end{eqnarray}
where $f^\pm_{R,b}$ is indeed given by (\ref{5cc}) and (\ref{5dd}).
\begin{figure}[!ht]
\begin{center}
\subfigure{\includegraphics[clip,width=6.8cm]{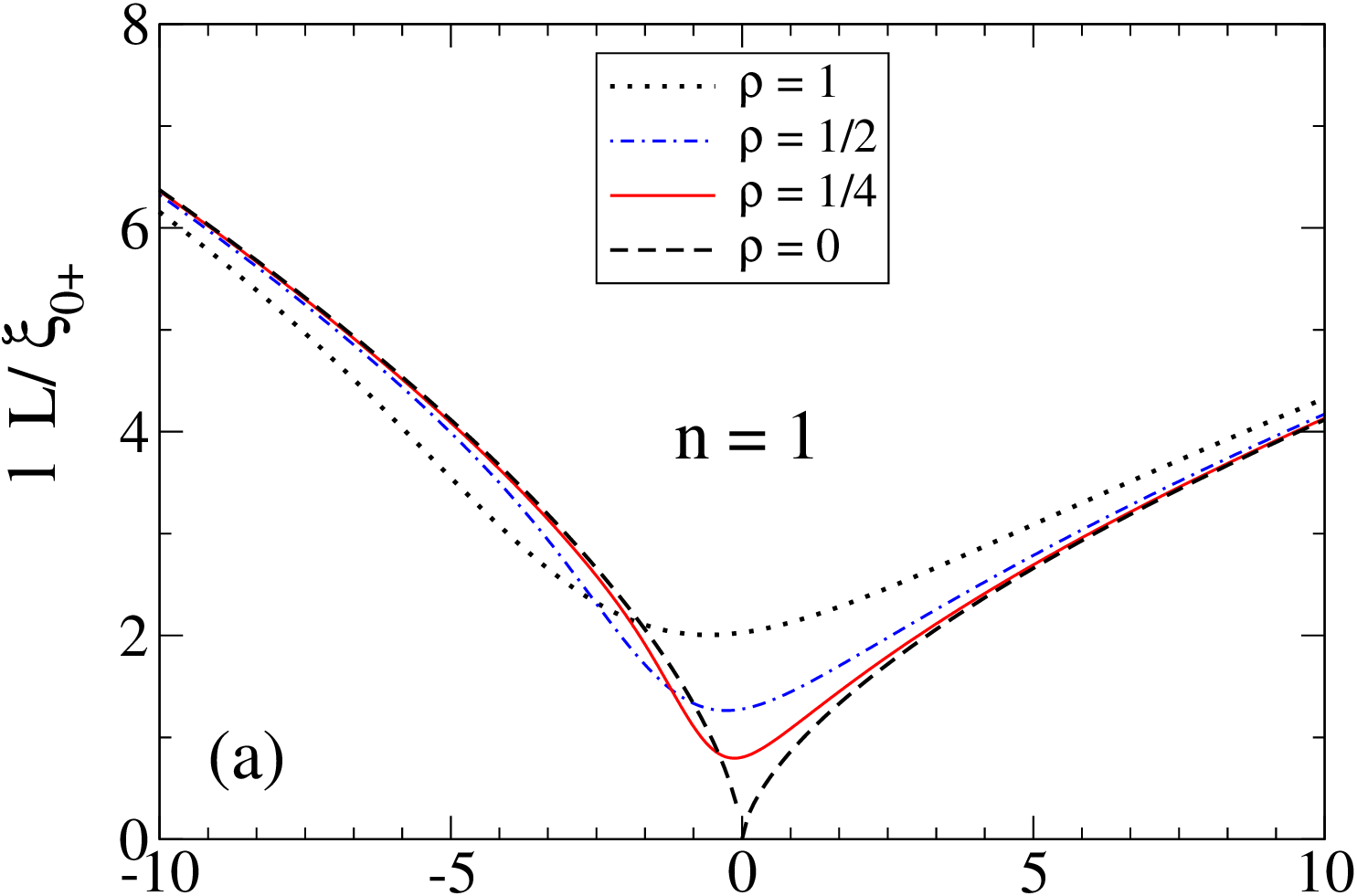}}
\subfigure{\includegraphics[clip,width=6.8cm]{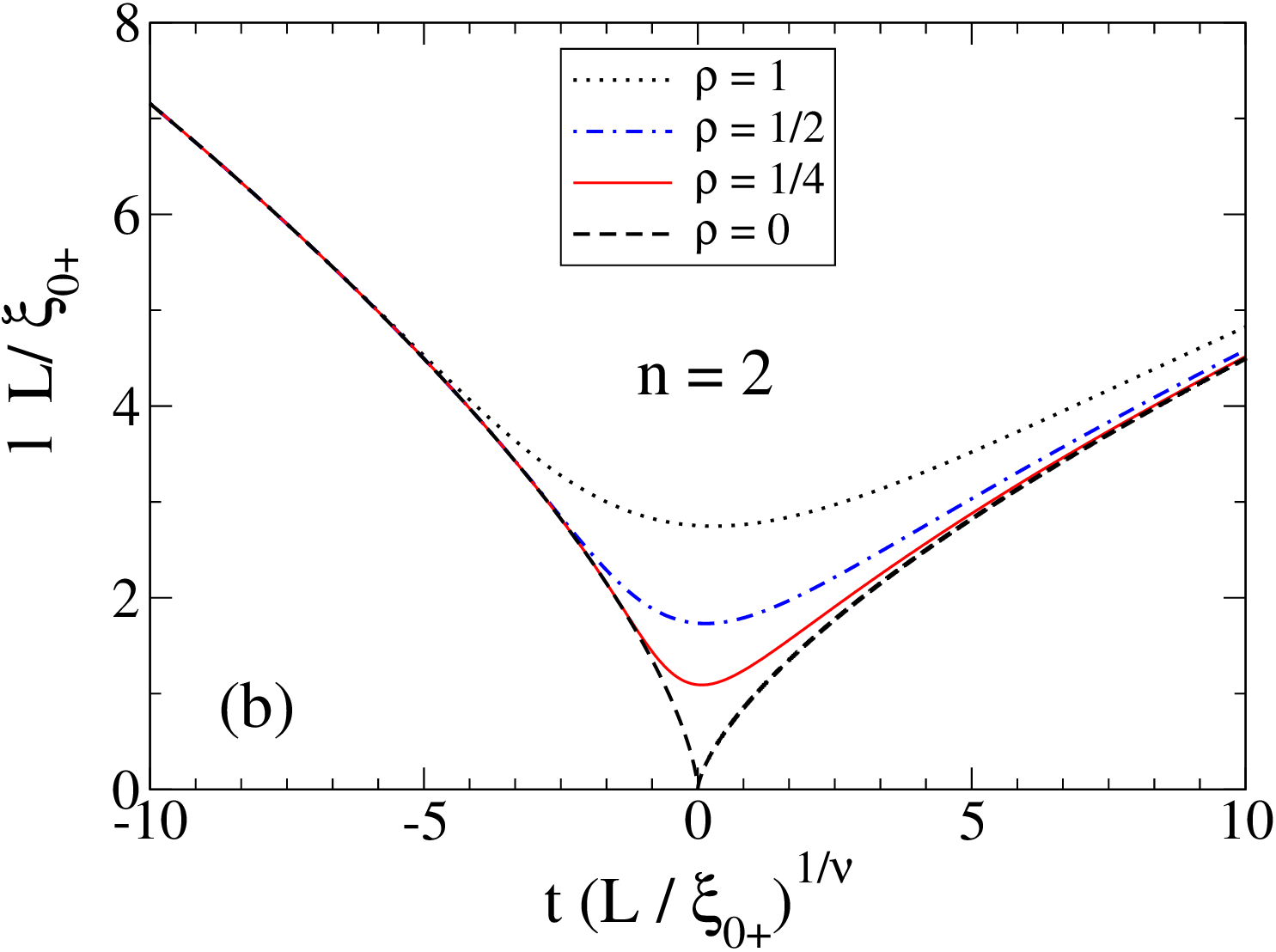}}
\end{center}
\vspace{0.9cm}
\caption{(Color online) Scaling function  $\tilde l =l L /\xi_{0+}$, (\ref{lasym}), of the flow parameter $l$ as a function of $\tilde x=t(L /\xi_{0+})^{1/\nu}$ (a) for $n=1$  and (b) for $n=2$  according to (\ref{implicit}) for several aspect ratios $\bar \rho = \rho>0$ of a slab geometry for $d=3$. The case $\rho=0$ (dashed lines) applies to bulk and film $(\infty^2\times L)$ geometry with $\tilde l_\pm$ given by (\ref{bulktildeflow}). The curves for $\rho>0$ provide a smooth crossover from the low- to the high-temperature region.}
\end{figure}
\vspace{0cm}
\renewcommand{\thesection}{\Roman{section}}
\renewcommand{\theequation}{5.\arabic{equation}}
\setcounter{equation}{0}
\section{Results below, at, and above ${\bf T_c}$}
\subsection{Finite-size scaling functions }
The asymptotic form of $f_{s}$
near $T_c$ is obtained from $f_R$, (\ref{5aaaL}) and (\ref{approx renorm fprime-s-l}), in the limit $l \ll 1$ or $ l\rightarrow 0$ as
\begin{eqnarray}
\label{6k}  f_R(r, u,L,\{\rho_\alpha\}, \mu) \;\,\; \longrightarrow  f_{s}(t,L,\{\rho_\alpha\})  = L^{-d}
F(\tilde x,\{\rho_\alpha\})\nonumber\\
\end{eqnarray}
with the scaling variable
$\tilde x $, (\ref{3jjx}). In this limit we have $u(l) \rightarrow\;u(0)=  u^*$ and
\begin{eqnarray}
 &&r(l)/(\mu^2l^2)\rightarrow \;Q^*\; t
\;l^{-1/\nu}=Q^*\tilde x (\mu l L)^{-1/\nu},\\
\label{6cc}
&&y(l) \rightarrow \; \tilde y = \tilde x \;Q^*\; \big[(\mu
l L)^{- \alpha /\nu} A_d \;{u^*}^{-1}\bar \rho\;^{1-d}\big]^{1/2}.\;\;\;\;\;\;\;\;
\end{eqnarray}
Equations (\ref{flowxx}), (\ref{ypsilonflowx}), and (\ref{5v}) imply asymptotically
\begin{eqnarray}
\label{lasym}
\mu l L \rightarrow \; \tilde l = \tilde l(\tilde x,\bar \rho)
\end{eqnarray}
where the scaling
function $\tilde l$ is determined implicitly by
\begin{subequations}
\label{implicit}
\begin{align}
\label{6ff} \tilde y + 12 \vartheta_{2,n}(\tilde y) =
\big[\tilde l^{d} A_d \;{u^*}^{-1} \bar \rho\;^{1-d}\big]^{1/2},
\\
\label{6gg} \tilde y =\; \tilde x\;Q^*\;\big[\tilde l^{- \alpha /\nu} A_d \;{u^*}^{-1}\bar\rho\;^{1-d}\big]^{1/2}.
\end{align}
\end{subequations}
These two equations also determine the scaling function $\tilde y = \tilde
y(\tilde x,\bar \rho)$. The quantity $l_{\rm T}$, (\ref{lT}), becomes asymptotically $l_{\rm T}(t,L,\bar \rho)\to \tilde l_{\rm T}(\tilde x,\bar \rho)$ with
\begin{eqnarray}
\label{lTasym}
\tilde l_{\rm T}(\tilde x,\bar \rho) = \tilde l^2-\;8\;\big[\tilde l^\varepsilon\; u^* A_d^{-1}\;\bar \rho\;^{d-1} \big]^{1/2}\;\vartheta_{2,n} (\tilde y).\;\;\;\;\;\;\;\;
\end{eqnarray}
In (\ref{6cc}) and (\ref{implicit}) we have used the hyperscaling relation
$ 2 - \alpha = d \nu $.
For $Q^* =Q (1, u^*, d)$  and for the $l\rightarrow0$ behavior of ${\cal B}(l)$ in
(\ref{5aaaL}) see Sec. III B. The resulting finite-size scaling function of the free energy density of the isotropic system in block geometry reads
\begin{eqnarray}
\label{scalfreeaniso} &&F(\tilde x,\{\rho_\alpha\})= -\; A_d \;\Big\{
\frac{\tilde l^d}{4d} \;- \frac{(n-1)}{\varepsilon}\Big[\frac{{\tilde l}^2_{\rm T}}{4 \tilde l^\varepsilon }-\frac{{\tilde l_{\rm T}}^{d/2}}{d}\Big] \nonumber\\ && + \;\nu\;{Q^*}^2 \tilde x^2 \tilde
l^{- \alpha/\nu} B(u^*)/(2\alpha)\Big\} +\bar \rho\;^{d-1}{\cal W}_n\big(\tilde y(\tilde x, \bar \rho)\big) \nonumber\\ && +\;\bar \rho\;^{d-1} \Big\{\frac{n}{2}\ln\Big[\frac{\tilde l^{\varepsilon/2} {[\Gamma(n/2)]}^{2/n}{u^*}^{1/2}  }{2 \pi^2 A_d^{1/2}} \;\bar \rho\;^{(d-1)/2}\Big]\nonumber\\ &&    + (1/2){\cal J}_0( {\tilde l}^2,  \{\rho_\alpha\})+  [(n-1)/2]{\cal J}_0( \tilde l_{\rm T}, \{\rho_\alpha\})\Big\}
\end{eqnarray}
where ${\cal J}_0$ and ${\cal W}_n$ are defined by (\ref{calJ3x}) and (\ref{calWx}).

Equation (\ref{scalfreeaniso}) is the central analytic result of this paper. It is valid for general $n\geq 1$ and $2<d<4$  in the range $L \gg \tilde a$ for  finite $\rho_\alpha$. It includes the cases of slab and film geometries (Sec. V.F)  as well as the case $n\to \infty$ in an approximate form (Sec. V.G). It describes the crossover from far below ($\tilde x \to -\infty$) to far above $T_c$ ($\tilde x \to \infty$) including the critical effects in the central finite-size regime $|\tilde x| \lesssim O(1)$ near $T_c$. In the low-temperature region, it describes the effects due to the Goldstone modes for $n>1$ and the qualitatively different behavior of systems with a one-component order parameter ($n=1$). It incorporates the correct bulk critical exponents $\alpha$ and $\nu$ and the complete bulk function $B(u^*)$ (not only in one-loop order).
The nonuniversal amplitude $\xi_{0+}$ contained in $\tilde x$ is the only system-dependent parameter, thus  the structure of (\ref{scalfreeaniso}) agrees with two-scale-factor universality at $h=0$ according to (\ref{twoscalf}). For finite $L$ and $\rho_\alpha$, $F(\tilde x, \{\rho_\alpha\})$  is an analytic function of  $\tilde x$ near $\tilde x=0$, in agreement with
general analyticity requirements.  Corrections to scaling are not included but can be taken into account by returning to (\ref{approx renorm fprime-s-l}). The scaling function  $\tilde l$ of the flow parameter for $d=3$ and $n=1,2$ is illustrated in Fig. 1 for a slab geometry ($\bar \rho = \rho$) with parameters specified in Sec. V.B.
The bulk part $F^\pm_b(\tilde x)$ of  $F$  is obtained from (\ref{scalfreeaniso}) in the limit of large $|\tilde x|$, as derived in Sec. V. B. This leads to $F^{ex} (\tilde x,\{\rho_\alpha\})$, (\ref{3kkx}), of the excess free energy density which determines the scaling function $X(\tilde x,\{\rho_\alpha\})$ of the Casimir force according to (\ref{3nn}).
For a slab geometry with aspect ratio $\rho$, the scaling functions $F(\tilde x, \rho)$ and
\begin{eqnarray}
\label{substisoX}
X(\tilde x,\rho)&=&(d-1)  F^{ex} (\tilde x,\rho)\nonumber\\ & -&
(\tilde x/\nu)\;
\partial  F^{ex}(\tilde x,\rho)/\partial\tilde x- \rho
\;\partial F^{ex}(\tilde x,\rho)/\partial \rho\;\;\;\;\;\;\;\;\;
\end{eqnarray}
are obtained from (\ref{scalfreeaniso}) after substituting (\ref{substiso})-(\ref{calJ3spec}). These results agree with Eqs. (8)-(13) of \cite{dohm2013}.
By definition, the functions $F^{ex}$ and $X$ have a weak singularity at $\tilde x=0$ arising from the subtraction of $F_b^\pm(\tilde x)$ which is nonanalytic at $\tilde x=0$ [see (\ref{3jjbulk})].
Our scaling functions with $n=1,2,3$ in three dimensions are shown in Figs. 2-11
for slab geometries in the range $0 \lesssim \rho\lesssim 1$. For comparison, MC data \cite{hucht2011,hasenbusch2010,hasenbusch2011, vasilyev2009,dan-krech} for the $d=3$ Ising, $XY$, and Heisenberg models as well as  theoretical predictions \cite{KrDi92a,GrDi07,wil-1,Jakub} are shown.
In the following  we discuss several aspects of our results.
\subsection{Bulk part of the free energy}
In performing the bulk limit of (\ref{scalfreeaniso}) we employ (\ref{calWbulk}),(\ref{4nn}), (\ref{bulklimitj}),  and the limits
\begin{eqnarray}
\label{bulktildeflow} \tilde l \longrightarrow \left\{
\begin{array}{r@{\quad\quad}l}
                \tilde l_+ = \;( \tilde x Q^{*})^\nu \;   & \mbox{for} \;\;\;  T > T_c\;, \\
\tilde l_- =  \;  (2 | \tilde x|Q^{*})^\nu & \mbox{for}
\;\;\;
                 T < T_c \;,
                \end{array} \right.\\
\label{bulktildeflowtrans} \tilde l_{\rm T} \longrightarrow \left\{
\begin{array}{r@{\quad\quad}l}
                \tilde l_+^2 = \;( \tilde x Q^{*})^{2\nu} \;   & \mbox{for} \;\;\;  T > T_c\;, \\
0 & \mbox{for}
\;\;\;
                 T < T_c \;,
                \end{array} \right.
\end{eqnarray}
[compare (\ref{bulk flow})].
The bulk part of  $F(\tilde x,\rho)$  is obtained large for $|\tilde x|\gg 1$ as $L^d f_{s}(t, L,\rho) \to F^\pm_b(\tilde{x})$ with $F^\pm_b$ given by (\ref{3jjbulk})
with the universal bulk amplitude ratios $Q_1$, (\ref{6oxx}),  and $A^-/A^+$, (\ref{6pxx}), in agreement with the bulk amplitudes (\ref{6m}) and (\ref{6n}). For comparison with \cite{pelissetto} we express $Q_1$ in terms of $R_\xi^+$ according to \cite{priv}
\begin{eqnarray}
\label{Rxi}
(R_\xi^+)^d= -\alpha(1-\alpha)(2-\alpha)Q_1.
\end{eqnarray}
As a test of these results we apply them to $d=3$ Ising, XY and Heisenberg systems with $n=1,2,3$ for which accurate numerical results are available. We employ the following values for $d=3$, \cite{dohm1985,larin,pelissetto,CDS1996,campostrinietal}:  $u^*=0.0404, 0.0362, 0.0327$, $B(u^*)= 0.502,1.005,1.508$, $\nu= 0.6301, 0.671, 0.7112$, $\alpha=2-3\nu$, and $Q^*=0.946, 0.939, 0.937$, for $n=1,2,3$, respectively. The values of $Q^*$ follow from Table I and Eq. (3.5) of \cite{krause} together with $u^*$ of \cite{larin}. Our results (\ref{6oxx})  and (\ref{6pxx}) yield
\begin{subequations}
\label{Rxiplus}
\begin{align}
\label{n1}
Q_1= -0.1093, \;\; \;\;R_\xi^+ = 0.272\;\;\;\;\;\; \; \mbox{for} \;\; \;n=1,
\\
\label{n2}
Q_1= 1.807,  \;\; \;\;R_\xi^+ = 0.363\;\;\;\;\;\; \; \mbox{for} \;\; \;n=2,
\\
\label{n3}
Q_1= 0.2607, \;\;  \;\;R_\xi^+ = 0.438\;\;\;\;\;\; \; \mbox{for} \;\; \;n=3,
\end{align}
\end{subequations}
\begin{subequations}
\label{Aminusplus}
\begin{align}
\label{An1}
A^-/A^+= 1.98\;\;\;\;\;\; \; \mbox{for} \;\; \;n=1,
\\
\label{An2}
A^-/A^+ = 0.935\;\;\;\;\;\; \; \mbox{for} \;\; \;n=2,
\\
\label{An3}
A^-/A^+= 0.516\;\;\;\;\;\; \; \mbox{for} \;\; \;n=3.
\end{align}
\end{subequations}
These results are in reasonable agreement with those
in the first row of Tables 11, 22, and 27 of \cite{pelissetto} (based on accurate high-temperature series analyses \cite{campostrinietal}) which yield $R_\xi^+=0.266, 0.355, 0.424$ and $A^-/A^+= 1.88, 0.943, 0.641$ for $n=1,2,3$, respectively. Only our value of $A^-/A^+$ for $n=3$, (\ref{An3}), deviates considerably from $0.641$ of \cite{pelissetto}.
Considering the fact that our present theory is an effective
finite-size theory  that is  not designed to produce highly
accurate bulk predictions, the bulk results (\ref{Rxiplus}) and (\ref{Aminusplus}) are acceptable. They are sensitive to the choice
of the the renormalization scheme and of the geometrical factor in defining the renormalized coupling, (\ref{5a}). Thus the reasonable quantitative agreement of our results (\ref{6oxx}) and (\ref{6pxx}) with established numerical results \cite{pelissetto} demonstrates the appropriateness of our fixed $d$ renormalization scheme with the choice of $A_d$, (\ref{geometric factor}), and lends credibility also to our finite-size RG approach.
\subsection{Analytic results below ${\bf T_c}$}

The analytic form of  $F^{ex}$ and $X$ for slab geometry well below $T_c$ is derived from (\ref{scalfreeaniso}) and (\ref{substisoX}) in App. C. The leading behavior for $-\tilde x \gg 1$ is
\begin{eqnarray}
\label{excess-scalfreeaniso18}
&&F_-^{ex}(\tilde x, \rho) = \frac{1}{2}\;{\cal G}_0((2|\tilde x| \;Q^*)^{2\nu }, \rho) \nonumber\\ && +\rho^{d-1} \Big[ - \frac{n-1}{2}\;  \nu(d-2) \ln (2|\tilde x| \;Q^*)+ \;\;c_n(\rho)\Big] \nonumber\\ &&  + \;O\big((n-2)|\tilde x|^{-d\nu/2}\big)+ \;O\big( \exp[-|\tilde x|^{d\nu}]\big),\;\,\,\,\,\;\,\,\,\,\,\,\,\;\,\,\,\,\\
\label{constcalCnx}
&&c_n(\rho)=   \frac{n-1}{2}\Big\{{\cal J}_0( 0,\rho)- \ln\Big[\frac{\pi^2\;  A_d}{ u^*\;\rho^{d-1}}\Big]\Big\}\nonumber \\&&+\ln \bigg[\frac{\Gamma(n/2)}{\pi^{1/2}}\bigg] +(n-2)\Big\{\ln2 - 12u^*/d  \\&&
+24u^*\alpha\big[1 / (64
u^*) \; + 1/(4d) \; + \; \nu B (u^*) / (8 \alpha)\big]\Big\},\nonumber \\
\label{Xbelow}
&&X_-(\tilde{x},\rho) = \nonumber \\ &&\frac{1}{2}\;\Big\{(d-1)   -
\frac{\tilde{x}}{\nu}
\frac{\partial }{\partial\tilde{x}}- \rho\;\frac{\partial }{\partial \rho}\Big\}\;{\cal G}_0( (2|\tilde x| \;Q^*)^{2\nu },\rho)\nonumber \\ && -\frac{n-1}{2}\rho^{d-1}\Big[1+ \rho\;\partial {\cal J}_0(0,\rho)/\partial \rho\Big]\nonumber \\&& + \;O\big((n-2)|\tilde x|^{-d\nu/2}\big)+ \;O\big( \exp[-|\tilde x|^{d\nu}]\big).
\end{eqnarray}
The function  ${\cal G}_0((2|\tilde x| \;Q^*)^{2\nu }, \rho) $ [see (\ref{calG3spec})] vanishes exponentially for $\tilde x \to -\infty$ (Eq. (A9) of \cite{dohm2011}). The low-temperature behavior differs fundamentally depending on whether Goldstone modes are absent ($n=1$) or present ($n>1$). We discuss these cases separately.

\begin{widetext}
%=======================================================
\begin{figure*}[!ht]
\begin{center}
\subfigure{\includegraphics[clip,width=7.0cm]{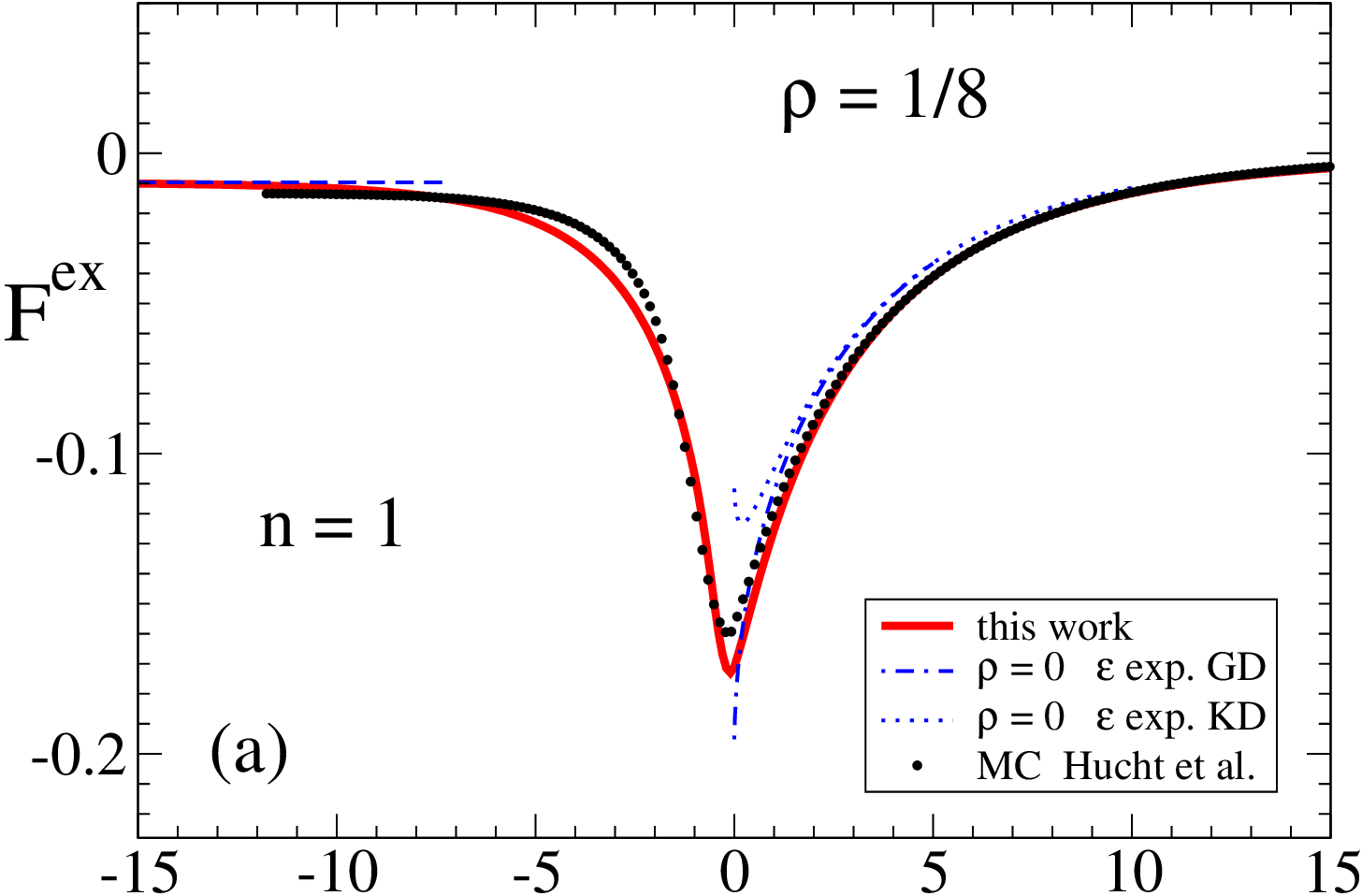}}
\subfigure{\includegraphics[clip,width=7.0cm]{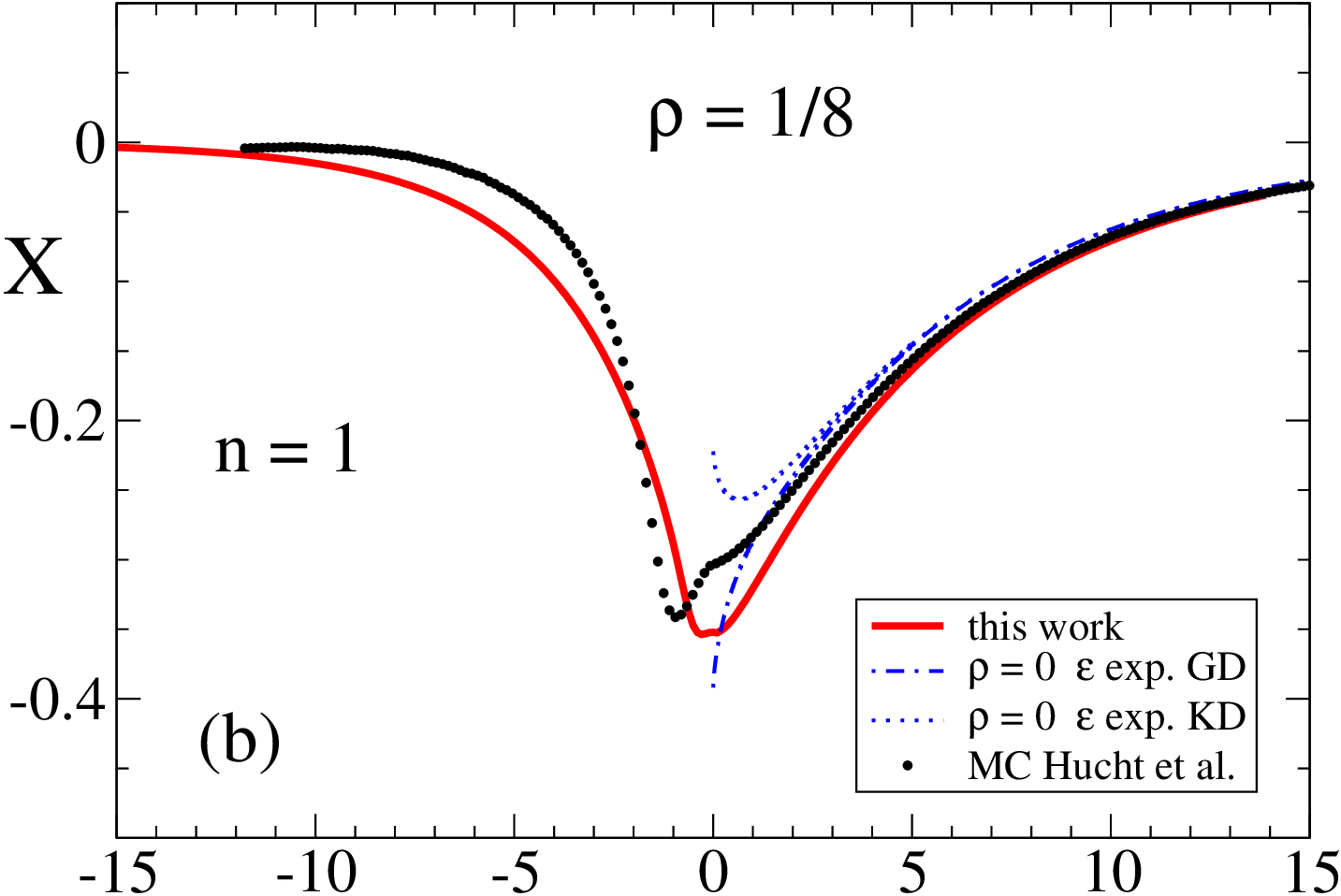}}
\subfigure{\includegraphics[clip,width=7.0cm]{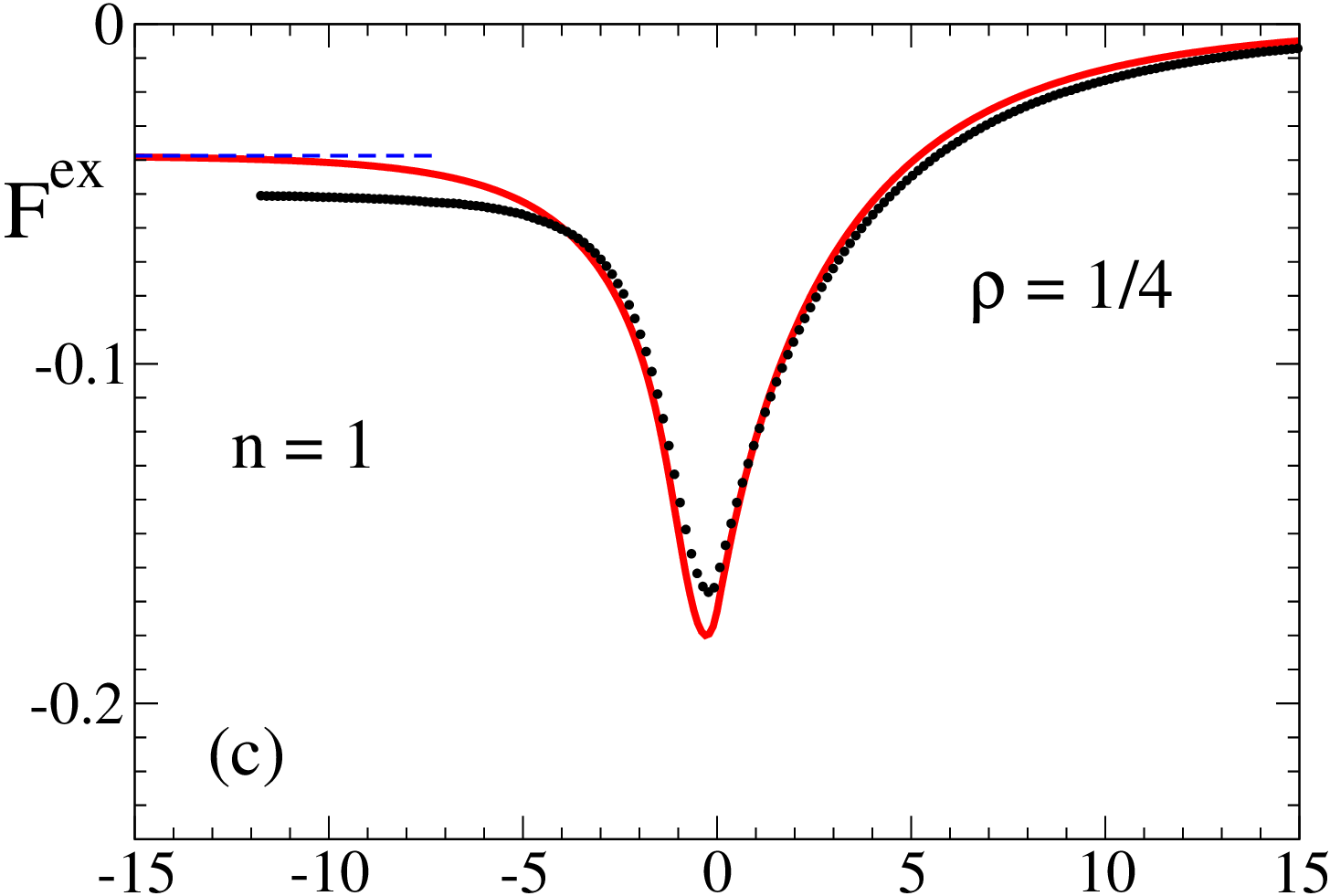}}
\subfigure{\includegraphics[clip,width=7.0cm]{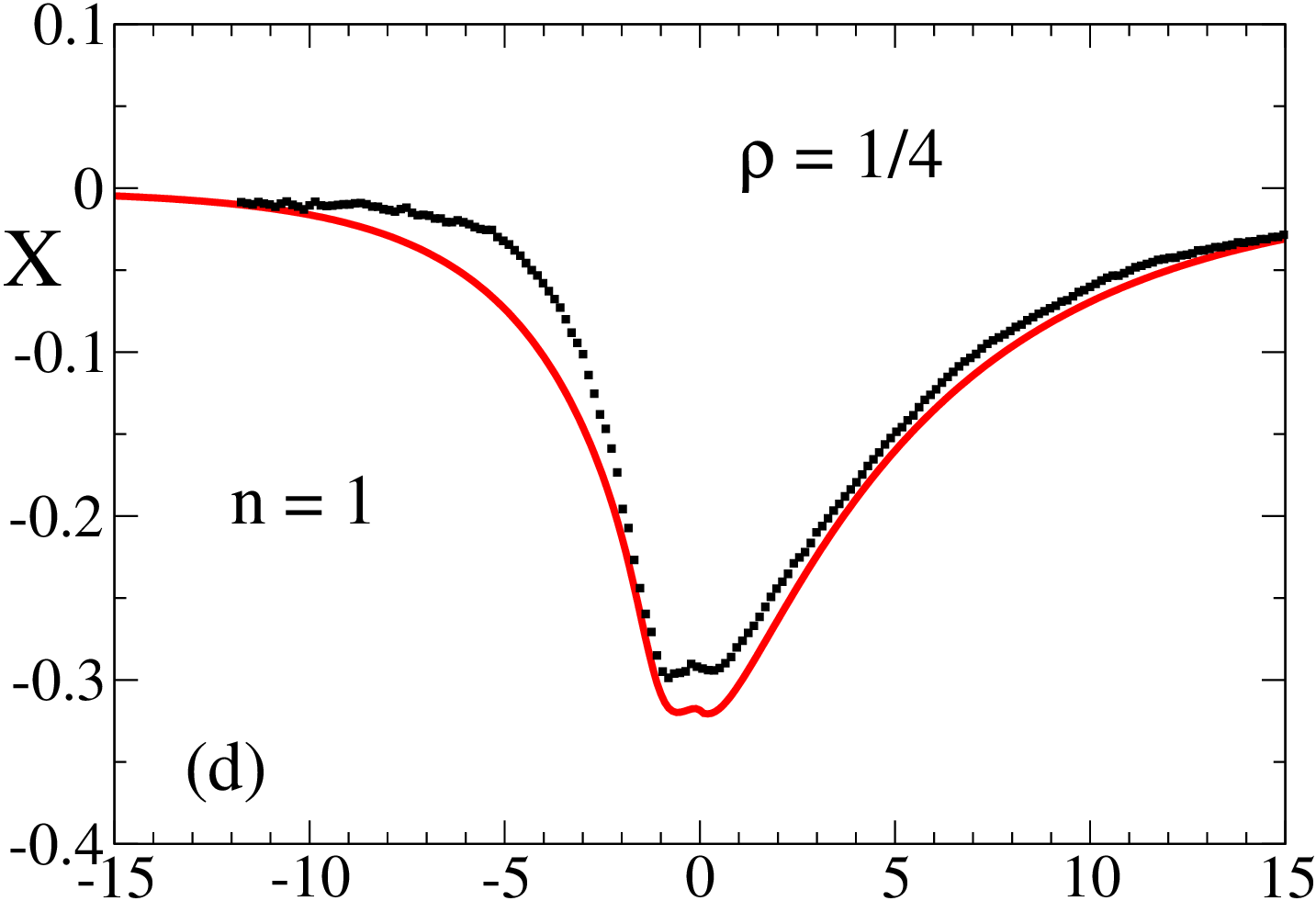}}
\subfigure{\includegraphics[clip,width=7.0cm]{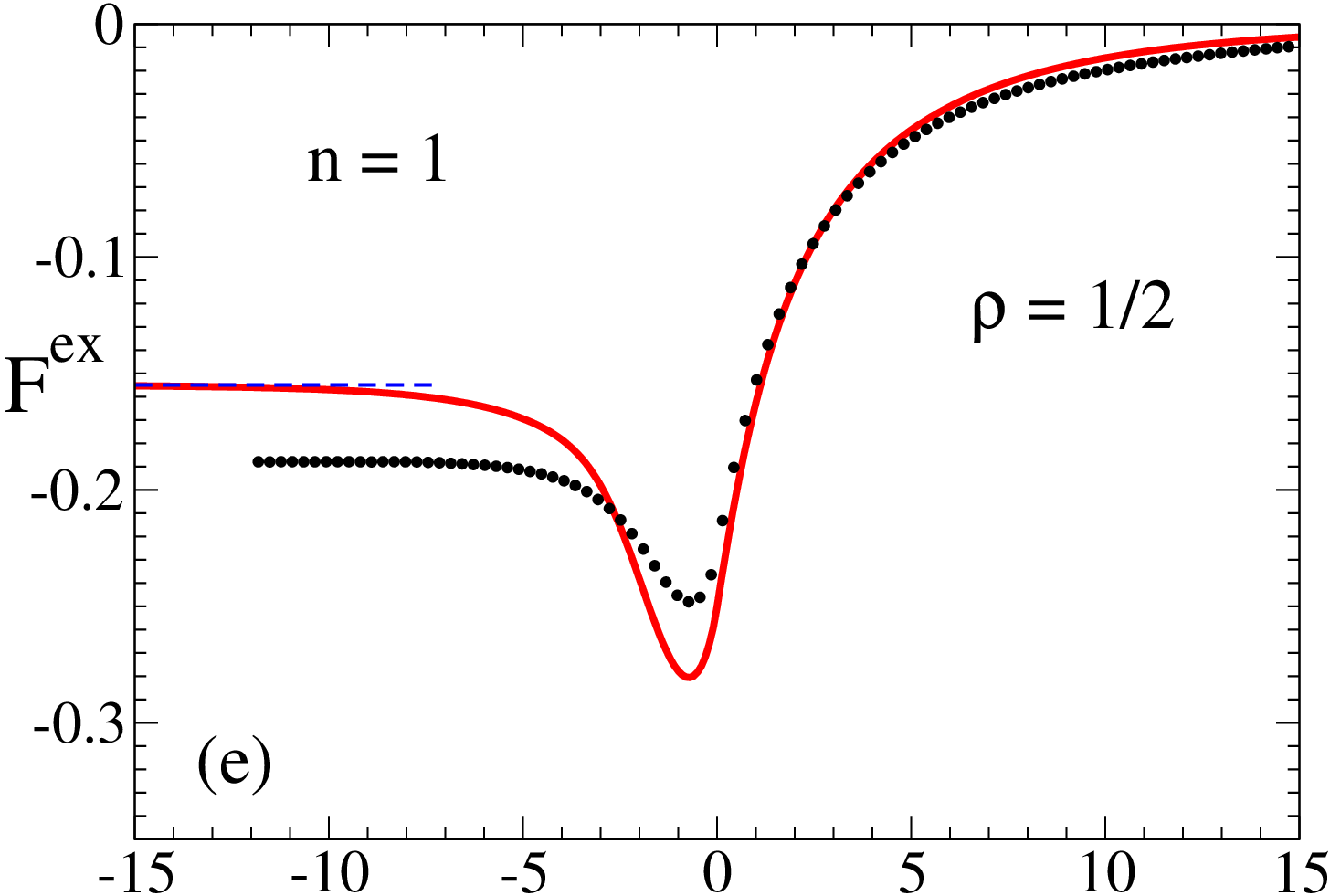}}
\subfigure{\includegraphics[clip,width=7.0cm]{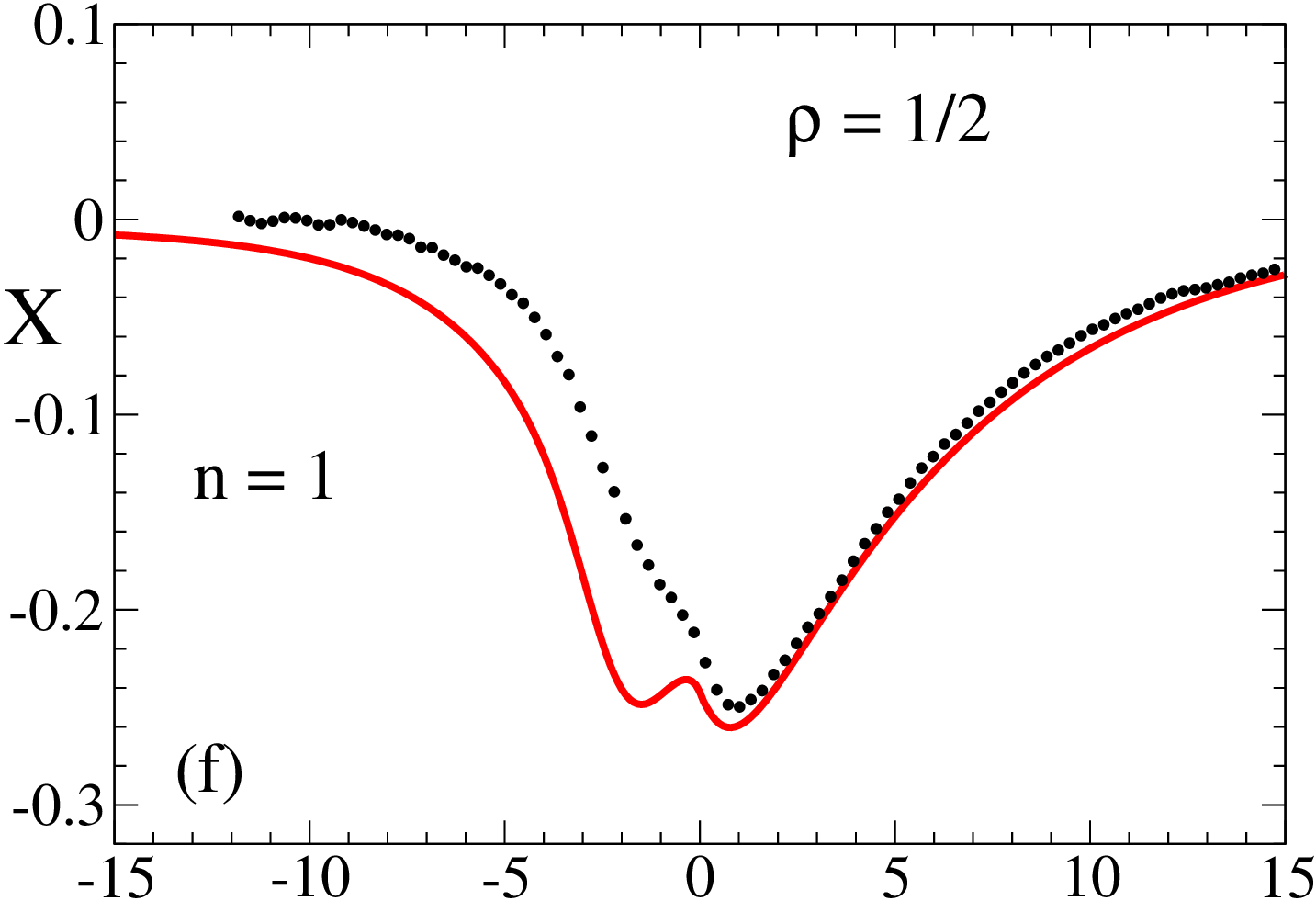}}
\subfigure{\includegraphics[clip,width=7.0cm]{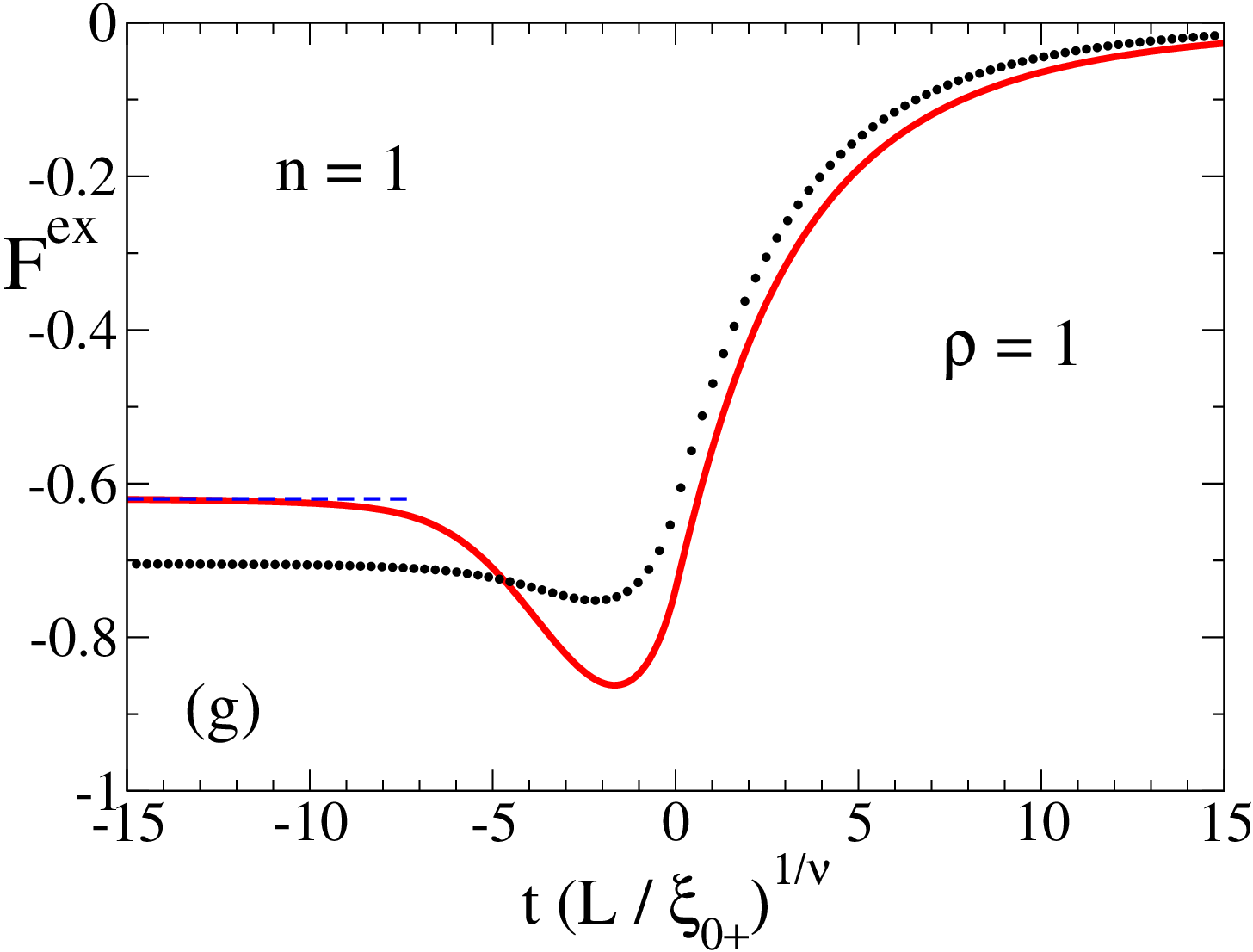}}
\subfigure{\includegraphics[clip,width=7.0cm]{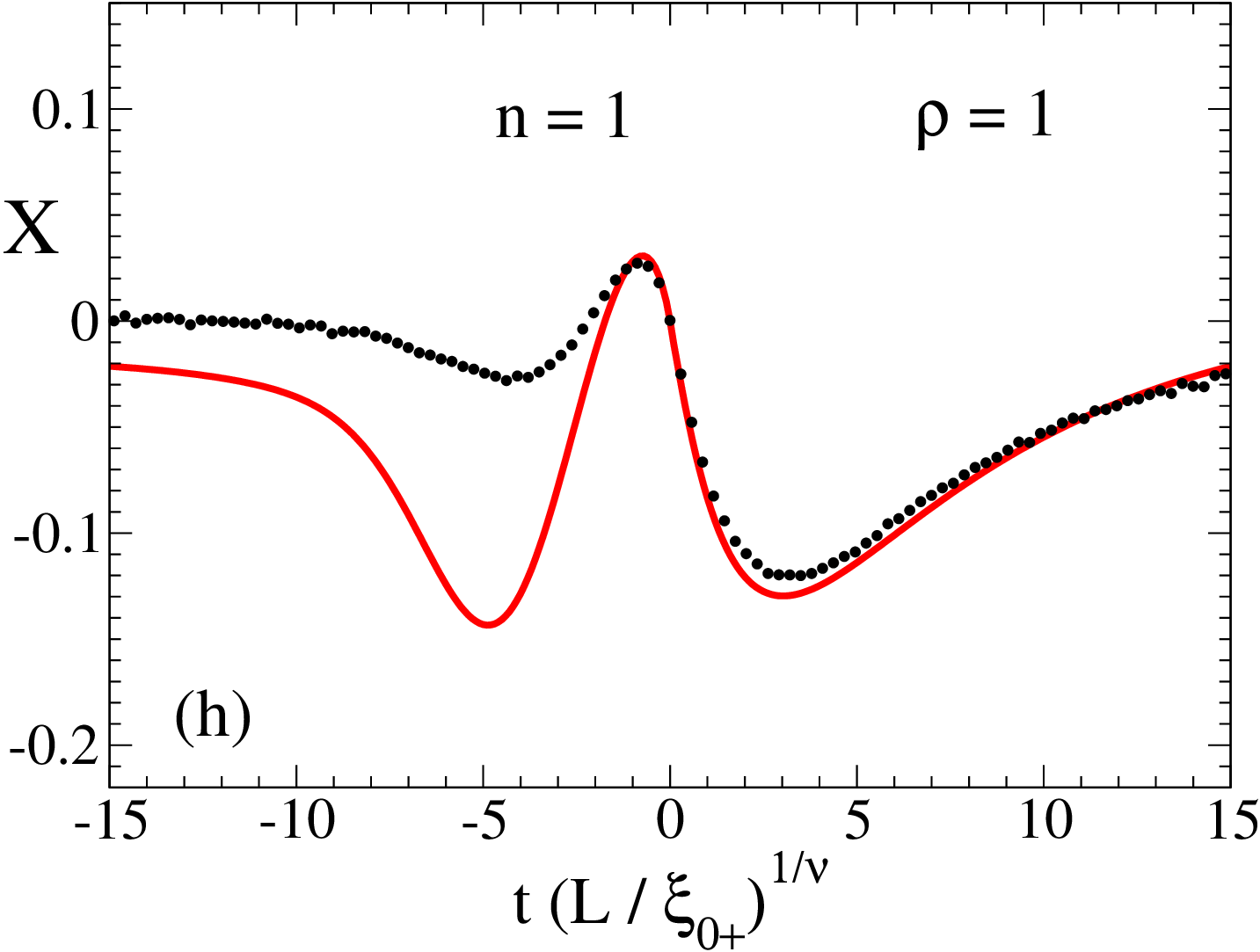}}
\end{center}
\caption{(Color online) Scaling functions $F^{ex}(\tilde x, \rho)$ in left panels and $X(\tilde x, \rho)$ in right panels in slab geometry for several aspect ratios $\rho$ as a function of $\tilde x = t (L/\xi_{0+})^{1/\nu}$  for $d=3, n=1$.  Solid lines: from (\ref{scalfreeaniso}) and (\ref{substisoX}). Dots: MC data  for  the $d=3$ Ising model from \cite{hucht2011}; they approach the finite low-temperature limit $-\rho^2 \ln 2$ as predicted analytically for the Ising model \cite{Privman-Fisher,dohm2010}. The solid lines in the left panels approach the finite limits $F^{ex}(-\infty,\rho)= -0.6199 \rho^2 $ (horizontal dashed lines), (\ref{lowTemplx}). The solid lines in the right panels go to zero for $\tilde x \to -\infty$. Dotted and dot-dashed lines in (a) and (b): $\varepsilon$ expansion for $\rho=0$ of \cite{KrDi92a,GrDi07}.}
\end{figure*}
\vspace*{0cm}
\end{widetext}

(i) {\it Case} $n=1$: For $d>2$ we obtain from (\ref{excess-scalfreeaniso18}) - (\ref{Xbelow})
\begin{eqnarray}
\label{lowTemplx}
&&\lim_{\tilde x \to -\infty}F_-^{ex}(\tilde x, \rho)= \;\rho^{d-1}\;\Big\{-\ln2 + 12u^*/d \nonumber\\&&
-24u^*\alpha\big[1 / (64
u^*) \; + 1/(4d) \; + \; \nu B (u^*) / (8 \alpha)\big]\Big\}, \;\,\,\,\,\;\,\,\,\,\,\,\,\\
\label{lowXn1}
&&\lim_{\tilde x \to -\infty}X_-(\tilde x,\rho)=0.
\end{eqnarray}
The finite low-temperature amplitude (\ref{lowTemplx}) for a {\it finite} geometry is in contrast to the vanishing low-temperature limit of $F^{ex}$ for {\it film} geometry for $n=1$ (Sec. V. F.).

The first term $-\rho^{d-1}\ln 2$ in (\ref{lowTemplx}) reproduces the exact analytic result for finite $d$-dimensional Ising models  at $T=0$ reflecting the twofold degeneracy of the ground state, with the two  homogeneous configurations of the spin-up  and spin-down phases at $h=0$ \cite{Privman-Fisher}. The additional terms  in (\ref{lowTemplx}) come from the algebraic decay $\propto (n-2)/\tilde y$ of the lowest-mode function $\vartheta_{2,n}(\tilde y)$ in (\ref{theta2largebelow}) for $n\neq 2$; they should be complemented by contributions of $O(u^*)$ that will arise from the $H^{(4)}$, (\ref{higher34}). Such contributions can be determined by a two-loop calculation within the ansatz proposed in Sec. VI. B of \cite{dohm2011} which should yield finite $O(u^*)$ corrections to Eq. (6.10) of \cite{dohm2011}.

Unlike the fixed-length spin variables $s_i = \pm 1$ of  Ising models, the "soft-spin" continuous variables $\varphi_i$ of the $\varphi^4$  Hamiltonian for $n=1$ (which belongs to the same universality class as Ising Hamiltonians) vary from $- \infty$ to $\infty$. Thus there exists no exact twofold degeneracy of the ground state of the $\varphi^4$ model in the presence of a {\it finite} four-point coupling $u_0$. We recall that fixed-length spin models and the $\varphi^4$ model become equivalent in the limit $u_0 \to \infty$ at fixed ratio $r_0/u_0$ \cite{cd2000-3}. But our finite-size calculations are performed at {\it finite} $u_0$. In general, there is no {\it a priori} reason to expect universality with regard to the low-temperature behavior
of different systems belonging to the same universality class.
Consequently, no quantitative agreement  between the MC data for the Ising model \cite{hucht2011, hasenbusch2009} and the results of our $\varphi^4$ model far below $T_c$ can be postulated even if the latter model could be treated exactly. This aspect was not taken into account in a comment in \cite{dohm2013} on  a presumed non-applicability  of Eq. (8) of \cite{dohm2013} to $F(\tilde x, \rho)$ for $n=1$ well below $T_c$. The fact that the low-temperature tails $\propto \rho^{d-1}$ of $F^{ex}$ of our approximate theory deviate from the Ising model limit $-\rho^{d-1} \ln 2$ in the left panels of Fig. 2 is not a shortcoming of our finite-size RG approach but should be attributed to the nonuniversal difference between a fixed-length spin and a soft-spin model. Our theory should be well applicable to the Casimir force of the $n=1$ universality class. Further studies should be devoted, however, to the issue of a possible nonuniversality of the low-temperature behavior of $f^{ex}$ for $n=1$. MC studies of  $F^{\text ex}$ for the $d=3$ $ \varphi^4$ model \cite{hasenbusch1999} rather than for the $d=3$ Ising would be desirable.

The constant terms $\propto \rho^{d-1}$ of $F^{ex}(-\infty,\rho)$ do not contribute to $X$ which is in good overall agreement with the MC data in the right panels of Fig. 2, except for the case $\rho=1$ well below $T_c$. The dotted and dot-dashed lines in Figs. 2(a) and (b) represent the $\varepsilon = 4-d$ expansion results \cite{KrDi92a,GrDi07} in film geometry which agree well with the MC data \cite{hucht2011} for $\rho\ll 1$ above $T_c$ but yield an unphysical cusp-like behavior at $T_c$.

(ii) {\it Case} $n > 1$: For $d>2$ a nontrivial prediction of our theory is the logarithmic divergence of $F_-^{ex}$,  with the leading term for $\tilde x \to -\infty$ for general $n>1$
\begin{eqnarray}
\label{logasym}
 - [(n-1)/2]  \;\nu(d-2) \;\rho^{d-1} \; \ln (2 |\tilde x|Q^*) \;\; \;\;\; \;\;\;
\end{eqnarray}
as follows from (\ref{excess-scalfreeaniso18}). Such a logarithmic divergence of $F_-^{ex}$ was found previously \cite{dohm2011} in the large-$n$ limit. Almost perfect agreement is found between our theory for $n=2$ and the MC data for  the $d=3$ $XY$ model \cite{hasenbusch2011} in Figs. 3(a), (c), and (g) in the low-temperature regime $\tilde x \lesssim -5$ where the logarithmic term dominates. It is remarkable that the divergent term (\ref{logasym}) agrees with the logarithmic term of the exact result (per component) (\ref{4hhx}) in the large-$n$ limit where $\nu= (d-2)^{-1}, Q^*=1$. The divergence in the low-temperature limit for a finite geometry is in contrast to the finite low-temperature limit in {\it film} geometry (Sec. V.F).

For a slab geometry with $\rho=1/6$ there is also excellent overall agreement of our prediction of $X(\tilde x,\rho)$ with the MC data \cite{vasilyev2009} for the $d=3$ $XY$ model as shown in Fig. 3(b). For larger values of $\rho \gtrsim 1/2$, our curves of $X$ for $n=2$ exhibit a local maximum slightly below $T_c$. Further theoretical and MC studies are necessary in order to clarify wether or not these local maxima are an artifact of our approximation. Such local maxima below $T_c$ do not appear in the exact results of the large-$n$ limit (Fig. 6, see also Figs. 3(b) and 4(b) of \cite{dohm2011}).

We briefly comment on the comparison between theories and MC data for $n=3$ in Fig. 4. There are systematic deviations of the MC data \cite{dan-krech,danprivate} from our predictions for $\rho=1/6$ for all temperatures and from the theoretical curves of \cite{KrDi92a,GrDi07}  for $\rho=0$ above $T_c$. A similar discrepancy exists for $n=2$ between the MC data in Fig. 3 of \cite{dan-krech} and those in Fig. 6 of \cite{vasilyev2009} for $\rho=1/6$ . As noted in \cite{vasilyev2009}, this  discrepancy for $n=2$ might be due to to the uncertainty in the normalization factor for the vertical scale of the MC data in \cite{dan-krech}. We suspect that the same uncertainty is the reason for the deviations of the $n=3$ MC data \cite{dan-krech} from the theoretical curves in Fig. 4.

\begin{widetext}
%=======================================================
\begin{figure*}[!ht]
\begin{center}
\subfigure{\includegraphics[clip,width=7.0cm]{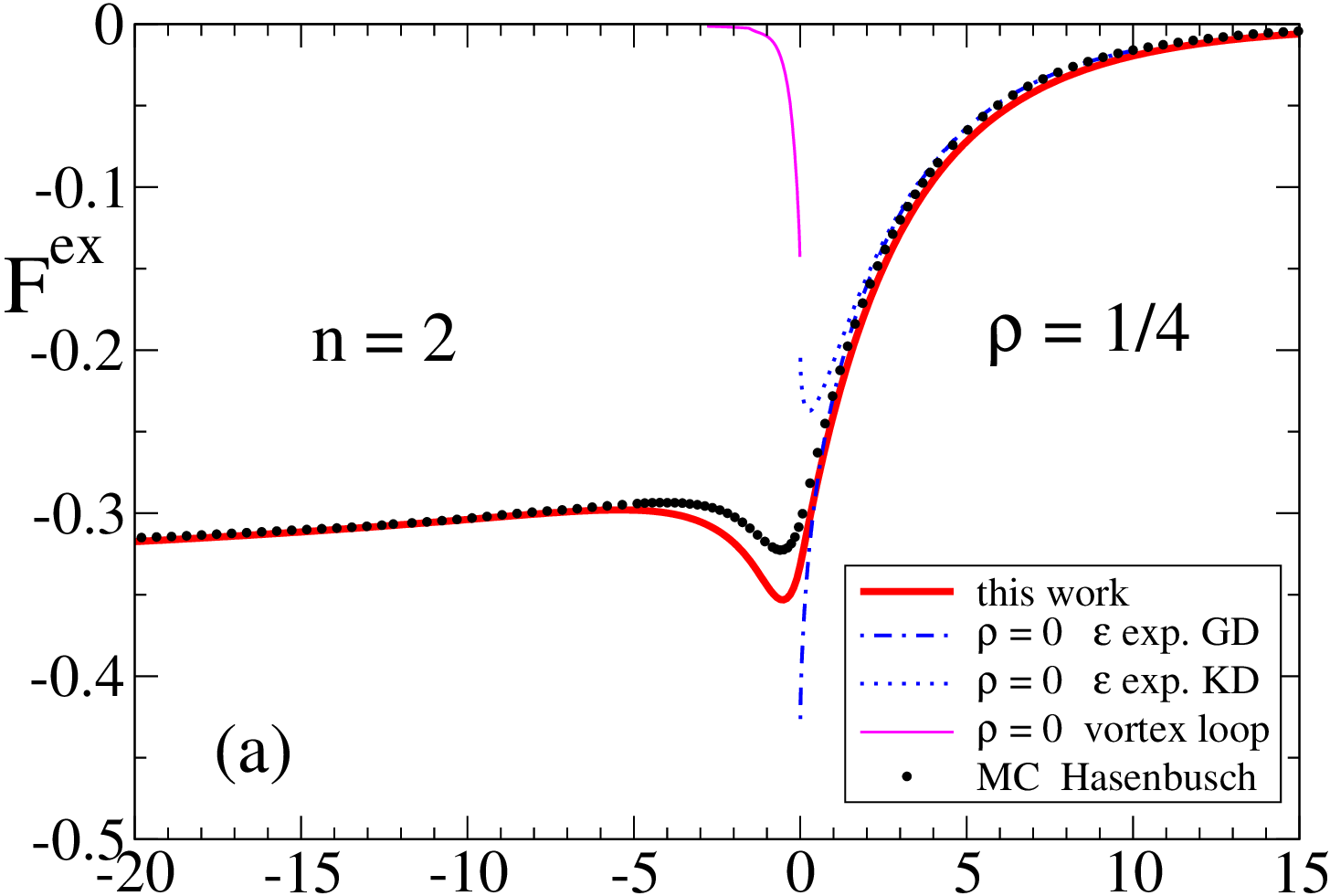}}
\subfigure{\includegraphics[clip,width=7.0cm]{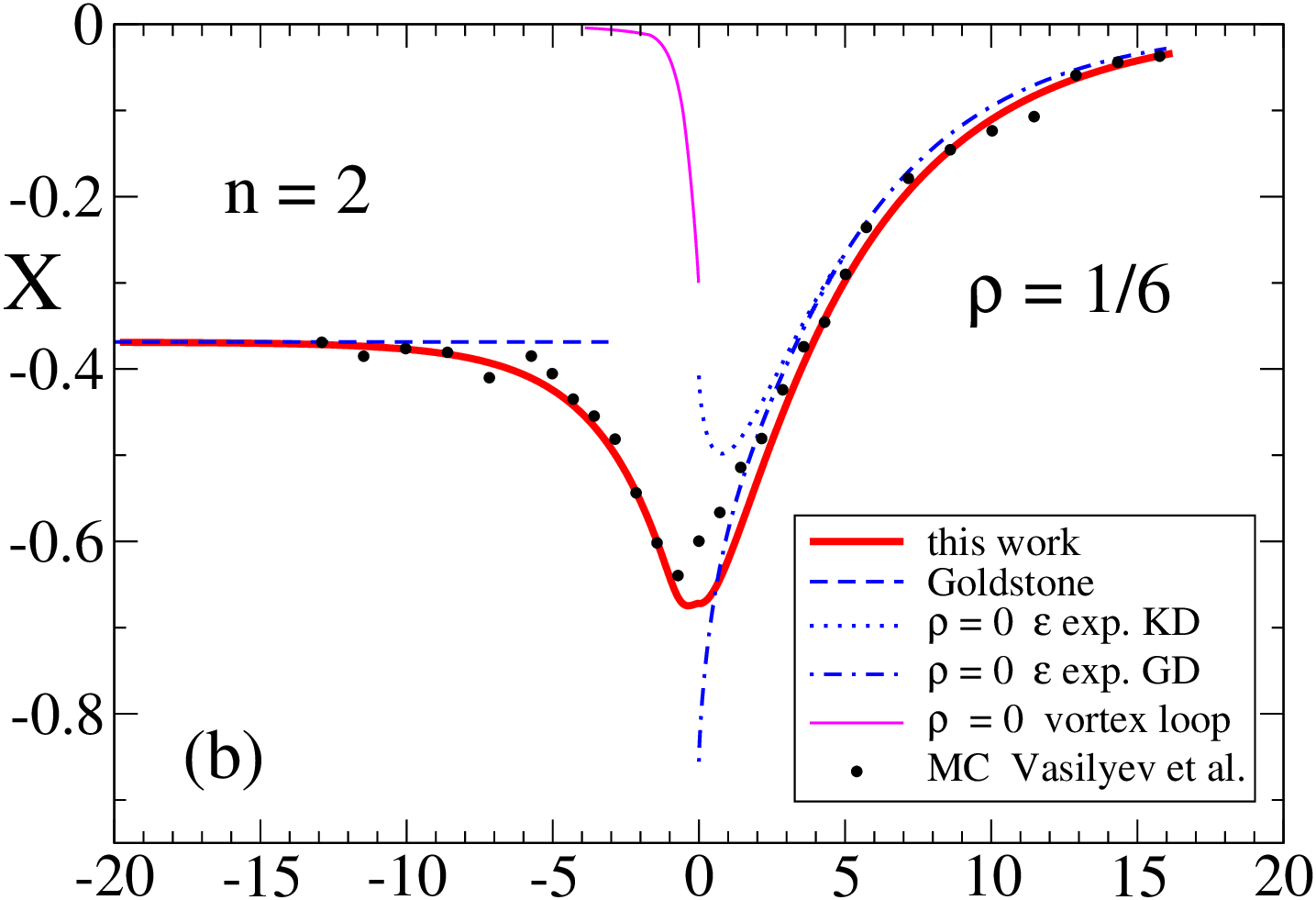}}
\subfigure{\includegraphics[clip,width=7.0cm]{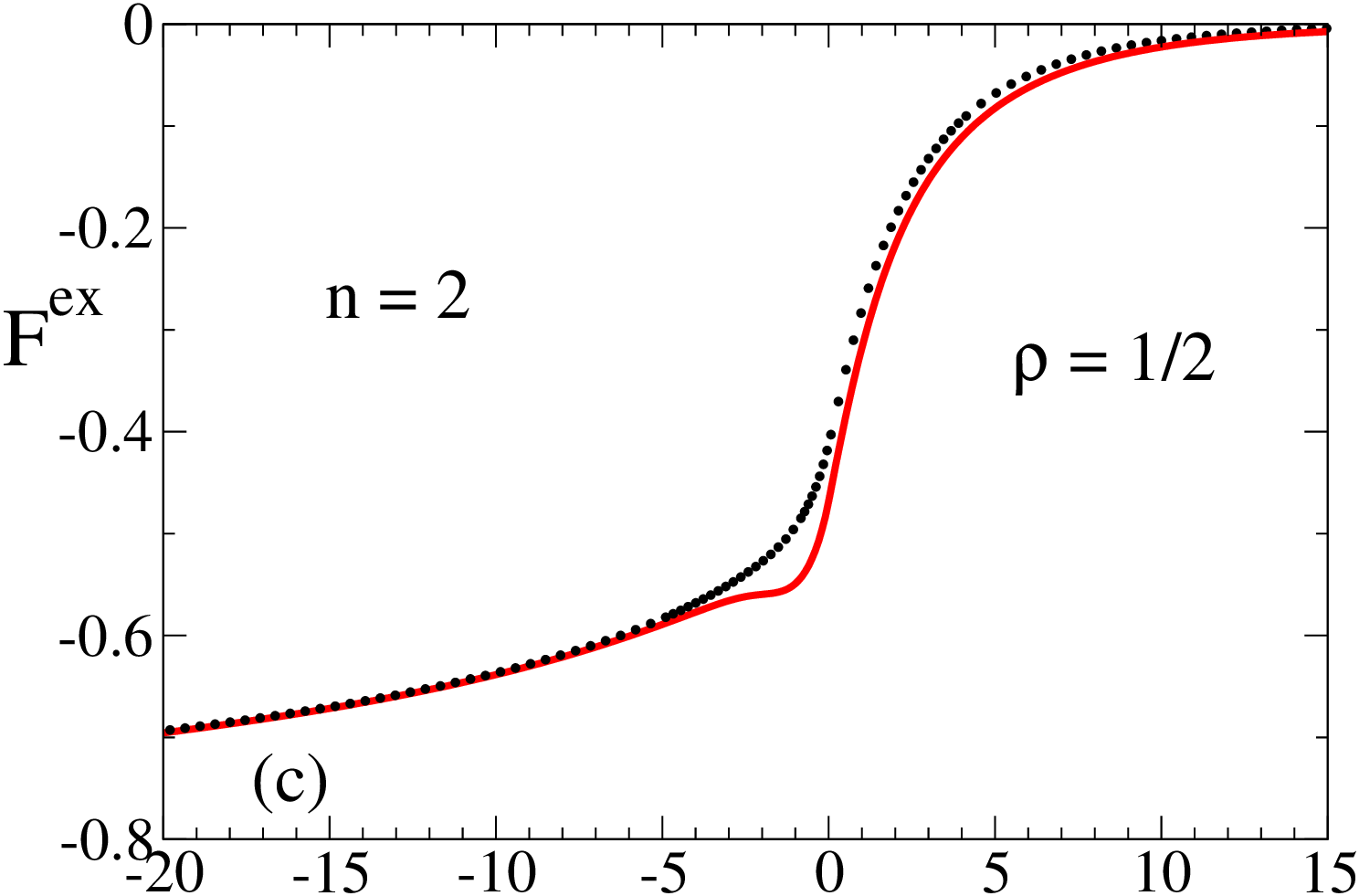}}
\subfigure{\includegraphics[clip,width=7.0cm]{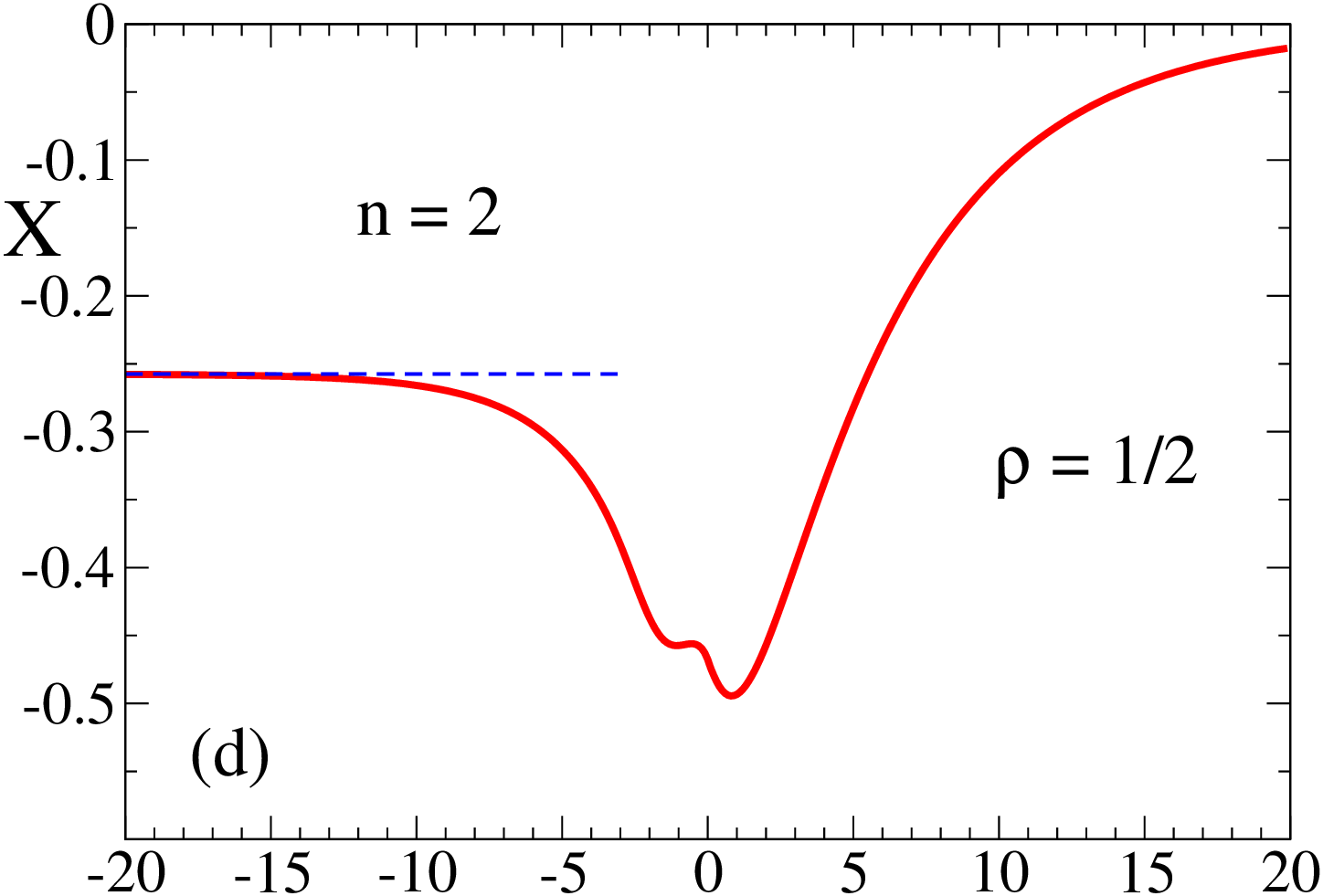}}
\subfigure{\includegraphics[clip,width=7.0cm]{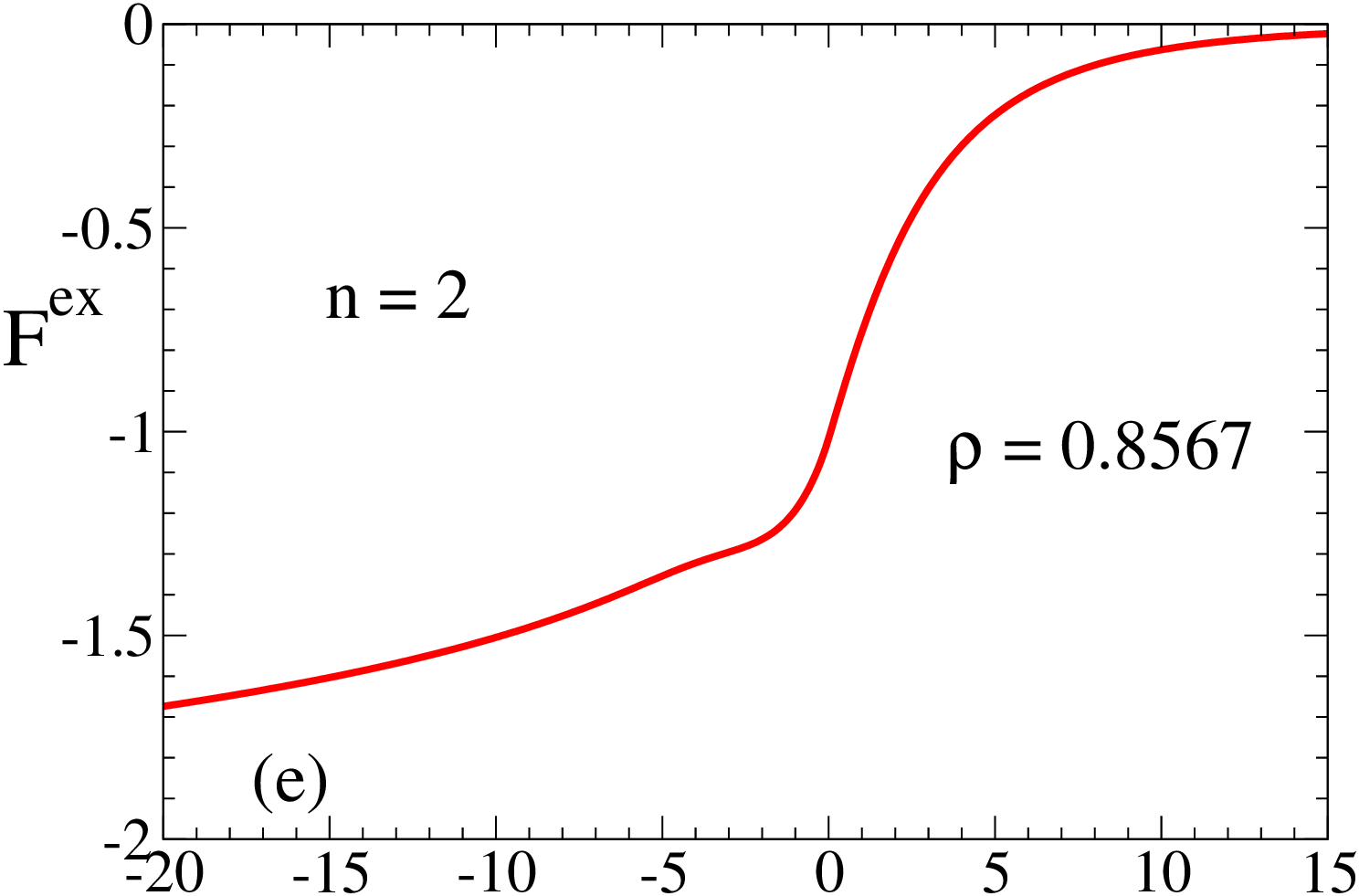}}
\subfigure{\includegraphics[clip,width=7.0cm]{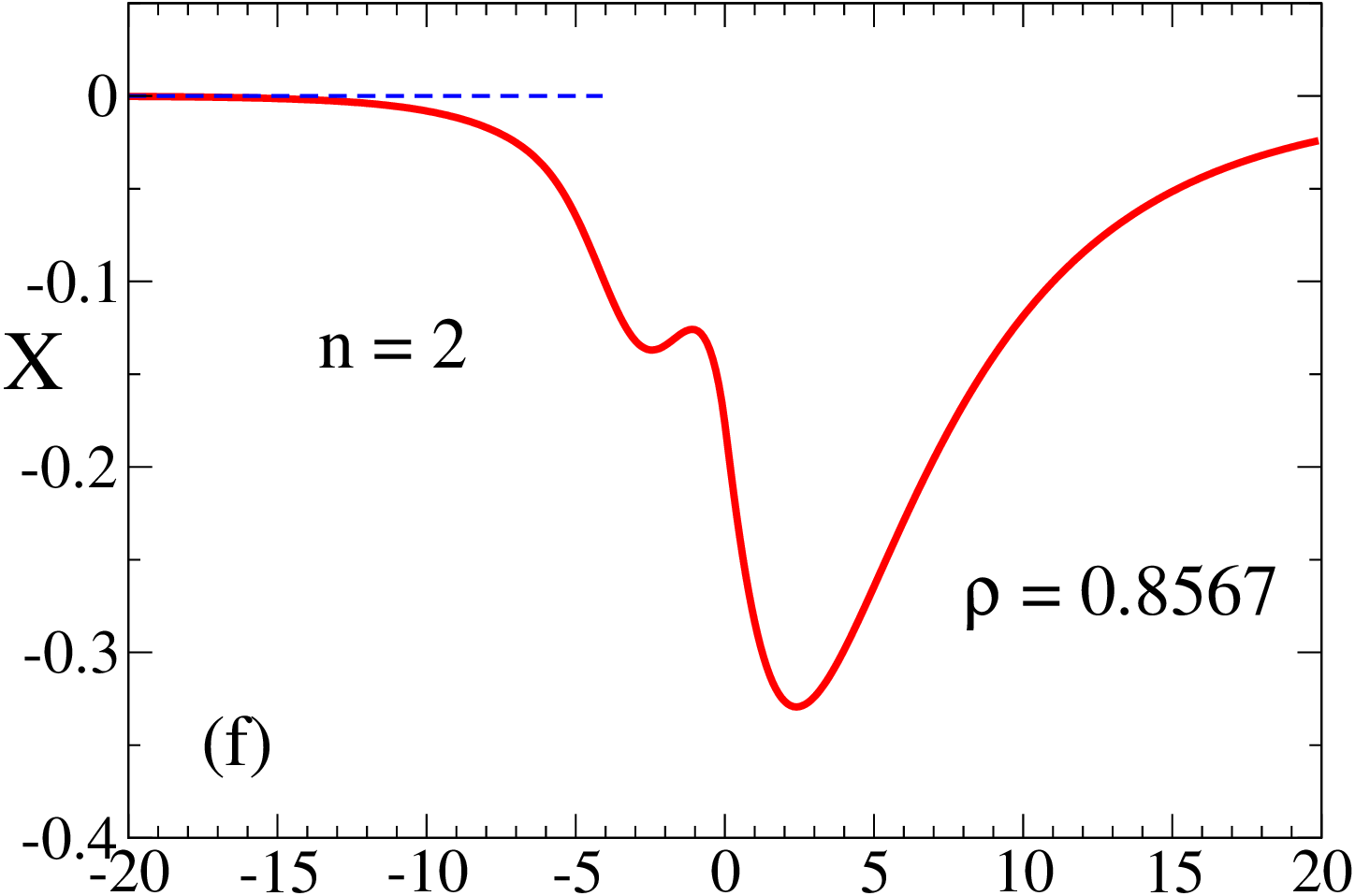}}
\subfigure{\includegraphics[clip,width=7.0cm]{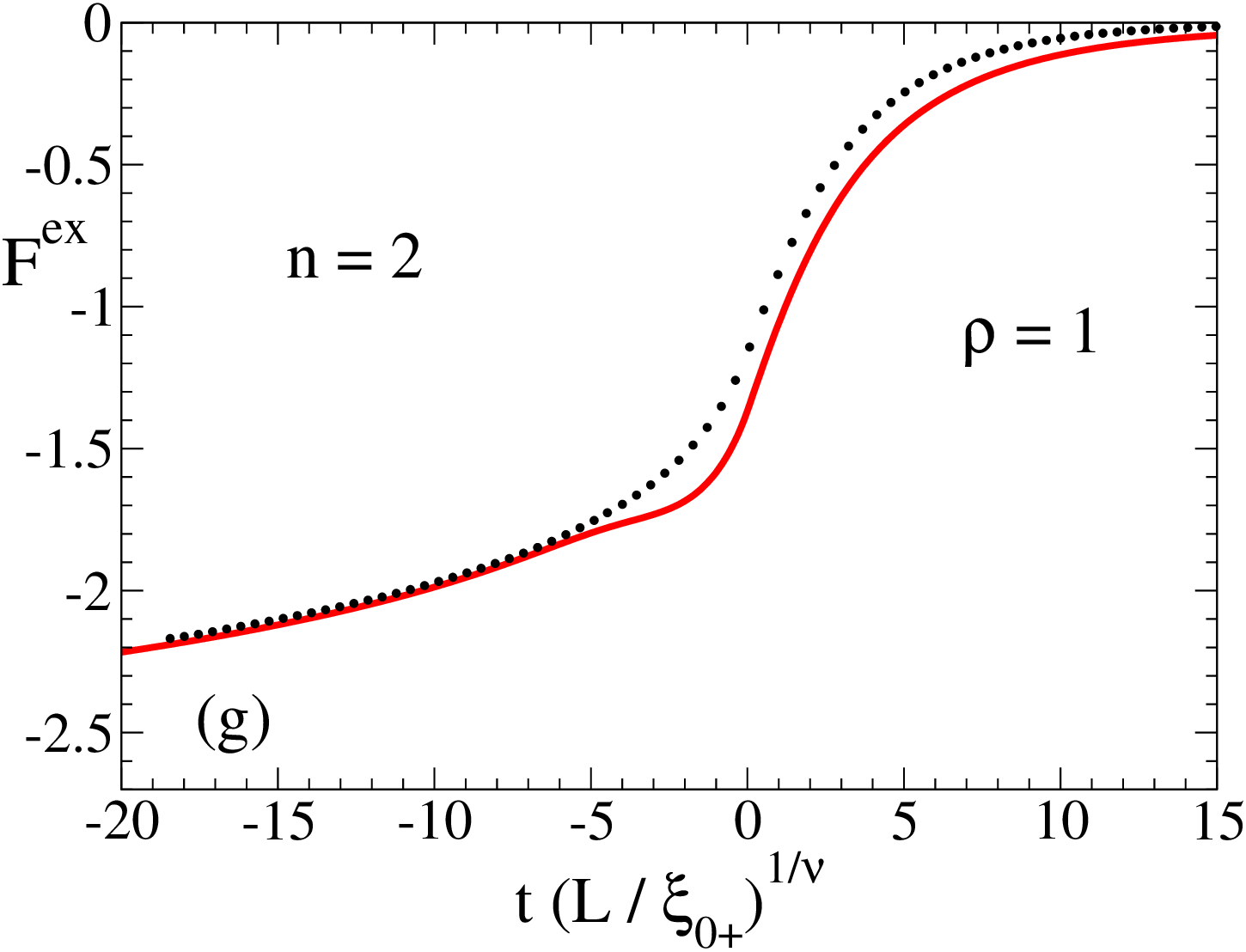}}
\subfigure{\includegraphics[clip,width=7.0cm]{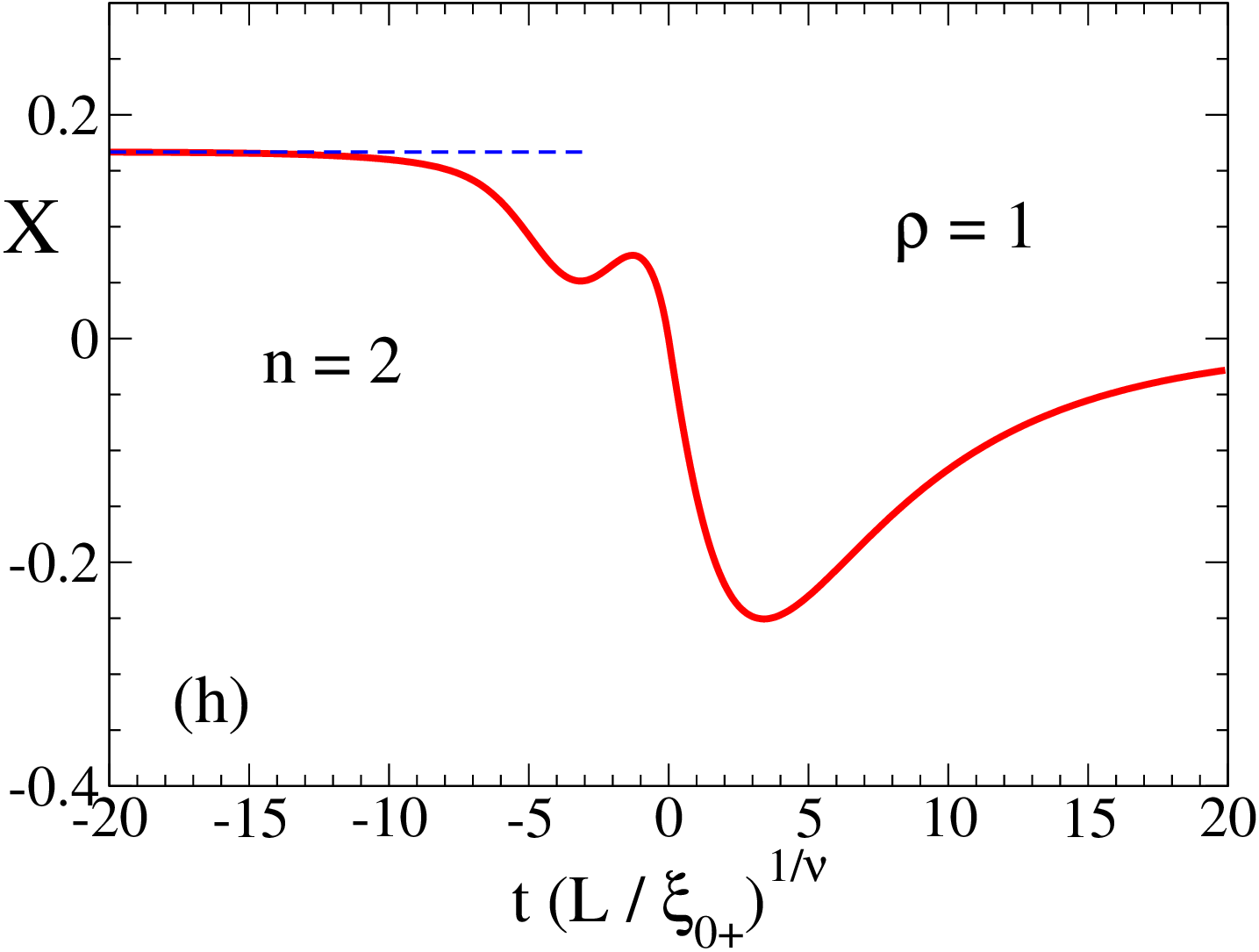}}
\end{center}
\caption{(Color online) Scaling functions $F^{ex}(\tilde x, \rho)$ in left panels and $X(\tilde x, \rho)$ in right panels in slab geometry for several aspect ratios $\rho$ as a function of $\tilde x = t (L/\xi_{0+})^{1/\nu}$  for $d=3, n=2$.  Solid lines: from (\ref{scalfreeaniso}) and (\ref{substisoX}). Dots: MC data  for  the $d=3$ XY model from \cite{hasenbusch2011} in (a), (c), (g) and from \cite{vasilyev2009} in (b).  Horizontal dashed lines in right panels: finite low-temperature limit $X(-\infty,\rho)$, (\ref{lowXn2}). The solid lines in the left panels diverge toward $-\infty$ for $\tilde x \to -\infty$. Dotted, dot-dashed, and thin lines in (a) and (b) for $\rho=0$ from $\varepsilon$ expansion \cite{KrDi92a,GrDi07} and from vortex-loop renormalization  \cite{wil-1}, respectively.  }
\end{figure*}
\vspace*{0cm}
\end{widetext}
\begin{widetext}
%=======================================================
\begin{figure*}[!ht]
\begin{center}
\subfigure{\includegraphics[clip,width=7.0cm]{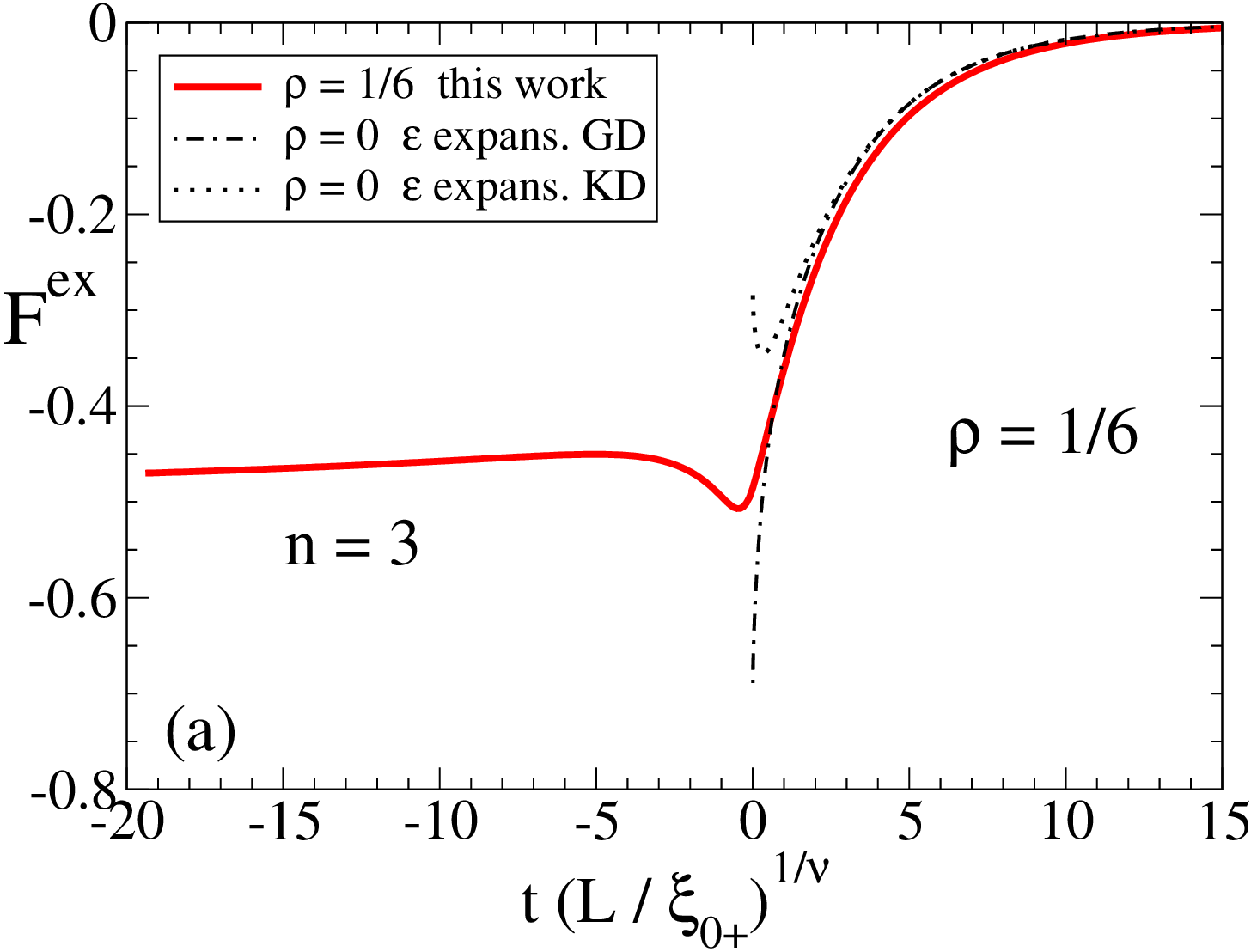}}
\subfigure{\includegraphics[clip,width=7.0cm]{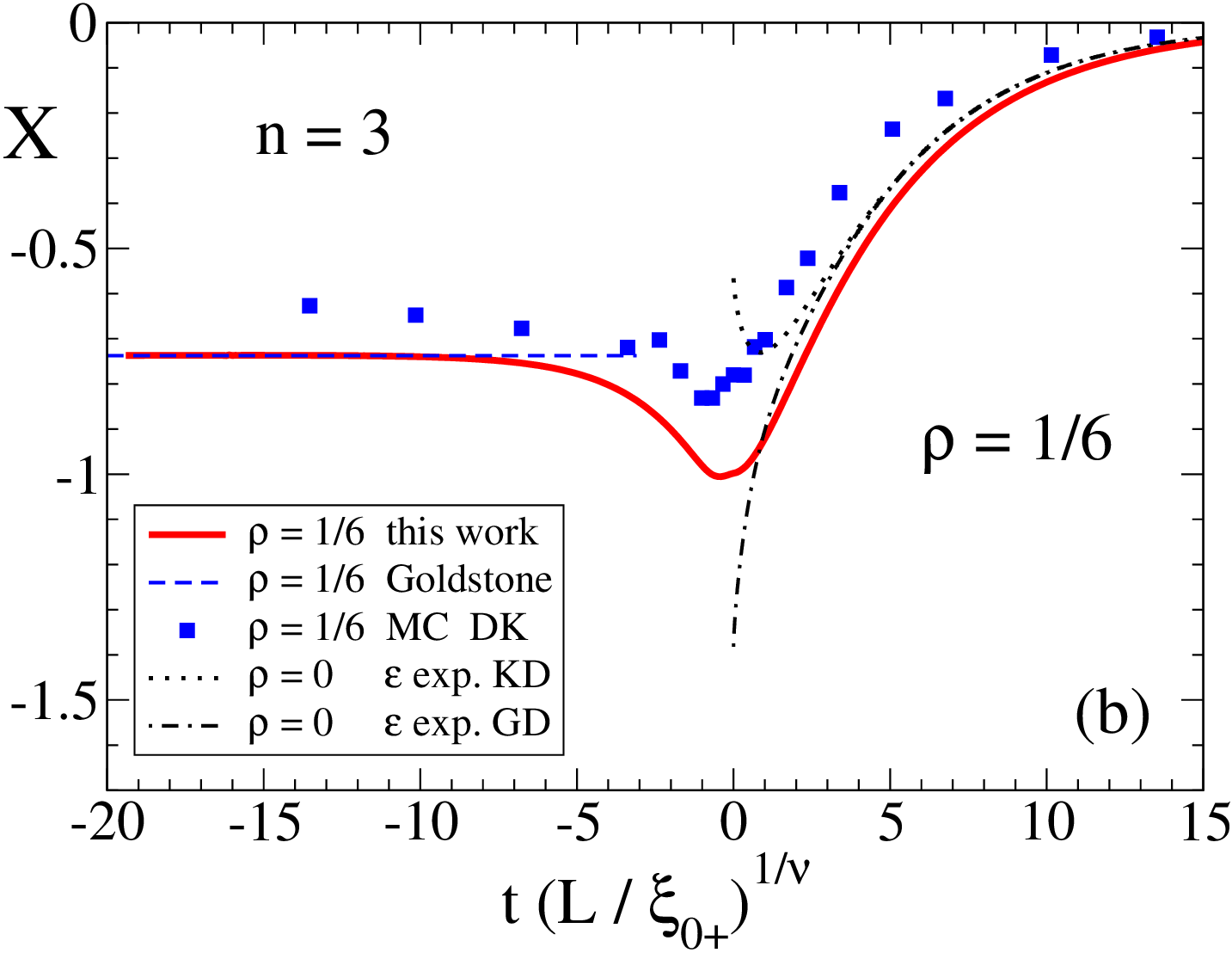}}
\end{center}
\vspace{0.5cm}
\caption{(Color online) Solid lines: Scaling functions  (a) of the excess free energy density and (b) of the Casimir force in slab geometry with $\rho=1/6$ for $d=3, n=3$. Solid lines: from (\ref{scalfreeaniso}) and (\ref{substisoX}). Dotted and dot-dashed lines: $\varepsilon$ expansion for $\rho=0$ of \cite{KrDi92a,GrDi07}. MC data in (b) for  the Heisenberg model from \cite{dan-krech,danprivate}. Horizontal dashed line in (b) represents the low-temperature limit $X(-\infty,1/6)=-0.7375$, (\ref{lowXn2}), due to the Goldstone modes.  The solid line in (a) diverges logarithmically toward $-\infty$ for $\tilde x \to -\infty$. The vertical scale of the MC data in (b) has been adjusted in \cite{dan-krech}, see text in the third paragraph after (\ref{logasym}).}
\end{figure*}
\vspace{0cm}
\end{widetext}
\vspace{0cm}
\begin{figure}
\includegraphics[clip,width=6.0cm]
{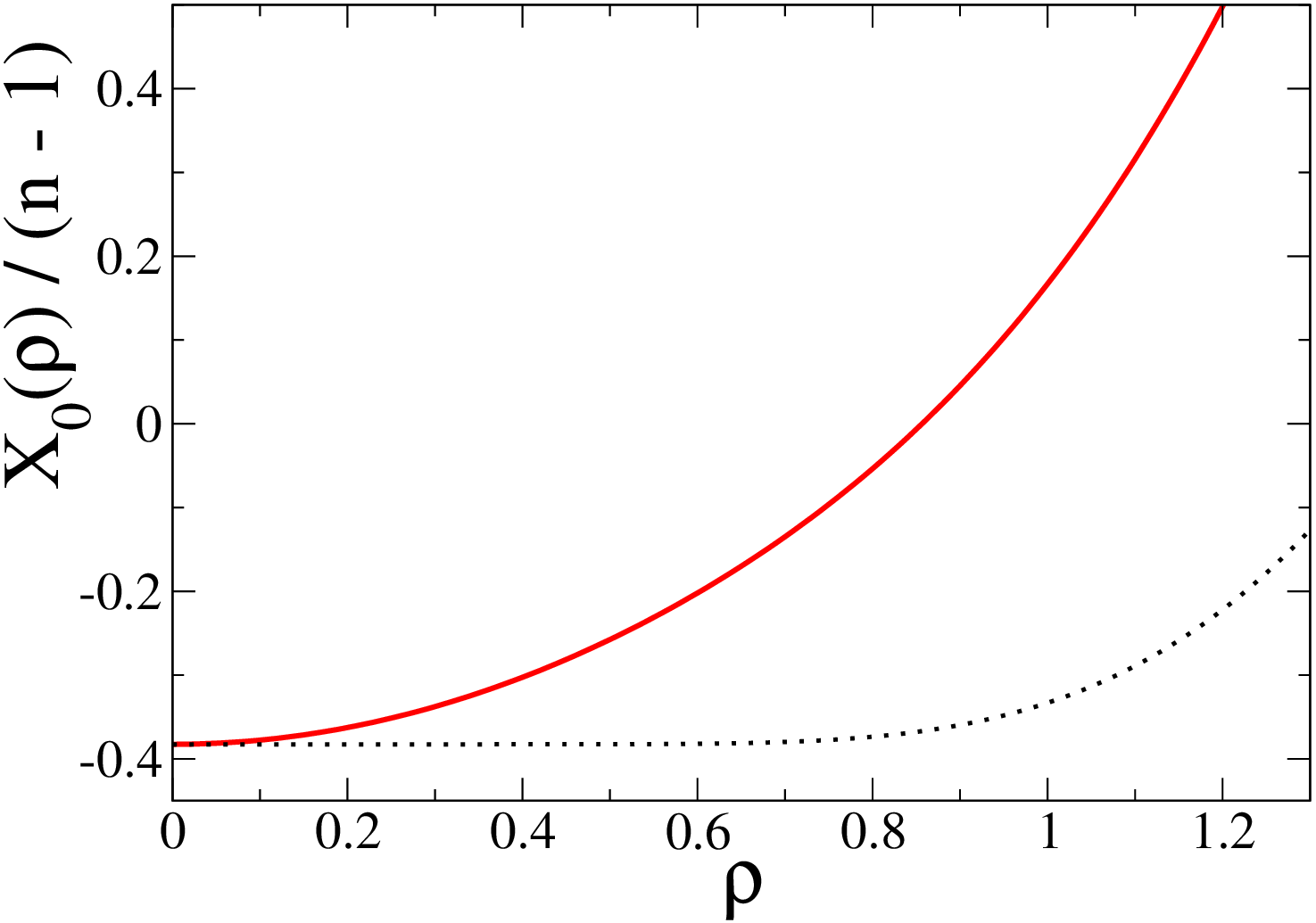}
\vspace{0cm}
\caption{(Color online) Low-temperature Casimir amplitude $X(-\infty,\rho)\equiv X_0(\rho)$, divided by $n-1$, for  $n>1,d=3$  in slab geometry as a function of the aspect ratio $\rho$.  Solid line: lowest-mode separation approach, (\ref{lowXn2}). It yields $ X_0(\rho_0)=0$ at  $\rho_0=0.8567$ for $n>1$ according to (\ref{lowXzeroiso}) and is exact in the limit $n\to \infty$ according to  (\ref{X-large-nx-low}) for arbitrary $\rho \geq 0$. Dotted line:  1-loop perturbation theory, (\ref{lowXoneloop}).}
\end{figure}
\vspace{0cm}
\vspace{0cm}
\begin{figure}
\begin{center}
\subfigure{\includegraphics[clip,width=6.6cm]{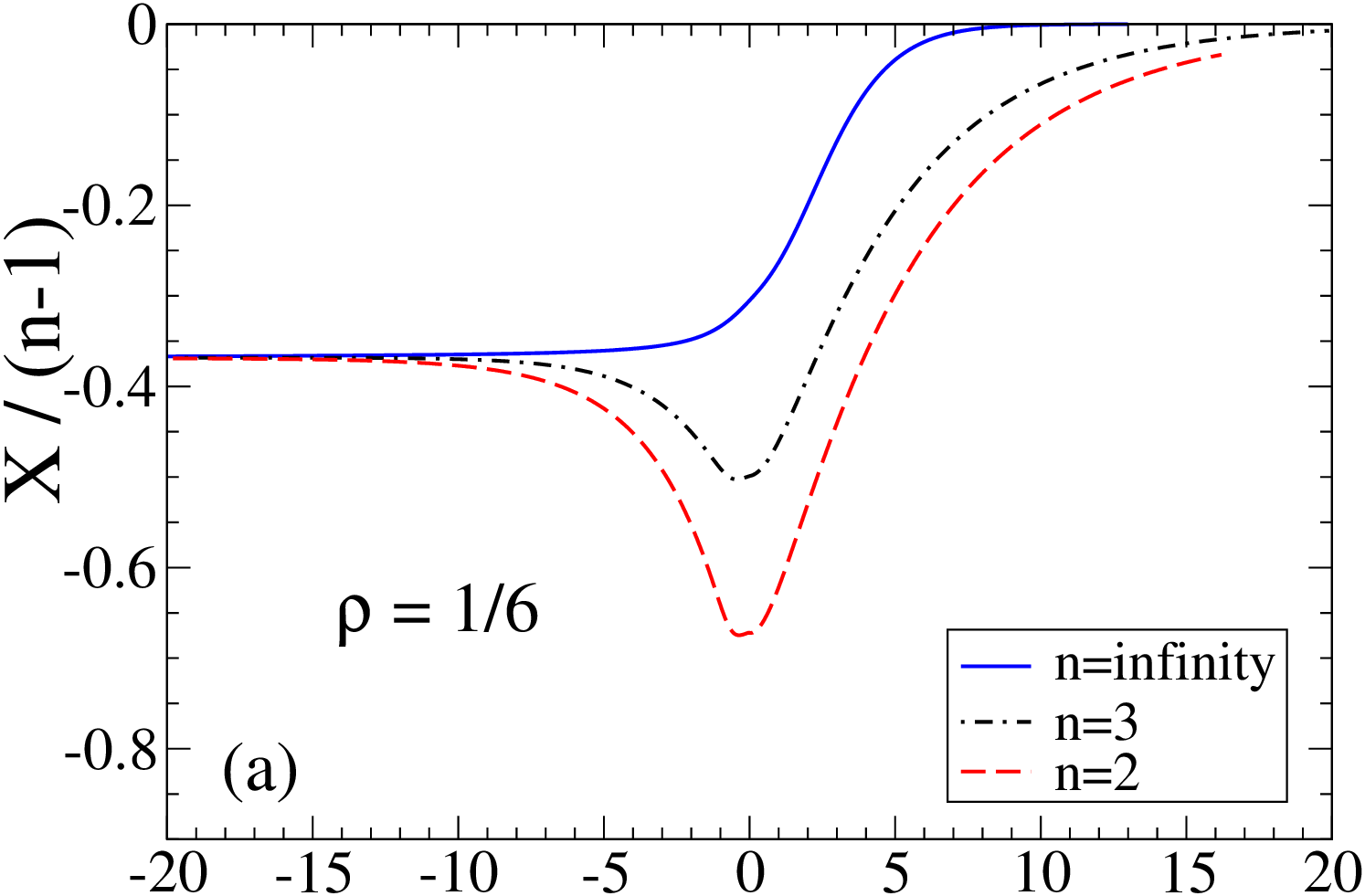}}
\subfigure{\includegraphics[clip,width=6.6cm]{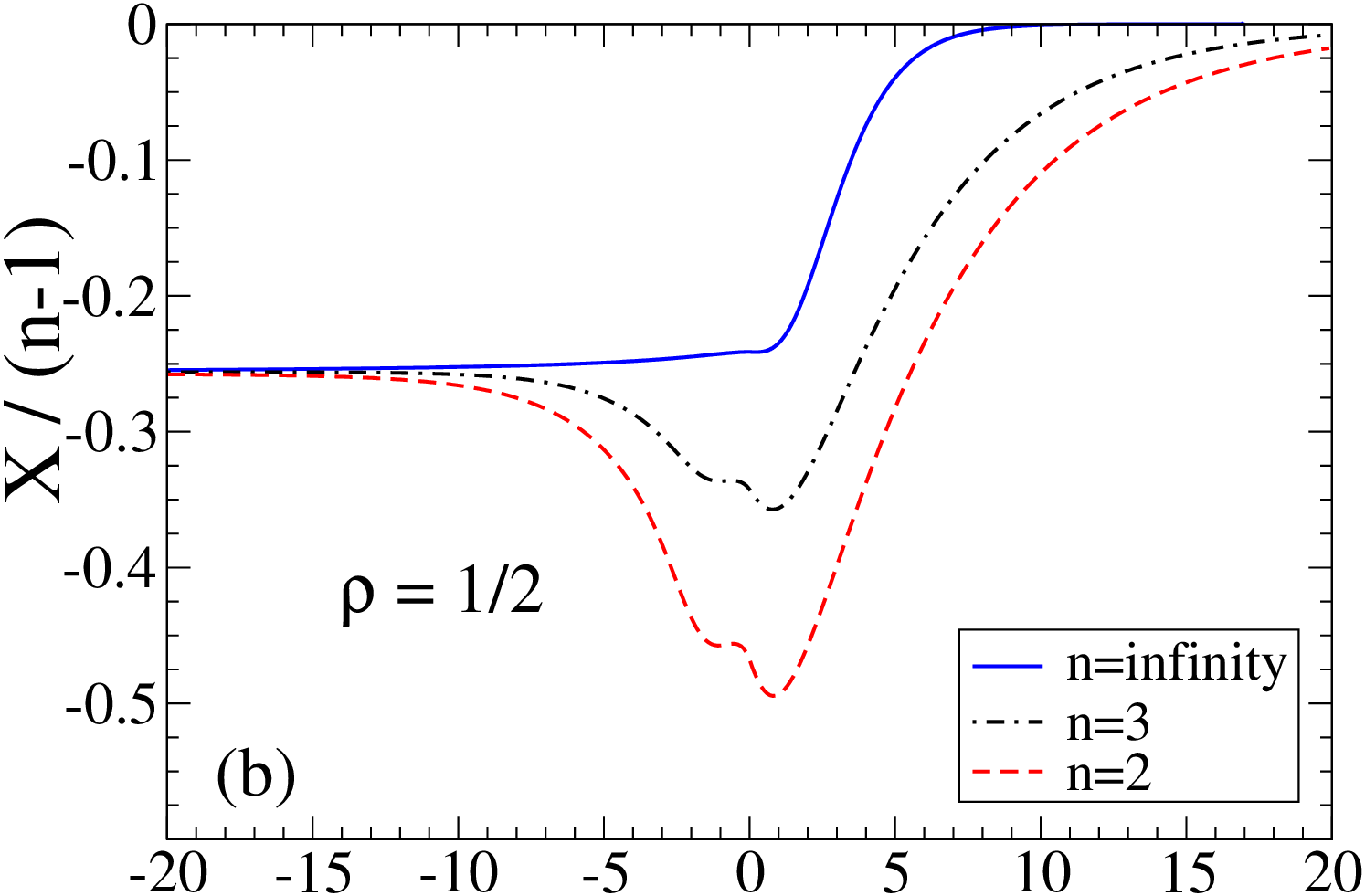}}
\subfigure{\includegraphics[clip,width=6.6cm]{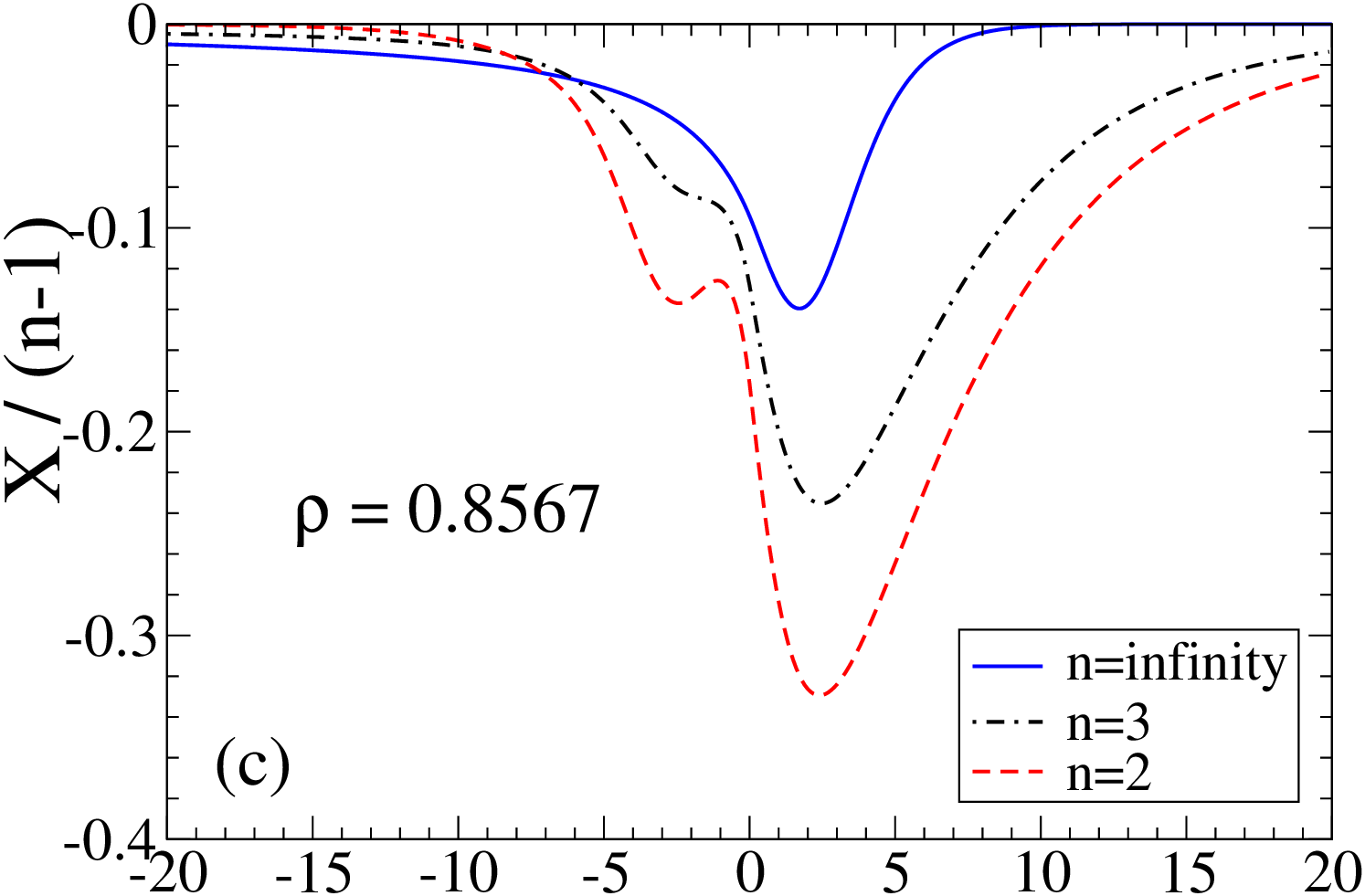}}
\subfigure{\includegraphics[clip,width=6.6cm]{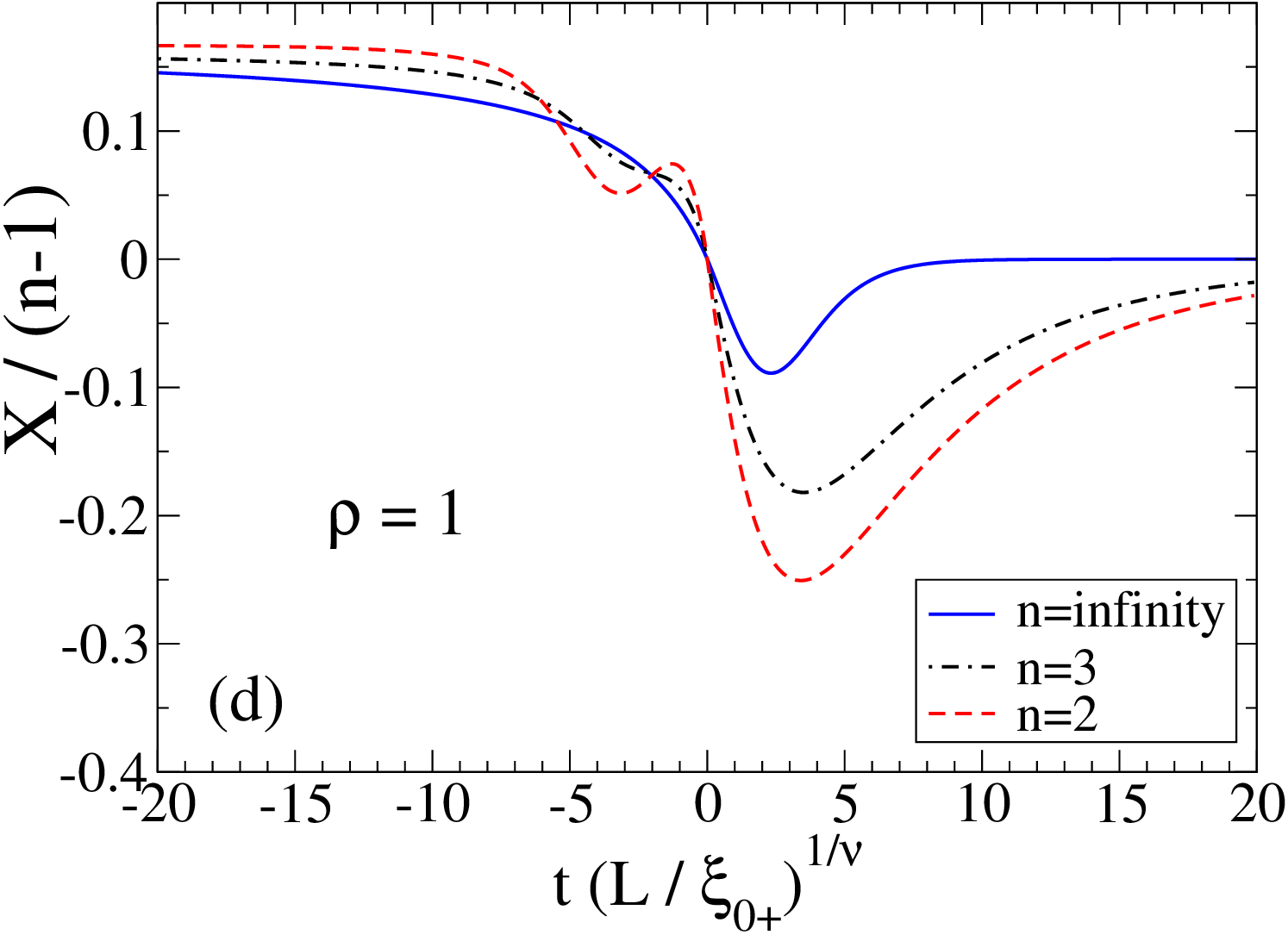}}
\end{center}
\vspace{0.8cm}
\caption{(Color online) Casimir force scaling function $X$  divided by $n-1$ for $n=2,3$ and $\hat X$ for $n=\infty$  in slab geometry for $d=3$ as a function of the scaling variables $\tilde x $, (\ref{3jjx}), and $\hat x$, (\ref{xiunend}), respectively, for several $\rho$. Solid line: exact result $\hat X(\hat x,\rho)$,  (\ref{X-large-nx-isoslab}). Dashed and dot-dashed lines for $n=2,3$: $\hat X(\tilde x,\rho)/(n-1)$ of the lowest-mode separation approach from (\ref{scalfreeaniso})  and (\ref{substisoX})(compare Figs. 3 and 4). All lines have the same low-temperature limit (\ref{lowXn2}) and (\ref{X-large-nx-low}) for each $\rho$. For $\rho = 0.8567$, the low-temperature amplitudes vanish for  $1<n\leq \infty$, see Fig. 5. For $\rho=1$, the amplitudes at $t=0$ vanish for $1\leq n\leq \infty$. }
\vspace{0cm}
\end{figure}
\vspace{0cm}
An interesting feature in Fig. 3 is the prediction of the change of sign of the low-temperature Casimir force from attractive for $\rho<0.8567$ to repulsive for $\rho>0.8567$. Analytically, this follows from (\ref{Xbelow}) which implies that the Casimir force scaling function has the following {\it finite} low-temperature limit for finite $n>1$ in a finite volume
\begin{eqnarray}
\label{lowXn2}
&& X(-\infty,\rho)=-[(n-1)/2]\;\rho^{d-1}\Big[1+ \rho\;
\partial {\cal J}_0(0, \rho)/\partial \rho\Big]\;\;\;\;\;\;\;\;\;
\end{eqnarray}
with  ${\cal J}_0(0, \rho)$ given by (\ref{calJ3spec}). The finite value (dashed lines in the right panels of Fig. 3) reflects the effect of the long-ranged fluctuations induced by the massless Goldstone modes corresponding to the long-ranged correlations (\ref{trans correl large}). From (\ref{lowXn2}) we obtain the Casimir force amplitude per component in the large-$n$ limit
in agreement with the exact result (\ref{X-large-nx-low}). The result (\ref{lowXn2}) differs from the Gaussian critical amplitude $X^G(0,\rho)$, (\ref{cascalJcritslab}), unlike the case  of film geometry for $n>1$ (Sec. V. F).
The amplitude (\ref{lowXn2}) divided by $n-1$ is plotted  for $d=3$ in Fig. 5 (solid line).
A vanishing of this amplitude for general $1<n\leq\infty$ in a slab geometry is predicted if
\begin{eqnarray}
\label{lowXzeroiso}
\rho\;\partial {\cal J}_0(0, \rho)/\partial \rho=-1
\end{eqnarray}
is satisfied according to (\ref{lowXn2}). This condition is exact
in the limit $n\to \infty$ according to  (\ref{X-large-nx-low}).  The solid line in Fig. 5 yields $ X(-\infty,\rho_0)=0$ at  $\rho_0=0.8567$ for general $n>1$ according to (\ref{lowXzeroiso}). For finite $n$, however, we expect $n$-dependent corrections of $O(u^*)$ to (\ref{excess-scalfreeaniso18})- (\ref{Xbelow}) and (\ref{lowXn2}) in a more complete theory. This can be tested by MC simulations for $\varphi^4$ and fixed-length spin models. In a finite block geometry the condition (\ref{lowXzeroiso}) becomes
$ \sum_{\beta=1}^{d-1}\rho_\beta\;
\partial {\cal J}_0(0, \{\rho_\alpha\})/\partial \rho_\beta=-1.$

In Fig. 6 we show for several $\rho$ how the low-temperature amplitude (\ref{lowXn2}) emerges for $\tilde x \to - \infty$ from the scaling function $X(\tilde x,\rho)/(n-1)$ for $n=2,3$ and from the exact result $\hat X(\hat x,\rho)$, (\ref{X-large-nx-isoslab}),  in the large-$n$ limit.
For $\rho=1$ our theory predicts a vanishing  Casimir force at $t=0$ for all $n$, see Fig. 6(d), as found previously for $n=1$ \cite{dohm2011,hucht2011} and for $n=\infty$ \cite{dohm2011}, in agreement with the analytic result for general $n$ \cite{hucht2011}, except for the Gaussian model (see Fig. 7(b) and App. A). One may suspect that the unexpected nonmonotonic portions of the dashed curves ($n=2$) for $\rho \gtrsim 1/2$ slightly below $T_c$ in Fig. 6 are an artifact of our approximation. In this context we note that also for $n=1, \rho=1$ the MC data shown in Fig. 2 (h) exhibit a nonmonotonic behavior slightly below $T_c$. Further studies are necessary in order to clarify this issue.
\subsection{Analytic results at ${\bf T_c}$}
At $T_c$  we obtain from (\ref{6ff}) -(\ref{substisoX}) for a slab geometry
\begin{figure}[!ht]
\begin{center}
\subfigure{\includegraphics[clip,width=6.6cm]{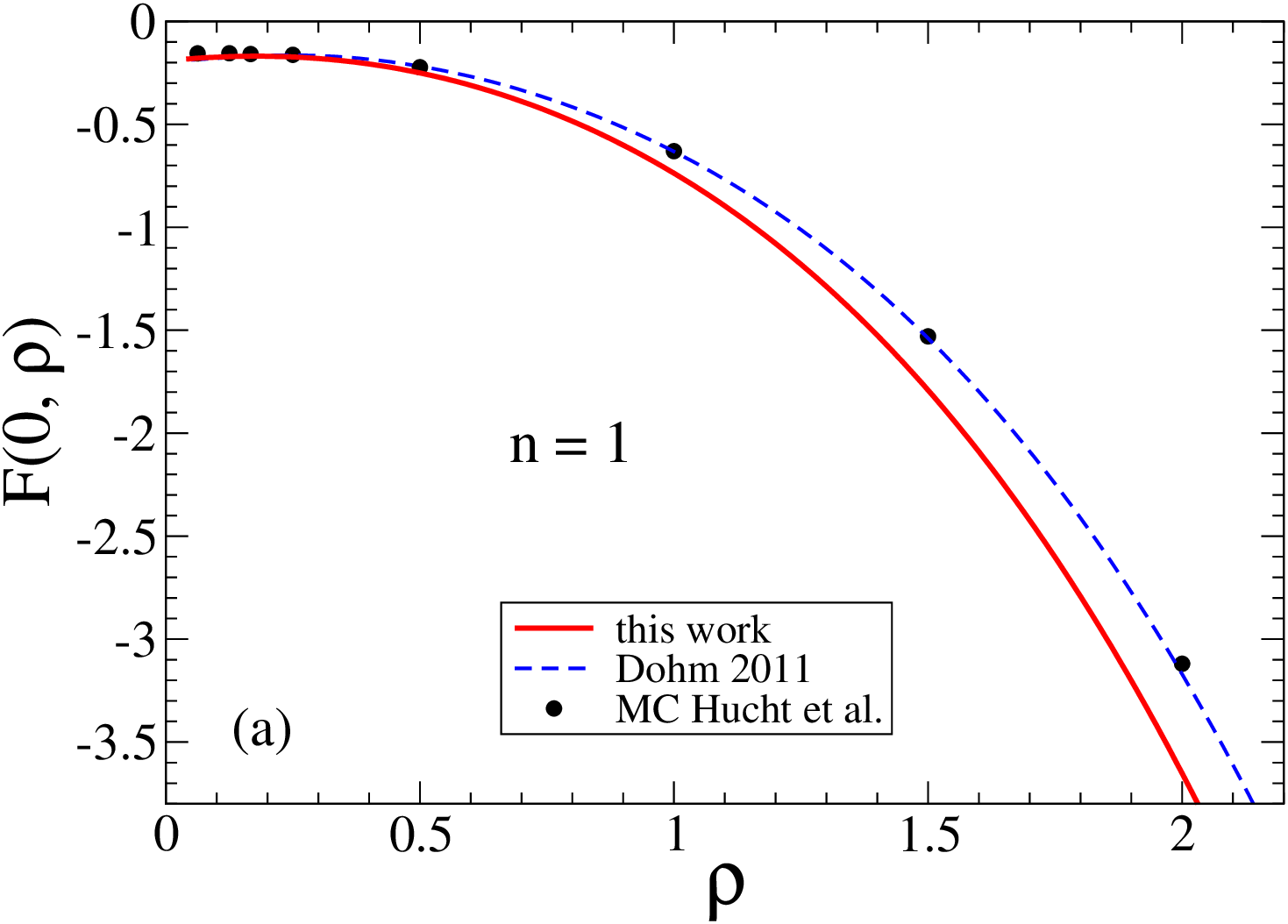}}
\subfigure{\includegraphics[clip,width=6.6cm]{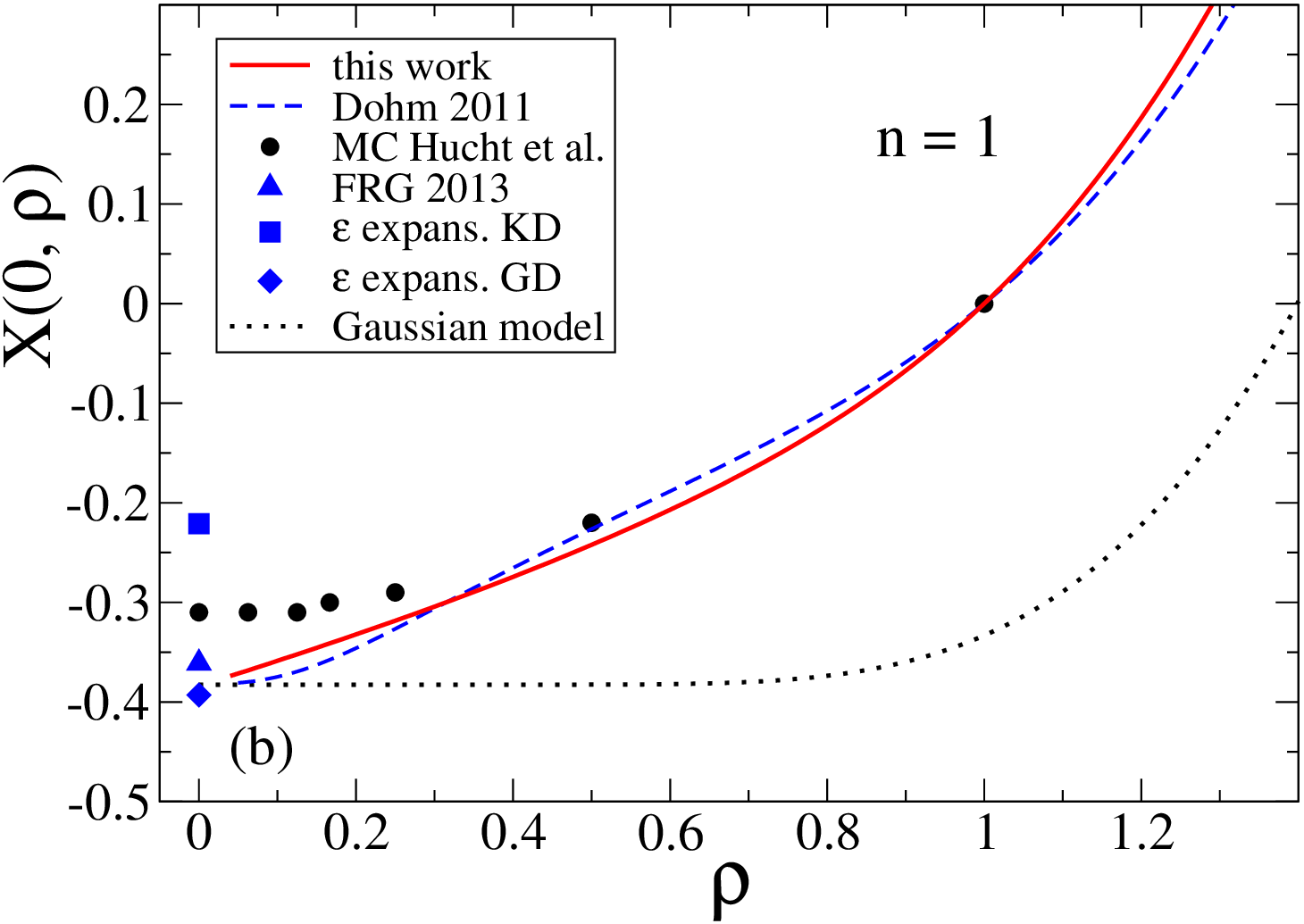}}
\end{center}
\vspace{0.5cm}
\caption{(Color online) Amplitude at $T_c$ of (a) the free energy density and (b) the  Casimir force for $d=3,n=1$  in slab geometry as a function of $\rho$. Solid lines: from (\ref{scalfreeanisoTc}) and (\ref{CasimiratTc}). Dashed lines: from Eqs. (5.1) and (5.10) of \cite{dohm2011}. Both dashed and solid lines in (b) yield $X=0$ for $\rho=1.$ Square, diamond, and triangel for $\rho=0$ in (b): from $\varepsilon$ expansion \cite{KrDi92a,GrDi07} and functional RG \cite{Jakub}. Dots: MC data of Table I of \cite{hucht2011}. Dotted line: Gaussian model, (\ref{cascalJcritslab}), and 1-loop perturbation theory, (\ref{cascalJcrit1loop}).}
\vspace{0cm}
\end{figure}
\begin{eqnarray}
\label{CasimiratTc}
&&X(0,\rho) =(d-1)  F(0,\rho) - \rho\;\partial F(0,\rho)/\partial \rho,\\
\label{scalfreeanisoTc}
&&F(0, \rho)= \rho^{d-1} \nonumber\\&&\times\;\Big\{-
u^* \left[\vartheta_{2,n} (0)\right]^2 \Big[\frac{36}{d} + 144\frac{(n-1)}{d\varepsilon}\;\Big(\frac{1}{3^{d/2}}-\frac{d}{36}\Big)\Big]\nonumber\\ && + \;{\cal W}_n(0) + (1/2){\cal J}_0( {\tilde l_c}^2, \rho)  +  [(n-1)/2)]{\cal J}_0( {\tilde l_c}^2/3, \rho)  \nonumber\\ && +\frac{n}{2}\ln\Big[\frac{\tilde l_c^2 \;{[\Gamma(n/2)]}^{2/n}  }{24 \pi^2\; \vartheta_{2,n} (0) }\Big]\Big\},\\
&&\tilde l_c^{d/2} =12 {u^*}^{1/2}  A_d^{-1/2}
\vartheta_{2,n}(0)\;\rho^{(d-1)/2},
\end{eqnarray}
with ${\cal W}_n(0)= -\;  \ln \big[(1/2)\Gamma(n/4) \big]$ {\bf } and
$\vartheta_{2,n} (0)= \;\Gamma\big((n+2)/4\big)/\Gamma\big(n/4\big)$.
The amplitudes (\ref{CasimiratTc}) and (\ref{scalfreeanisoTc}) are compared in Fig. 7 for $d=3$ with MC data for $n=1$ \cite{hucht2011}
and with earlier predictions for $n=1$ \cite{dohm2011,KrDi92a,GrDi07,Jakub}.
The $\rho$-dependence at $T_c$ for $\rho \gtrsim 0.2$ is best described by the earlier version of our theory for $n=1$ \cite{dohm2011} (dashed lines in Fig. 7, for a discussion of this version see also Sec. V. H and Fig. 11). Note that $X(0,1)$ vanishes for $n=1, 2,3, \infty$, as shown in Figs. 2 (h), 3 (h), 6 (d), 7 (b), and 9, in agreement with the analytic proof of \cite{hucht2011}; this does not hold for the Gaussian model [dotted line in Fig. 7(b)], as explained in App. A.
\subsection{Analytic results above ${\bf T_c}$}
For general $n$, Eq. (\ref{scalfreeaniso})  yields the leading behavior for large $\tilde x$  in a slab geometry (App. C)
\begin{eqnarray}
\label{excess+scalfreeanisoasymx}
&&F_+^{ex} (\tilde x,\rho) \approx  (n/2)\;{\cal G}_0\big( (\tilde x \;Q^*)^{2\nu }, \rho\big)\nonumber\\ &&+\;u^*\nu\;[-3 n^2 + 6 n B(u^*)] \;\rho^{d-1}\;,\;\,\,\,\;\;\,\,\,\;\,\;\;\,\,\,\;\\
\label{Xabove}
&&X_+(\tilde{x},\rho) \approx
\nonumber \\
\label{Xabove2}
&& -\frac{n}{2}\rho^{d-1}\Big\{2+\Big[
\frac{\tilde x}{\nu}
\frac{\partial}{\partial \tilde x}+ \rho
\frac{\partial}{\partial \rho}\Big]{\cal J}_0\big((\tilde x\;Q^*)^{2\nu}, \rho\big)\Big\}.\;\;\;\;\;\;\;\;\;
\end{eqnarray}
As shown in Figs. 2 and 3, these results are in good agreement with the  available MC data for $n=1,2$ well above $T_c$. The functions ${\cal G}_0$ and $X_+$ go exponentially to zero for $\tilde x \to \infty$. On the basis of 1-loop perturbation theory (App. B) and of the exact result for $n=\infty$  (App. A)  one expects   $F_+^{ex}$ to decay to {\it zero} for large $\tilde x$ rather than to a (small) finite constant  of $O(u^*)$. This constant in (\ref{excess+scalfreeanisoasymx}) comes from the algebraic decay $\propto n \tilde y^{-1}$ of $\vartheta_{2,n}(\tilde y)$ in (\ref{theta2largeabove}). Because of the smallness of $u^*$ for $n=1,2,3$ it is not visible in Figs. 2-4 but becomes of $O(\rho^{d-1})$ in the large-$n$ limit (Sec. V.G). We expect that in a more complete calculation well above $T_c$ including all terms up to $O(u^*)$ the finite constant in (\ref{excess+scalfreeanisoasymx}) will be canceled and the exponential  decay of the function  $F_+^{ex}$ to zero for $\tilde x \to \infty$ will be confirmed. The constant $\propto \rho^{d-1}$ does not contribute to $X_+$. Eq. (\ref{Xabove2}) agrees with ordinary perturbation theory in one-loop order, (\ref{cascalJplus}), together with $Q^*= 1 + O({u^*}^2)$.

In the scaling results (\ref{excess+scalfreeanisoasymx}) and (\ref{Xabove2}) there is no dependence on the lattice constant $\tilde a$. This is, however, not valid for sufficiently large  $L /\xi_+$ or sufficiently large $\tilde x$ at fixed $\rho_\alpha$ corresponding to a calculation of $\Delta(r, \{L_\alpha\})$, (\ref{bb9xy}), for the case $r L_\alpha^2 \gg 1$. It has been shown \cite{cd1999,dohm2008,dohm2011,kastening-dohm} that there exists a violation of universality and finite-size scaling in an "exponential regime" where the scaling functions are exponentially small and where an explicit dependence on $\tilde a$ exists at any finite  $  \xi_+ < \infty$ with $ \xi_+/\tilde a \gg 1$, i.e., even arbitrarily close to $T_c$.
In this regime (compare Fig.2 of \cite{dohm2011}),
the {\it exponential} ("true") bulk correlation length (mentioned in Sec. III. C) in the direction of one of the cubic axes  \cite{dohm2008,cd2000-2,fish-2}
$\xi_{\bf e}  = (\tilde a/2)
\left\{{\rm arsinh} \left[\tilde a/(2
\xi_+)\right]\right\}^{-1}$
must be employed. (The latter expression is a one-loop result for finite $n$ and is exact for $n \to \infty$  \cite{cd2000-2}.)
For the case of a block geometry and for a nearest-neighbor interaction, this regime is characterized by
$\label{calGnull} L_\alpha \gtrsim  24  (\xi_+)^3 / \tilde a^2$,
$\alpha=1,...,d$.
Similar remarks apply  to the exponential tail of $X_-$, (\ref{Xbelow}), for $n=1$ below $T_c$ \cite{dohm2011}, and to the scaling functions above $T_c$ in film geometry (Sec. V.F) and in the large-$n$ limit. In Sec. X of \cite{dohm2008} it was shown that nonuniversal nonscaling finite-size effects on the excess free energy density exist in an exponential regime even in fully isotropic continuum systems as described by the isotropic $\varphi^4$ field theory.
\subsection{Film limit}
We consider the $\infty^{d-1}\times L$  film  geometry as the limit  $L_\parallel \to \infty$  of a finite ($L_\parallel^{d-1}\times L$) slab geometry at fixed $L$.
A film critical point exists at a finite temperature $0< T_{c,\text{film}}(L) < T_c$
below the bulk critical temperature $T_c=\lim_{L\to \infty}T_{c,\text{film}}(L)$ for $ n=1,d>2$, for $n=2,d\geq 3$, and for $n>2, d>3$. For $n=2, d=3$, this is a Kosterlitz-Thouless transition. No finite $T_{c,\text{film}}(L)$ exists for $n>2$ in $d\leq3$ dimensions.
Our theory does not capture an $L$ dependent shift of the film critical temperature since the flow parameters $l_+$ and $l_-$, as determined by (\ref{rprimelong}), (\ref{rprimetrans}), (\ref{lasym}), and (\ref{bulktildeflow}), vanish at the same temperature $T_c$ for both the bulk and film system. Similarly, no such a shift was captured  in the earlier  analytic calculations  \cite{KrDi92a,wil-1,GrDi07} for film geometry with periodic BC.
To perform the film limit we start from (\ref{scalfreeaniso}) together with the substitutions (\ref{substiso})-(\ref{calJ3spec}) as well as with the result  (\ref{calWbulk}) in the large-volume limit. In addition we need the following film limit of the function ${\cal J}_0$ together with $\rho K(\rho^2z) \to (\pi/z)^{1/2}$ for $\rho \to 0$ at fixed $L$ for $d>1$,
\begin{eqnarray}
\label{filmlimitj}
&&\lim_{\rho \rightarrow 0}V^{-1}{\cal J}_0(x,\rho)=L^{-d}\;{\cal G}_{0,{\text film}}(x),\\
\label{GfilmNullx}
&&{\cal G}_{0,{\text film}}(x)={\cal G}_0(x,0) =
\int^\infty_0 \frac{dy}{ y} \exp \left(- \frac{x y}{4
\pi^2} \right) \nonumber\\&&  \times \left(\pi/y\right)^{(d-1)/2}\left[\left(\pi/y\right)^{1/2} \; - \; K(y) \right].
\end{eqnarray}
\begin{widetext}
\begin{figure*}[!ht]
\begin{center}
\subfigure{\includegraphics[clip,width=7.0cm]{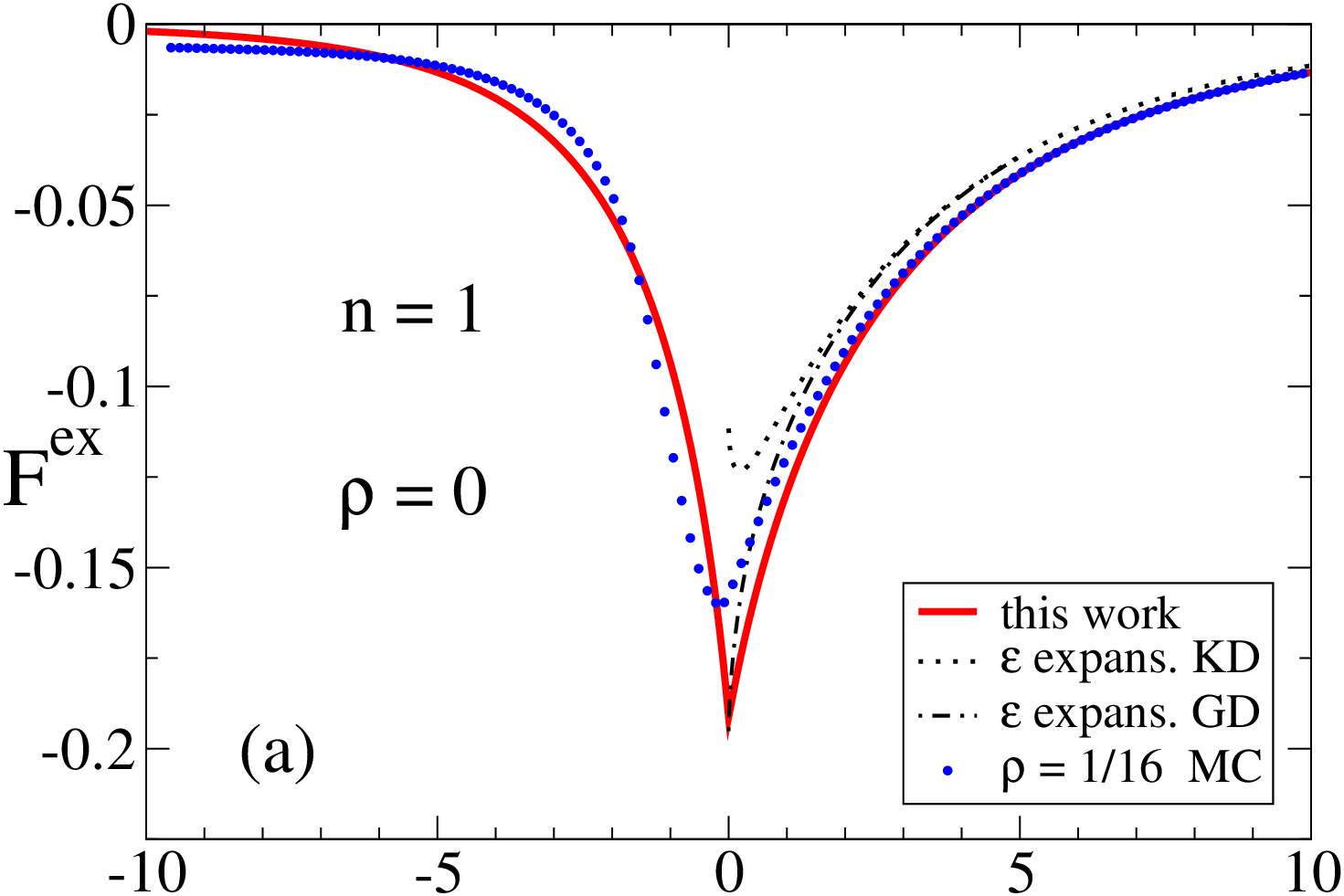}}
\subfigure{\includegraphics[clip,width=7.0cm]{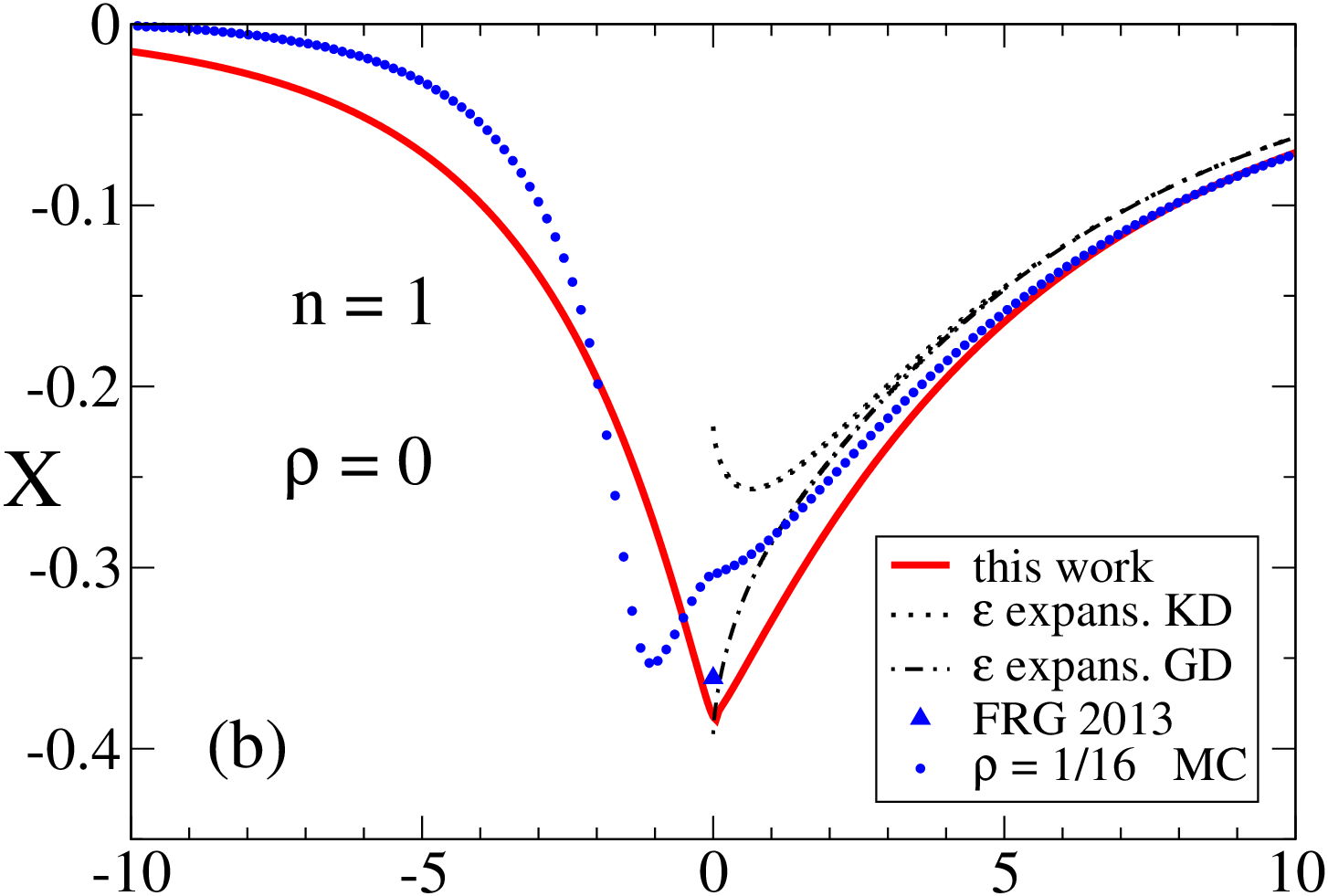}}
\subfigure{\includegraphics[clip,width=7.0cm]{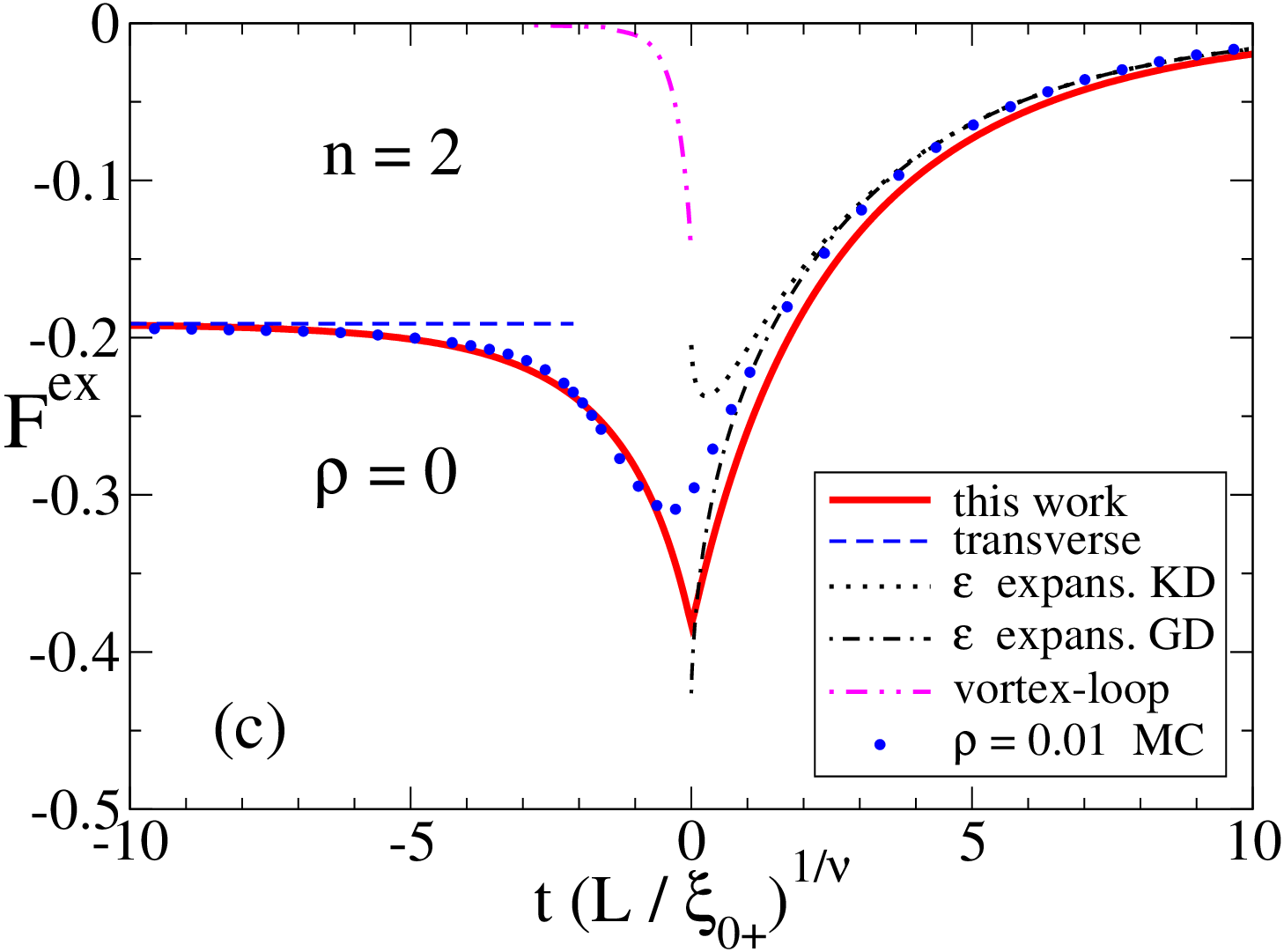}}
\subfigure{\includegraphics[clip,width=7.0cm]{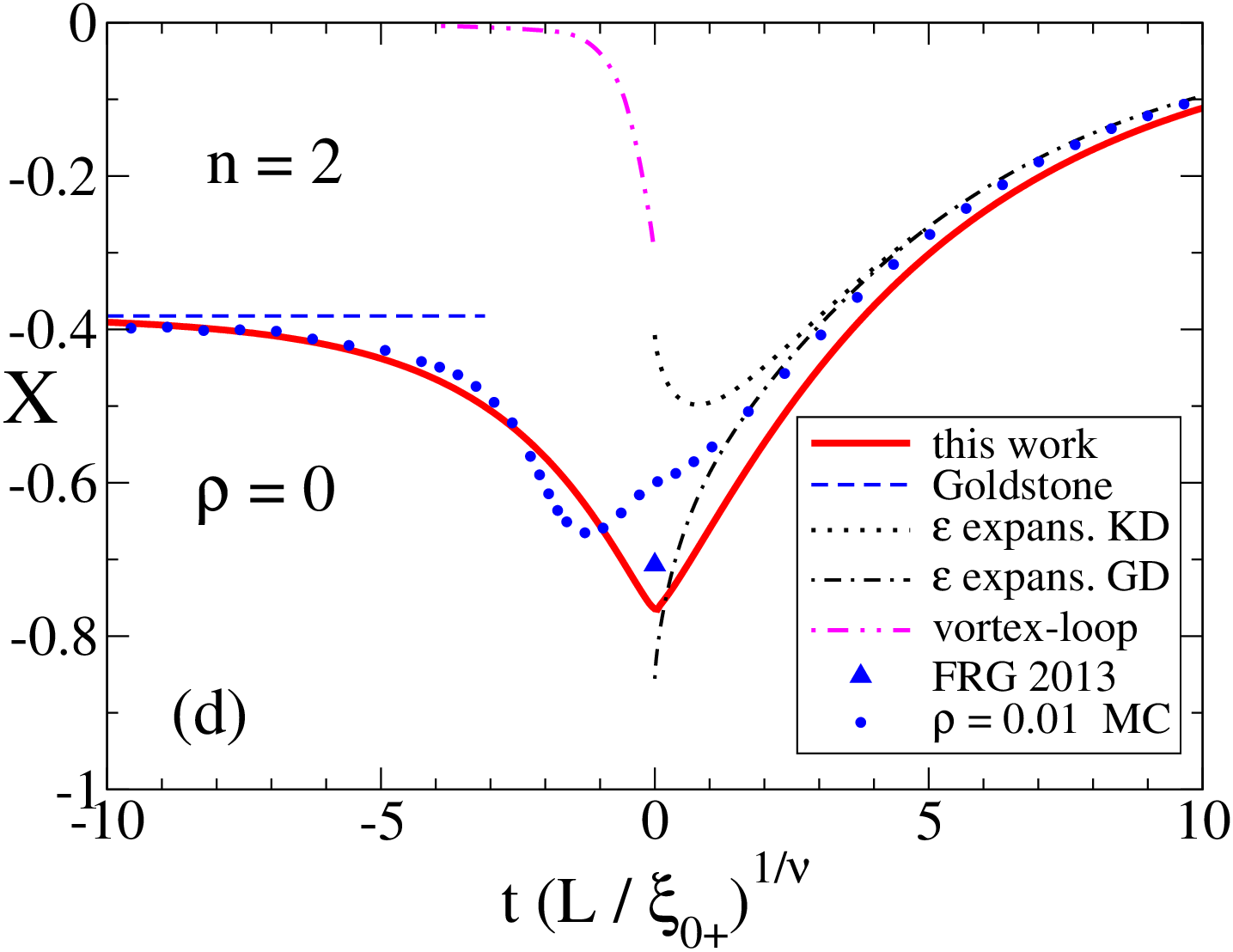}}
\end{center}
\caption{(Color online) Scaling functions $F^{ex}_{\text film}(\tilde x)$ and $X_{\text film}(\tilde x)$ in a $d=3$ film ($\infty^2\times L$) geometry ($\rho=0$) as a function of $\tilde x = t (L/\xi_{0+})^{1/\nu}$  for $n=1$ and $n=2$. Thick solid lines from (\ref{fexfilmscalingplus})-(\ref{Casimirfilmscaling}).  MC data  in (a) and (b) for  the $d=3$ Ising model from \cite{hucht2011} for $\rho=1/16$ (Fig.3 L=16) and in (c) and (d) for the $d=3$ XY model from \cite{hasenbusch2010} for $\rho=0.01$.  $\varepsilon$ expansion results for $\rho=0$  from \cite{KrDi92a} (dotted lines) and \cite{GrDi07} (dot-dashed lines). Double-dot-dashed lines in (c) and (d): vortex-loop renormalization \cite{wil-1}. Dashed lines in (c) and (d) from (\ref{casimirlowxx}) with ${\cal G}_{0,film}(0) = -\;\zeta(3)/\pi$. Triangels in (b) and (d) from functional RG \cite{Jakub}.}
\end{figure*}
\end{widetext}
This leads to the scaling forms of the excess free energy density $f^{ex,\pm}_{{\text film}}(t,L) =  L^{-d}F^{ex,\pm}_{{\text film}}(\tilde x)$ and of the Casimir force $F^{\pm}_\text{Cas,film}(t,L)=L^{-d}X^{\pm}_\text{film}(\tilde x )$ of the film system
\begin{eqnarray}
\label{fexfilmscalingplus}
&&F^{ex,+}_{{\text film}}(\tilde x)= \;(n/2)\;{\cal G}_{0,film} \big({Q^*}^{2\nu} \tilde x^{2\nu}\big),\\
\label{fexfilmscalingminus}
&&F^{ex,-}_{{\text film}}(\tilde x)=\nonumber\\&& \;(1/2)\;{\cal G}_{0,{\text film}} \big({Q^*}^{2\nu} |2\tilde x|^{2\nu}\big)+[(n-1)/2]\;{\cal G}_{0,{\text film}} \big(0\big),\;\;\;\;\;\;\;\;\;\;\\
\label{Casimirfilmscaling}
&&X^{\pm}_{film}(\tilde x)=(d-1)F^{ex,\pm}_{film}(\tilde x)-
\frac{\tilde x}{\nu}
\frac{\partial F_{film}^{ex\pm}(\tilde x)}{\partial\tilde x},
\end{eqnarray}
above ($+$) and below ($-$) $T_c$
with the scaling variable $\tilde x$, (\ref{3jjx}). The same results can be derived by renormalized one-loop perturbation theory  \cite{dohm2017II}.
Apart from the factor $Q^* = 1 + O({u^*}^2)$, Eq. (\ref{fexfilmscalingplus}) agrees  with  Eq. (7.5) of \cite{kastening-dohm}. For $\tilde x =0$ we have
\begin{eqnarray}
\label{calGnullfilmnull} {\cal G}_{0,film}(0) ={\cal G}_0(0,0) =-2\;\pi^{-d/2}\Gamma(d/2)\zeta(d)
\end{eqnarray}
according to Eq. (4.6a) of \cite{kastening-dohm} and Eq. (5.7) of \cite{KrDi92a} (up to a sign misprint there).
Although (\ref{GfilmNullx})-(\ref{calGnullfilmnull}) remain finite in the formal limit $d\to2$, they do not yield reliable results for $d=2$, as expected. In particular they do not reproduce the exact results obtained for the $d=2$ Ising model in a ($\infty\times L$) strip geometry \cite{rud2010}.

In Fig. 8 our results are compared for $d=3$ with MC data \cite{hasenbusch2010,hucht2011} and with earlier theories \cite{KrDi92a,wil-1,GrDi07,Jakub}. All theoretical curves show cusp-like non-analyticities at $\tilde x=0$ for finite $L$ which can be interpreted as the signature of an unrenormalized film singularity at an unshifted film transition temperature. Apart from this shortcoming, our theory
is in reasonable agreement with the MC data away from $T_c$.  Our amplitudes at $T_c$ are slightly closer to the result of the functional renormalization group \cite{Jakub} than the amplitudes predicted by the $\varepsilon$  expansion \cite{KrDi92a,GrDi07}. The derivation of the  simple form of our results (\ref{fexfilmscalingplus}) -(\ref{Casimirfilmscaling}) requires  considerably less computational effort both above and below $T_c$ than the derivation of the more complicated higher-loop $\varepsilon$ expansion results above $T_c$ \cite{KrDi92a,GrDi07}. This is an advantage of using the minimal subtraction approach at fixed dimension. The result of the vortex-loop RG technique  \cite{wil-1} [double-dot-dashed lines in Figs. 8(c) and (d)] is far from the data. It yields vanishing scaling functions for $\tilde x \to -\infty$ since it neglects the dominant influence of the Goldstone modes \cite{commentDantchev}.

We add the following comments.
(i) In the low-temperature limit $\tilde x \to - \infty$, the longitudinal part of  (\ref{fexfilmscalingminus}) decays exponentially
(compare Eq. (4.7) of \cite{kastening-dohm}),
thus
$F^{ex,-}_\text{film}(\tilde x)$ and
$X^-_\text{film}(\tilde x)$ vanish exponentially for $n=1$ in the low-temperature limit, in agreement with the MC data.
For $n>1$, however, the Goldstone modes yield a finite amplitude
$X^-_\text{film}(- \infty)=(d-1) F^{ex,-}_\text{film}(- \infty)$   which is similar to the Gaussian critical amplitude  $X^G_\text {film}(0)$, (\ref{CasimirGaussAmp}),
\begin{eqnarray}
\label{casimirlowxx}
\frac{X^-_\text{film}(-\infty )}{n-1}= \frac{X^G_\text {film}(0)}{n}=\frac{d-1}{2}\;{\cal G}_{0,\text {film}}(0)
\end{eqnarray}
[dashed line in Fig. 8(d)]. A corresponding relation holds for $F^{ex,-}_\text{film}(- \infty)/(n-1)={\cal G}_{0,\text {film}}(0)/2$.
From  (\ref{casimirlowxx}) we obtain the amplitude per component in the large-$n$ limit
in agreement with the exact result  (\ref{Xinfinfilmlow}) for $d\leq 3$.

(ii) Near $T_c$ the functions $F^{ex,\pm}_{{\text film}}$ and $X^{\pm}_{film}$ have cusp-like shapes which arise from the small-$x$ behavior
\begin{eqnarray}
\label{GfilmNullsmall3}
{\cal G}_{0,film}(x) = -\zeta(3)/\pi - [x(\ln  x - 1)]/(4\pi) + O( x^{3/2})\;\;\;\;\;\;
\end{eqnarray}
for $d=3, x>0$ (compare Eq. (4.4a) of \cite{kastening-dohm}). Similar to earlier theories \cite{KrDi92a,wil-1,GrDi07}, this does not correctly describe the film critical behavior at the (shifted) film transition where $F^{ex,-}_\text{film}$ and $X^-_\text{film}$ should display only weak singularities for $n=1$ and $n=2$ \cite{vasilyev2009}. The latter singularities are not detected by the smooth MC data in Fig. 8 at finite $\rho = 1/16$ and   $\rho = 0.01$ (see also Figs. 6 and 13 of \cite{vasilyev2009} at $\rho = 1/6$).  A further shortcoming of (\ref{fexfilmscalingplus})-(\ref{Casimirfilmscaling}) and of earlier theories \cite{KrDi92a,GrDi07} is the prediction of a singularity of the film system at $T_c$ even for  $n>2, d\leq 3$ although in this case there should be no film transition at all at a finite temperature. Such problems are circumvented in our lowest-mode separation approach  for finite  geometries where no artificial singularities are present.

\vspace{0cm}
\begin{figure}
\includegraphics[clip,width=6.0cm]
{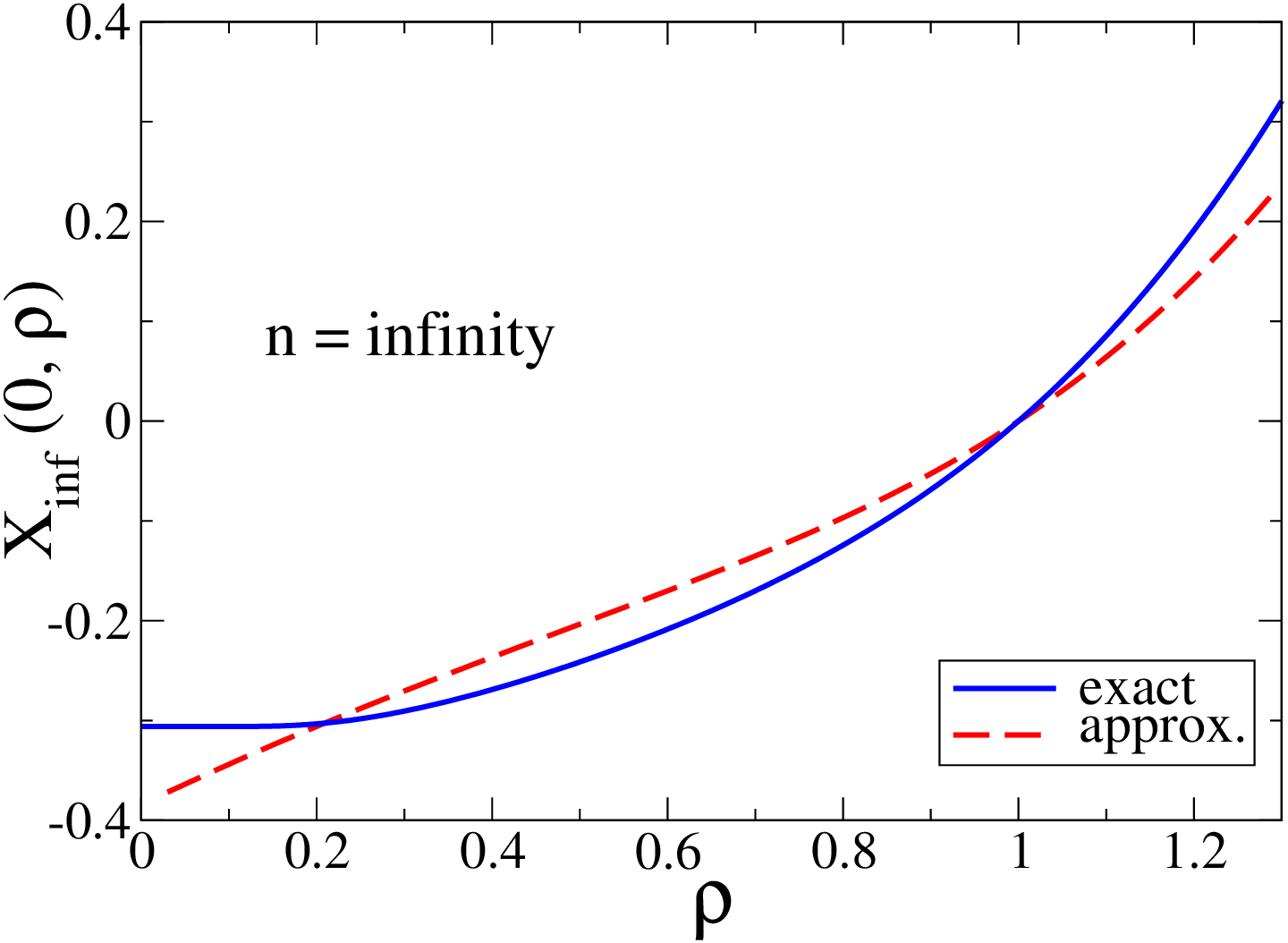}
 \vspace{0cm}
 \caption{(Color online) Casimir amplitude at $T_c$  for $n=\infty$, $d=3$  as a function of the aspect ratio $\rho$. Solid line: exact result (\ref{X-large-nx-isoslab}) with $\hat x=0$. Dashed line: $X_{inf}(0,\rho)$ of the lowest-mode separation approach,  (\ref{XbeiTclargen}). Both lines coincide at $\rho=0.209$ and yield a vanishing amplitude at $\rho=1$. Compare Fig. 5(b) in \cite{dohm2011} and Fig. 7(b) of this paper. }
\end{figure}
\vspace{0cm}
\vspace{0cm}
\begin{figure}
\begin{center}
\subfigure{\includegraphics[clip,width=6.0cm]{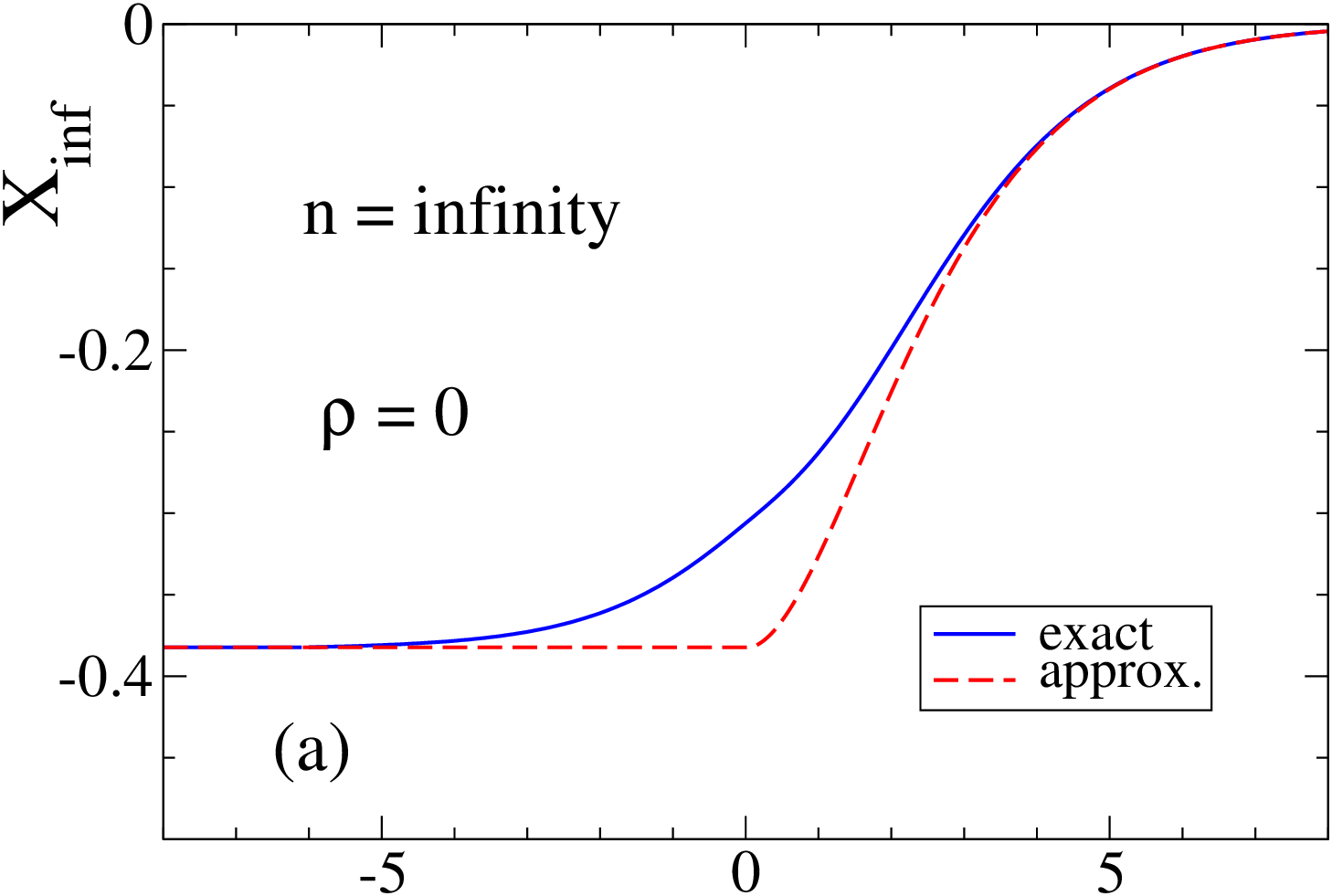}}
\subfigure{\includegraphics[clip,width=6.0cm]{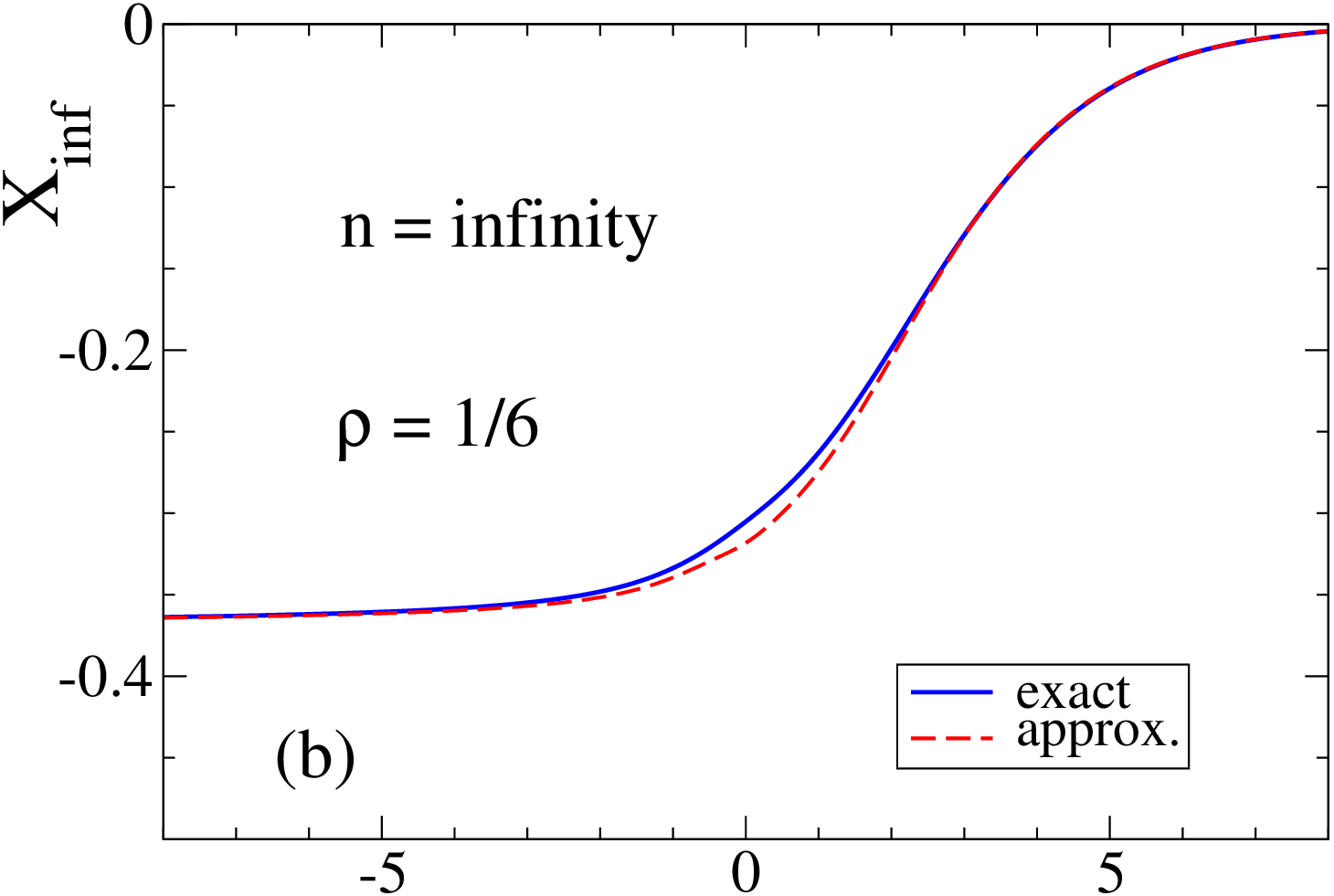}}
\subfigure{\includegraphics[clip,width=6.0cm]{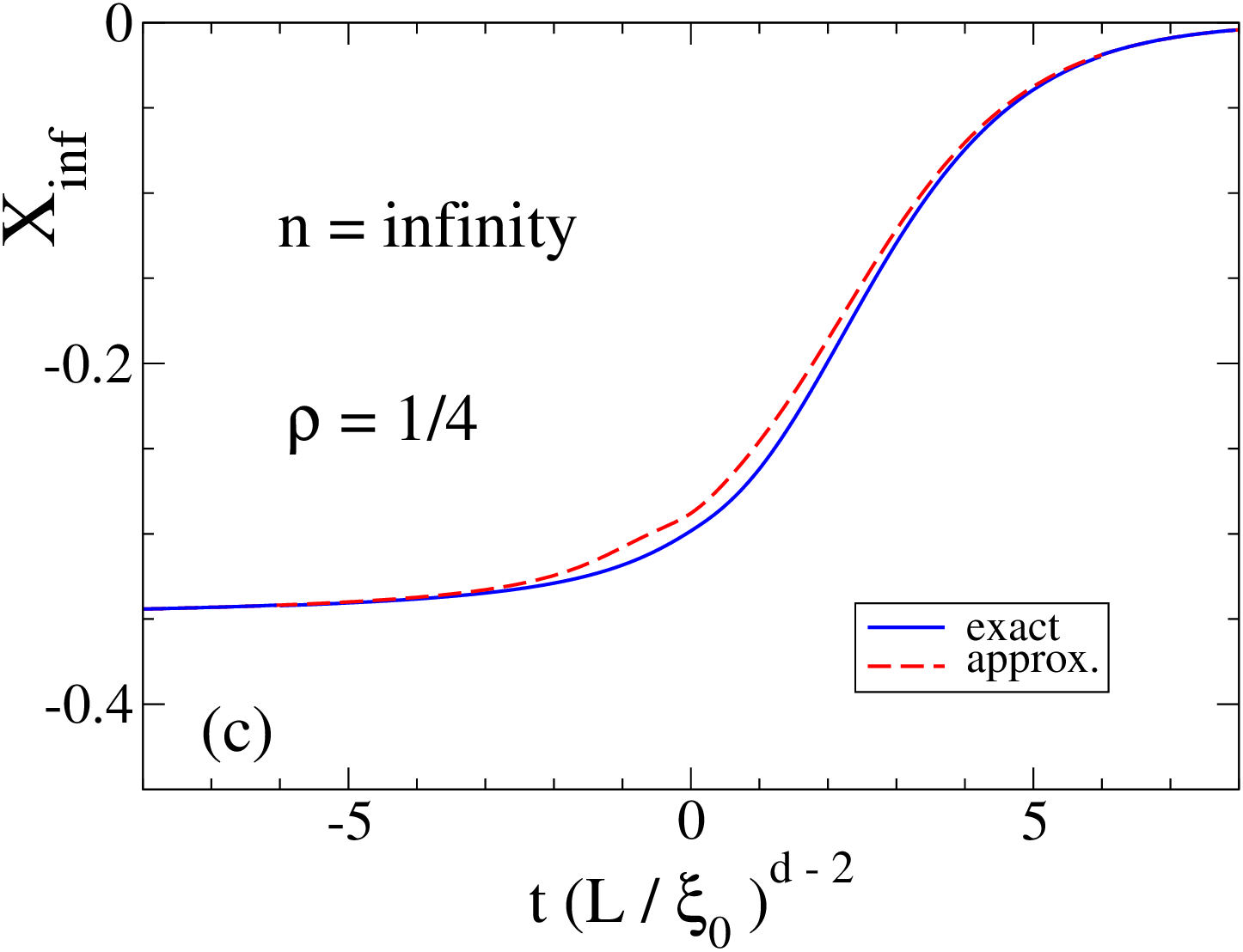}}
\end{center}
\vspace{0.5cm}
\caption{(Color online)  Casimir force scaling function for $n=\infty,d=3$ as a function of $\hat x = t (L/\xi_0)^{d-2}$  in film geometry $ (\rho=0)$ and in slab geometries with $ \rho=1/6, 1/4$.  Solid lines: exact theory, (\ref{X-large-nx-isoslab}). Dashed lines: $X_{inf}(\hat x,\rho)$ of the lowest-mode separation approach derived from (\ref{excessfreelargen}) and (\ref{3jjfilmlargenx}). For each $\rho$, the solid and dashed lines have the same low-temperature limit (\ref{X-large-nx-low}) and (\ref{Xinfinfilmlow}). }
\vspace{0cm}
\end{figure}
\vspace{0.9cm}
\subsection{ Large-${\bf n}$ limit}
In Figs. 9 and 10 we compare for $d=3$ the large-$n$ limit $X_{inf}(\hat x, \rho)=\lim_{n \to \infty}X(\tilde x, \rho)/n$ of the result $X(\tilde x, \rho)$ of our lowest-mode separation approach (dashed lines) with the exact result $\hat X(\hat x, \rho)$, (\ref{X-large-nx-isoslab}) (solid lines),  with $\hat x$ given by (\ref{xiunend}).
As shown in Fig. 9, there is exact agreement at $T_c$ between $X_{inf}(0, \rho)$ and $\hat X(0, \rho)$ for $\rho=0.209$ and $\rho=1$ and good agreement for other values of $\rho$ in the range $0.1 \lesssim \rho \lesssim 1.2$. In Fig. 10 the scaling functions are shown in the range $-8 \leq \hat x \leq 8$. For $\rho= 1/6,1/4$, excellent agreement is found in the whole temperature region. The agreement deteriorates for larger values of $\rho \gtrsim 1/2$. In the high- and low-temperature limits, however, exact agreement is found for all $\rho\geq 0$, i.e., $X_{inf}(\infty, \rho)=\hat X(\infty,\rho)=0$ and  $X_{inf}{(-\infty, \rho)}= \hat X(-\infty,\rho)$,  (\ref{X-large-nx-low}). The dashed line in Fig. 10 (a) for $\rho=0$ (film geometry) follows  from
\begin{eqnarray}
\label{3jjfilmlargenx}
F^{ex}_{inf}(\hat x,0)= \left\{
\begin{array}{r@{\quad \quad}l}
\; (1/2)\;{\cal G}_{0,{\text film}} (\hat x^{2/(d-2)}),\quad          &  \;\hat x > 0, \\
\; (1/2) \;{\cal G}_{0,{\text film}} (0),& \;\hat x < 0 ,\;\;\;
               \end{array} \right.\;\;
\end{eqnarray}
where $F^{ex}_{inf}(\hat x, \rho)=\lim_{n \to \infty}[F(\tilde x, \rho)-F^\pm_b(\tilde x)]/n$. Eq. (\ref{3jjfilmlargenx}) follows from (\ref{fexfilmscalingplus}) and (\ref{fexfilmscalingminus}) in the large-$n$ limit.
Below we present the derivation of the results for $\rho >0$ shown in Figs. 9 and 10.

An important role is played by the function $\vartheta_{2,n} (y)$, (\ref{thetax}). The large-$n$ limit of this function is given by Eq. (A6) of \cite{Esser} for  $-\infty \leq z \leq \infty$
\begin{subequations}
\label{largen}
\begin{align}
\label{thetalargen}
\lim_{n \to \infty}n^{-1/2}\vartheta_{2,n} (n^{1/2}z)=g(z),
\\
\label{gvonz}
g(z)={[z/2+(4+z^2/4)^{1/2}]}^{-1}.
\end{align}
\end{subequations}
According to (\ref{6ff}) and (\ref{6gg}), the scaling function of the flow parameter $\tilde l_\infty(\hat x, \rho)=\lim_{n \to \infty}\tilde l(\tilde x,\rho)$ and the scaling function $z(\hat x,\rho)=\lim_{n \to \infty}\tilde y(\tilde x,\rho)/n^{1/2}$ are determined implicitly by
\begin{subequations}
\label{6largen}
\begin{align}
\label{6fflargen}
z + 12\; g(z) =
\big[\tilde l_\infty^{d} A_d \; \rho^{1-d}4/\varepsilon\big]^{1/2},\;\;\;\;\;\;
\\
\label{6gglargen}
z =\; \hat x\;\big[\tilde l_\infty^\varepsilon A_d \;\rho^{1-d}4/\varepsilon\big]^{1/2}.
\end{align}
\end{subequations}
Here  we have employed the fixed point values given after (\ref{renormalization-constants-large-n}), and the ratio of critical exponents $\lim_{n \to \infty} \alpha/\nu=-\varepsilon$.
The scaling function (\ref{lTasym}) has a finite limit $ l_{\rm T\infty}(\hat x,\rho)=\lim_{n \to \infty} l_{\rm T}(\tilde x_,\rho)$,
\begin{subequations}
\label{Tlargen}
\begin{align}
\label{lTlargen}
l_{\rm T\infty}(\hat x,\rho)= \tilde l_\infty(\hat x,\rho)^2\;f(z),
\\
\label{fTlargen}
f(z) = 1-8/[12+z/g(z)],
\end{align}
\end{subequations}
with $z= z(\hat x,\rho)$. From (\ref{scalfreeaniso}), (\ref{3jjbulk}), (\ref{6oxx}), and (\ref{6pxx}) we then find
\begin{eqnarray}
\label{excessfreelargen}
&&F^{ex}_{inf}(\hat x,\rho)= \tilde l_\infty^d \frac{A_d}{\varepsilon}\;  \Big[\frac{ f(z)^2}{4}-\frac{ f(z)^{d/2}}{d}\Big] \nonumber\\&&+\; \rho^{d-1}\Big\{\frac{1}{2} {\cal J}_0(l_{\rm T\infty},\rho) + \frac{1}{2}\ln\Big[\frac{\tilde l_\infty^{\varepsilon/2}\varepsilon^{1/2}\rho^{(d-1)/2}}{4 \pi^2 A_d^{1/2}}\Big]\nonumber\\&& \;\;\;\;\;\;\;\;\;- \frac{1}{4} +\frac{1}{2}\int_0^{z}dz'\;g(z')\Big\}\; -\;{\cal A}_b^\pm( \hat x)
\end{eqnarray}
with $z= z(\hat x,\rho)$ where the bulk part ${\cal A}_b^\pm(\hat x)=\lim_{n \to \infty}(1/n)F^\pm_b(\tilde x)$ is
\begin{eqnarray}
\label{3jjbulklargen}
 &&{\cal A}_b^\pm(\hat x)= \left\{
\begin{array}{r@{\quad \quad}l}
                         \; \hat Q^+_{1}\; \hat x^{d/(d-2)}\quad          & \mbox{for} \;T > T_c, \\
                         \; \hat Q^-_{1}(-\hat x)^{d/(d-2)}& \mbox{for} \;T <
                 T_c ,\;\;\;\;\;
                \end{array} \right.\;\;\;\;
\end{eqnarray}
\begin{subequations}
\label{Qlargen}
\begin{align}
\label{Qpluslargen}
\hat Q^+_{1} = \; - \;  A_d/(4d)  \; ,
\\
\label{Qminuslargen}
\hat Q^-_{1} = \;-\; 2^{d /(d-2)} A_d/(16 \varepsilon).
\end{align}
\end{subequations}
In deriving (\ref{excessfreelargen}) we have used the integral representation (\ref{calWthetay}) of ${\cal W}_n(\tilde y)$ in the form
\begin{eqnarray}
\label{calWthetax}
\frac{{\cal W}_n(n^{1/2}z)}{n}= \frac{{\cal W}_n(0)}{n}+\frac{1}{2}\int_0^{z}dz'\;\frac{\vartheta_{2,n}(n^{1/2}z')}{n^{1/2}}\;\;\;\;\;\;
\end{eqnarray}
together with (\ref{thetalargen}).
Employing (\ref{calG0decom}) we find  from  (\ref{excessfreelargen}) the amplitudes  at $T=T_c$
\begin{eqnarray}
\label{XbeiTclargen}
&&X_{inf}(0, \rho)=(d-1)F^{ex}_{inf}(0,\rho)-\rho\;
\partial F^{ex}_{inf}(0, \rho)/\partial \rho, \;\;\;\;\;\;\\
\label{freebeiTclargen}
&&F^{ex}_{inf}(0,\rho)\;=\; -\;3^{\varepsilon/2}\;\rho^{d-1}/d \;+\; (1/2)\;{\cal G}_0( {\tilde l_{c\infty}}^2/3, \rho), \;\;\;\;\;\;\;\;\;\\
&&\tilde l_{c\infty} =\lim_{n \to \infty}\tilde l_c= \big[9\varepsilon A_d^{-1}\rho^{d-1}\big]^{1/d},
\end{eqnarray}
where ${\cal G}_0(x,\rho)$ is defined in (\ref{calG3spec}).
For $\rho \to 0$, Eq. (\ref{excessfreelargen}) yields  the same result as (\ref{3jjfilmlargenx}).

We note the following shortcomings of our approximate result (\ref{excessfreelargen})-(\ref{Qlargen}). (i) For $\hat x \to -\infty$, $ F^{ex}_{inf}$  diverges logarithmically in  agreement with the exact result  (\ref{4hhx}), where, however, the subleading constant ${\cal C}_\infty$, (\ref{constcalC}), is replaced by
$ {\cal C}_{inf}={\cal C}_\infty -3 \rho^{d-1}\varepsilon/[8(d-2)] $ , as follows from the constant term $\rho^{d-1} c_n/n$ of (\ref{excess-scalfreeaniso18}) in the large-$n$ limit.
(ii) For $\hat x \to \infty$, $F^{ex}_{inf}$ has a finite limit $-3\rho^{d-1}\varepsilon /[4(d-2)]$, as follows from the constant term in (\ref{excess+scalfreeanisoasymx}) in the large-$n$ limit, whereas the exact result $\hat F^{ex,+}$ vanishes  for $\hat x \to \infty$ \cite{dohm2011}.
The finite constants $\propto \rho^{d-1}$ mentioned above do not contribute to the scaling function $X_{inf}$ of the Casimir force, as noted above and in \cite{dohm2011}. These constants are the consequence of the algebraic (rather than exponential) $z$ dependence of the function $g(z)$ for large $|z|$, similar to the effect of the algebraic $y$ dependence of $\vartheta_{2,n}(y)$ for large $|y|$, $n\neq 2$, discussed above.
(iii) In the range $0.15 \lesssim \rho \lesssim  0.4$, the $\rho$-dependence of $F^{ex}_{inf}(0,\rho)$, (\ref{freebeiTclargen}), agrees well with the exact result $\hat F(0,\rho)$ at $T_c$ given by (\ref{3.9xx}), but the agreement deteriorates for larger $\rho$.
(iv) The bulk amplitudes (\ref{Qlargen}) differ appreciably from the exact bulk values in the large-$n$ limit given in (\ref{bulkF}).
\vspace{2.8cm}
\begin{figure}[!ht]
\begin{center}
\subfigure{\includegraphics[clip,width=6.6cm]{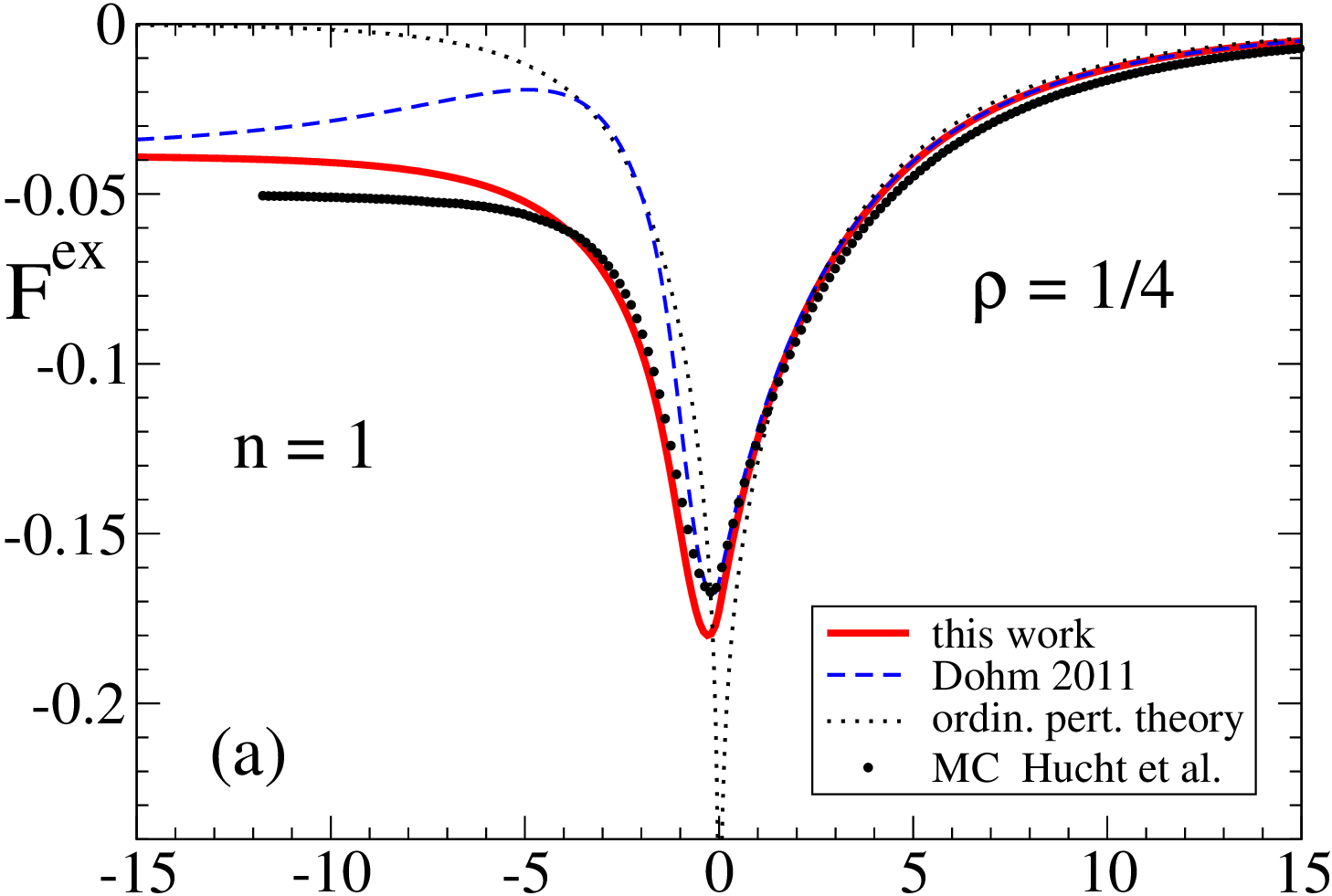}}
\subfigure{\includegraphics[clip,width=6.6cm]{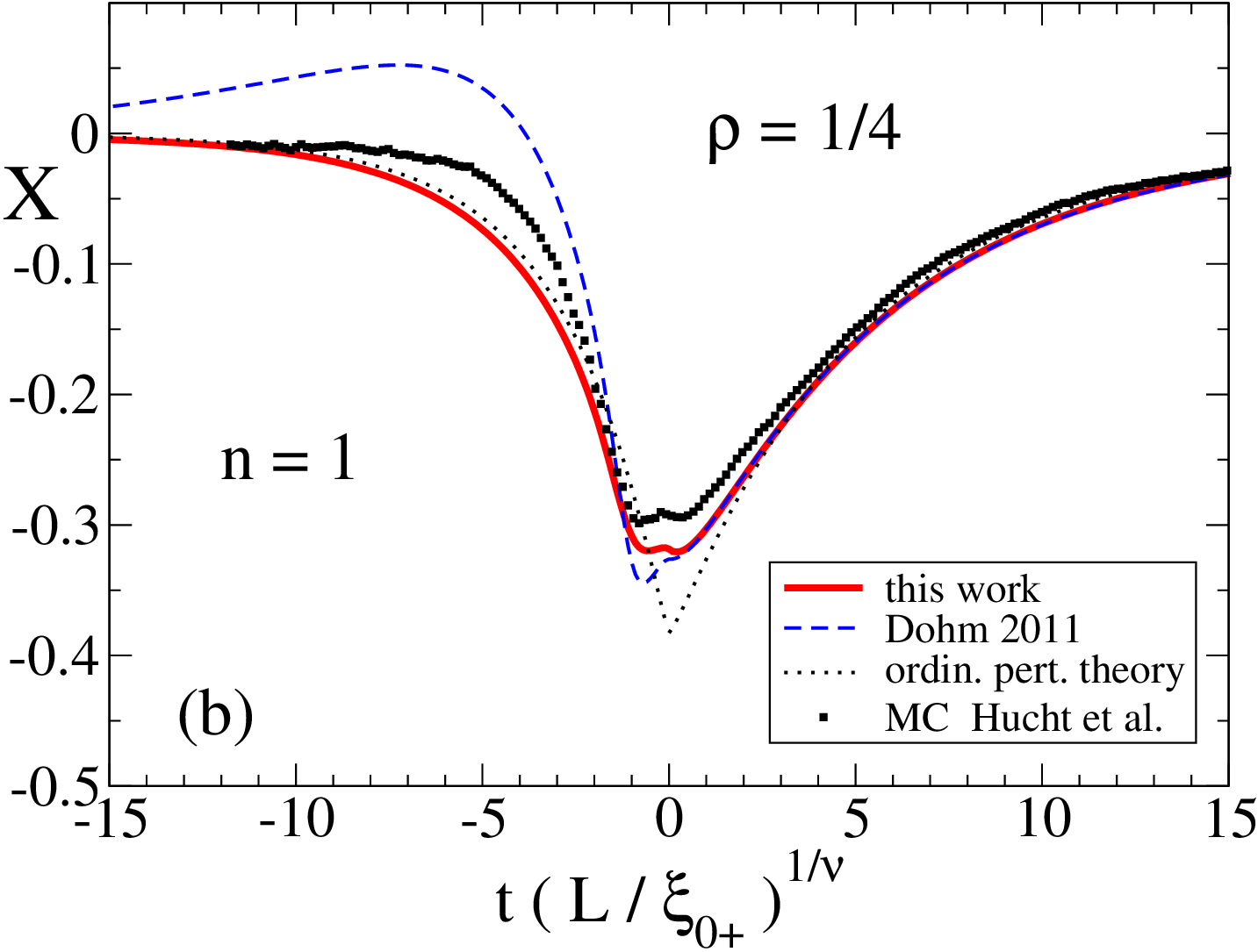}}
\end{center}
\vspace{0.9cm}
\caption{(Color online) Scaling functions
 $F^{ex}(\tilde x, \rho)$ and
$X(\tilde x, \rho)$ in slab geometry for $\rho=1/4$ as a function of $\tilde x = t (L/\xi_{0+})^{1/\nu}$  for $n=1,d=3$.  Solid lines from (\ref{scalfreeaniso}) and (\ref{substisoX}). Dashed lines from Eq. (4.55) of \cite{dohm2011}, compare Fig. 9 of \cite{dohm2011}. MC data  for  the Ising model from \cite{hucht2011}. Dotted lines: 1-loop perturbation theory, (\ref{VIIkoberhalbx})-(\ref{block-excess-below-ngleich1}). }
\end{figure}
\vspace{0cm}
\subsection{ Comparison with earlier results}
For the case $n=1$, an expansion of was made in \cite{dohm2008,dohm2009,dohm2011}  of the longitudinal  higher-mode contribution ${\bare{\Gamma}}_{\rm L}({{\bm\Phi}}^2)$, (\ref{Gammalongbare}),  around the (unrenormalize) lowest-mode average of ${{\bm\Phi}}^2$. On the level of the renormalized theory this means that the longitudinal part $\Gamma_{\rm L}({{\bm\Phi}}^2)$, (\ref {Gamma-longx}), was expanded up to second order in ${{\bm\Phi}}^2- {M}^2$,
\begin{subequations}
\label{GammaLongExp}
\begin{align}
\label{GammaLongExpansion}
\Gamma_{\rm L}({{\bm\Phi}}^2)=\Gamma_{\rm L}({M}^2)+C^{(1)}_{\rm L}({{\bm\Phi}}^2- {M}^2)\nonumber\\+\;(1/2)\;C^{(2)}_{\rm L}{({{\bm\Phi}}^2- {M}^2)}^2+O\big[{({{\bm\Phi}}^2- {M}^2)}^3\big],
\\
\label{GammaLongCoeff1}
C^{(1)}_{\rm L}=\big[\partial \Gamma_{\rm L}({{\bm\Phi}}^2)/\partial {{\bm\Phi}}^2\big]_{{{\bm\Phi}}^2={M}^2},
\\
\label{GammaLongCoeff2}
C^{(2)}_{\rm L}=\big[\partial^2 \Gamma_{\rm L}({{\bm\Phi}}^2)/\partial^2 {{\bm\Phi}}^2\big]_{{{\bm\Phi}}^2={M}^2}.
\end{align}
\end{subequations}
The contributions $\propto C^{(1)}_{\rm L}$ and   $\propto C^{(2)}_{\rm L}$ lead to terms  in Eqs. (4.55) of \cite{dohm2011} and (6.10) of \cite{dohm2008} which do not appear in our Eq. (\ref{scalfreeaniso}).

In the central finite-size regime near $T_c$, which was of primary interest in  \cite{dohm2008,dohm2009,dohm2011}, good agreement was indeed found between Eq. (4.55) of \cite{dohm2011}, Eq. (6.12) of \cite{dohm2008}, and various MC data of the $d=3$ Ising model \cite{Mon,vasilyev2009,hasenbusch2009,hucht2011}. (Note that our function ${\cal J}_0$  differs from the function $J_0$ used in \cite{dohm2008,dohm2011}. The relations between these functions are given in App. A of \cite{dohm2017II}.) The good agreement at $T_c$ is confirmed also in our Fig. 7 (dashed lines). It was noted in \cite{dohm2011}, however, that significant deviations from MC data existed well below $T_c$ and small but systematic deviations well above $T_c$. For this reason separate perturbation calculations were performed in \cite{dohm2011} outside the central finite-size regime which complemented the results of the lowest-mode separation approach.

The disadvantage of the expansion  (\ref{GammaLongExpansion}) is that it introduces several terms in the low-temperature region that are incomplete in the sense of a systematic expansion in powers of  $u^*$. This leads to a temperature dependence well below $T_c$ that is not quantitatively reliable for $\tilde x \lesssim -2$, as shown for the example $\rho=1/4$ by the dashed lines in Fig. 11. Our simpler approximation made in (\ref{distributionapprx}) avoids such unsystematic terms of higher order in $u^*$, except for the term of $O(u^*)$ in (\ref{excess+scalfreeanisoasymx}).

As far as the transverse  higher-mode contribution is concerned, one may consider to make a second-order expansion of $\Gamma_{\rm T}({{\bm\Phi}}^2)$ around $M^2$ similar to that in (\ref{GammaLongExpansion}). This is inappropriate, however, because of the divergence of the coefficient $C^{(2)}_{\rm T}$ in the bulk limit $V\to \infty$ not only for $T \to T_c$ but also for arbitrary $T<T_c$.
This Goldstone divergence originates from the vanishing of  $\bar r_{\rm T}$ in the bulk limit for any $T<T_c$ according to (\ref{rprimetrans}). On the other hand, the quantity $\Gamma_{\rm T}({{\bm\Phi}}^2)$ should remain finite and well behaved in the bulk limit of an exact theory below $T_c$ (in contrast to the true Goldstone divergencies of the longitudinal and transverse bulk susceptibilities).  Similar spurious Goldstone singularities are known from bulk perturbation theory for general $n$ below $T_c$ at vanishing external field \cite{Burnett,str1999,str2003}. As shown up to 4-loop order \cite{Burnett,str1999,str2003}, this problem can be solved  satisfactorily within a RG perturbation theory at finite $h$ where the limit $ h \to 0$ is taken {\it at the end} of the calculation; then all spurious Goldstone divergencies cancel among themselves. Performing such a project within finite-size theory is  beyond the scope of this paper.

An alternative approach of treating finite-size effects in the presence of both Goldstone and critical modes has been presented in \cite{CDS1996}. It is not straightforward to extend this approach to the free energy because the effective Hamiltonian of \cite{CDS1996} would require new higher-order {\it additive} renormalizations. It seems possible, however, to restrict the calculation to $f^{\text ex}$ and $F_{\text Cas}$ which would require only multiplicative renormalizations. This would be an interesting project.
\renewcommand{\thesection}{\Roman{section}}
\renewcommand{\theequation}{6.\arabic{equation}}
\setcounter{equation}{0}
\section{Dimensional crossover in the large-${\bf n}$ limit: exact results for ${\bf d>3}$ }
Exact results for the free energy and Casimir force in an $\infty^{d-1} \times L$ film geometry with periodic BC in the large-$n$ limit have been presented for $d\leq 3$  \cite{danchev1996,cd2004,dohm2011} where no finite film transition temperature $T_{cf}(L)$ exists. Here we extend the analysis to $3<d<4$. This is of conceptional interest since an exact description can be given for the dimensional crossover from the $d$ dimensional transition at bulk $T_c$ to the $d-1$ dimensional transition at the film critical temperature  $0<T_{cf}(L)<T_c$ for $d>3$.

Starting point are Eqs. (3.1)-(3.3) of \cite{dohm2011} for slab geometry where we take the film limit $\rho=L/L_\parallel\to 0$ at fixed $L$ corresponding to $V^{-1} \sum_{\bf k} \to L^{-1} \sum_p \int_{\bf q} $ with $\mathbf k\equiv(\mathbf q , p)$. Here  $\int_{\bf q}$ is a $(d-1)$ dimensional integral %with finite lattice cutoff
[see (\ref{kintegral})], and the sum $\sum_p$ runs over $p=2\pi m/L, m = 0, \pm1, \pm2, ...$ up to  $\pm\pi/\tilde a$. The resulting film free energy density per component $f_{film}$ can be decomposed into a non-singular bulk part $\hat f_{b,ns}(t)$ and a film part $\hat f_{film}(t,L)$,
\begin{eqnarray}
\label{ffilmdecomp}
&&f_{film}(t,L) = \hat f_{b,ns}(t) + \hat f_{film}(t,L),\\
\label{ns}
&&\hat f_{b,ns}(t)
= - \frac{\ln (2\pi)}{ 2 \tilde a^d} - \frac{r_0^2}{16 u_0 n} + \frac{1}{2} \int_{\bf k} \ln \{[  \delta
\widehat K (\mathbf k)] \tilde a^2 \},\;\;\;\;\;\;\;\\
\label{sfilm}
&&\hat f_{film}(t,L)=\frac{(r_0- r_{0c})\hat \chi_{f}^{-1}}{8 u_0 n}- \frac{\hat \chi_{f}^{-2}}{16 u_0 n} \nonumber\\&&+(1/2)\;\widehat{\cal I}_3(\hat \chi_{f}^{-1})+ (1/2)\;\Delta_{film}(\hat \chi_{f}^{-1},L),\\
\label{4bxfilm}
&&\hat \chi_{f}(t, L)^{-1}= r_0 -r_{0c} + \; 4 u_0n  \frac{\partial}{\partial \hat \chi_{f}^{-1}}\;\widehat{\cal I}_3(\hat \chi_{f}^{-1}) \nonumber\\&&+ \; 4 u_0 n \partial\;\Delta_{film}(\hat \chi_{f}^{-1},L)/\partial (\hat \chi_{f}^{-1}),\\
\label{deltafilm1}
&&\Delta_{film}(r,L)= L^{-1} \sum_p
\int_{\bf q} \ln
\{[r + \delta \widehat K (\mathbf q , p)] \tilde a^2\}\nonumber\\&&
 \;\;\;\;\;\;\;\;\;\;\;\;\;\;\;\;\;\;\;\;\;\;\;\;- \int_{\bf k} \ln \{[r + \delta \widehat K (\mathbf k)]
\tilde a^2\}.
\end{eqnarray}
For $\widehat{\cal I}_3$ see (\ref{bulk3}) and (\ref{bulkintsing}). For $ r_0 -r_{0c}\leq 0$, both bulk quantities $\hat \chi_b^{-1}=\lim_{L \to \infty} \hat\chi_f^{-1}$ and $\hat f_{b,s}=\lim_{L \to \infty} \hat f_{film}$ vanish. This holds for arbitrarily large negative $r_0 -r_{0c}$ including the low-temperature limit $r_0 -r_{0c}\to -\infty$. The vanishing of  $\hat \chi_b^{-1}$ is the characteristic of the massless Goldstone modes that are the origin of long-range correlations and the non-vanishing Casimir force for all $T$ below $T_c$. For finite but large $L$, $\hat \chi_f^{-1}$ is small not only in the critical region  but also in the entire low-temperature region $T \leq T_c$.
For  $L/\tilde a \gg 1$, $\tilde a^2 \hat\chi_f^{-1}\ll 1$,  $ L^2 \hat\chi_f^{-1} \lesssim O(1)$ we obtain
(see App. A of \cite{dohm2017II} and Eq. (3.11) of \cite{dohm2011})
\begin{eqnarray}
\label{Deltafilmx1}
&&\Delta_{film}(\hat\chi_f^{-1},L)=  L^{-d} \; {\cal G}_{0,film}
\big(\hat\chi_f^{-1}  L^2  \big), \\\;\;\;\;\;\;\;
\label{AbleitDeltafilm}
&&\partial\;\Delta_{film}(r,L)/{\partial r}= - L^{2-d}   \; {\cal G}_{1,film}(rL^2),\;\;\;\;\;\;\;\;\;\\
\label{GfilmEinsx}
&&{\cal G}_{1,{\text film}}(z)
=-\partial {\cal G}_{0,film}(z)/\partial z,
\end{eqnarray}
where  ${\cal G}_{0,film}$ is defined by (\ref{GfilmNullx})
which is identical with $2 {\cal I}^{(p)}_d(z)$ defined in Eq. (4.2a) of \cite{kastening-dohm}.  It is finite for $z\geq 0$ in $d>1$ dimensions.
For small $\hat \chi_f^{-1}$ the terms  $\propto \hat \chi_f^{-2}$ in (\ref{sfilm}) and $\hat \chi_f^{-1}$
in (\ref{4bxfilm}) are negligible. The remaining terms can be expressed in terms of $P_f=L\hat \chi_f(t,L)^{-1/2}$ and $\hat F_{film}= L^d \hat f_{film}(t,L)$ for $2<d<4$  as
\begin{eqnarray}
\label{3.10}
&& \hat F_{film} \; = \;  \; \frac{A_d}{2 \varepsilon}
\left[\hat x P_f^2 \; - \; \frac{2}{d} \; P_f^d
\right]  + \frac{1}{2}\;{\cal G}_{0,film}( P_f^2),\;\;\;\;\;\\
\label{Pimplizit}
&&P_f^{d-2} \; = \; \hat x \; - \;  (\varepsilon/A_d)\;{\cal G}_{1,film}( P_f^2),\\
\label{xiunend}
&&\hat x= t(L/\xi_{0})^{1/\nu_\infty}, \;\;\;\;\nu_\infty= (d-2)^{-1},\\
\label{corrinf}
&&\xi_{0}=[4u_0nA_d/(a_0\varepsilon)]^{\nu_\infty}.
\end{eqnarray}
In the bulk limit we obtain
$\hat \chi_f\to \hat \chi_b$,  $\hat F_{film}\to \hat F_{b \infty} =L^d \hat f_{b,s}$
for $2<d<4$,
\begin{eqnarray}
\label{chiunendlich}
&& \hat \chi_b = \left\{ \begin{array}{r@{\quad\quad}l}
                  \xi^2_{0}t^{-\gamma_\infty}\hspace{1.0cm} & \mbox{for} \; T\gtrsim T_c\;, \\ \infty & \mbox{for} \;
                 T\leq T_c\;,
                \end{array} \right.\\
\label{bulkF}
&& \hat F_{b \infty}(\hat x)= \left\{ \begin{array}{r@{\quad\quad}l}
                  Y_\infty\hat x^{d\nu_\infty}\hspace{1.0cm} & \mbox{for} \; T\gtrsim T_c\;, \\ 0& \mbox{for} \;
                 T\leq T_c\;,
                \end{array} \right.
\end{eqnarray}
where $\gamma_\infty= 2\nu_\infty=2/(d-2)$, $Y_\infty=(d-2)A_d/(2d\varepsilon)$.
The Casimir force scaling function $\hat X_{\text{film}}
=  - L^d\partial[L
\hat f^{ex}_\text{film} ]/\partial L$ with $\hat f^{ex}_\text{film}=\hat f_\text{film}-\hat f_{b,s}$ is obtained from
(\ref{3.10}), (\ref{Pimplizit}), and  (\ref{bulkF}) as
\begin{eqnarray}
\label{Xinftyfilmx}
\hat X_{film}(\hat x)&=&
\hat F_{b\infty}(\hat x)+ [(d-1)/2]{\cal G}_{0,film}(P_f^2) \nonumber\\&+& (A_d/\varepsilon)\Big[(1/2)\hat x P_f^2- (d-1)P_f^d/d\Big]  \;\;
\end{eqnarray}
The condition for film criticality is $\hat \chi_{f}(t, L)^{-1} \to 0$
at finite $L$ corresponding to $P_f\to 0$. For $d\leq 3$, $P_f$ is finite for $-\infty < \hat x <\infty$, and $P_f\to 0$ for $\hat x \to -\infty$ since ${\cal G}_{1,{\text film}}(P_f^2)$ diverges for $P_f\to 0$. This implies that no film transition exists at a finite value of $\hat x$, and $\hat X_{film}(\hat x)$ is an analytic function of $\hat x$ for $-\infty < \hat x< 0$  with the finite low-temperature amplitude
\begin{eqnarray}
\label{Xinfinfilmlow}
\hat X_{film}(-\infty)=
 [(d-1)/2]\;\;{\cal G}_{0,film}(0).
\end{eqnarray}
Thus, for the $\varphi^4$ model with periodic BC in the large-$n$ limit for $d\leq 3$,
the finite-size scaling regime below $T_c$ is not limited by non-scaling effects.
This is in contrast to the case of free BC in film geometry where recent numerical and analytic studies of several models \cite{diehl2012,DanRud} in the ($d=3,n=\infty)$ universality class do not unambiguously answer the question concerning the possible existence of a nonscaling regime for the Casimir force far below $T_c$.

For
$d>3$, ${\cal G}_{1,{\text film}}(P_f^2)$ has a finite limit
\begin{eqnarray}
\label{GfilmNullNull}
{\cal G}_{1,{\text film}}(0)=-(1/2)\pi^{d/2}\Gamma\big[(d-2)/2\big]\zeta(d-2)
\end{eqnarray}
for $P_f\to 0$. This implies that $P_f$ is finite only in the range $\hat x^* < \hat x <\infty$, with $P_f \to 0$ for  $\hat x \to \hat x^* $ where (\ref{Pimplizit}) yields the finite negative value
\begin{eqnarray}
\label{xfilmc}
x^*=
(\varepsilon/A_d)\;{\cal G}_{1,{\text film}}(0)<0\;\;\;\;\;\;\;\;\;\;\;\;
\end{eqnarray}
of the scaling variable (\ref{xiunend}) for $3<d<4$, corresponding to the fractional shift
\begin{eqnarray}
\label{shiftfilm}
t_{cf}= [T_{cf}(L)-T_c]/T_c &=&x^*(L/\xi_{0})^{-1/\nu_\infty}<0\;\;\;\;\;\;\;\;\;\;\;\;
\end{eqnarray}
of the film critical temperature. For $d>3$ and $t\geq t_{cf}$ it is appropriate to rewrite (\ref{Pimplizit}) as
\begin{eqnarray}
\label{Pimplizitstern}
P_f^{d-2} \; &=& \; \hat x \; - \;x^* - (\varepsilon/A_d)\;{\cal H}( P_f^2),\\
\label{calH}
{\cal H}(z)&=&{\cal G}_{1,{\text film}}(z)-{\cal G}_{1,{\text film}}(0)\nonumber\\
\label{Hfilm}
&=&\frac{1}{4
\pi^2}
\int^\infty_0 dy  \left[\exp \left(- \frac{z y}{4
\pi^2} \right)-1 \right]\left(\frac{\pi}{y}\right)^{(d-1)/2} \nonumber\\&  \times & \left\{\left(\pi/y\right)^{1/2} \; - \; K(y) \right\}.
\end{eqnarray}
Now (\ref{Pimplizitstern}) and (\ref{3.10})
determine $P_f$ and $\hat F_{film}$
as functions of $\hat x  - x^* \propto T-T_{cf}\geq 0$
and provide an exact description of the crossover from well above bulk $T_c$ to the film critical region near  $T_{cf}(L)$ at finite $L$ below $T_c$ including the bulk critical behavior for $T\approx T_c, L\to \infty $. This corresponds to the change from $P_f\gg 1$ over $P_f=P_c >0$  to $P_f=0$. The value of $P_c$ at bulk $T_c$ is determined implicitly by (\ref{Pimplizit}) for $\hat x =0$,
\begin{eqnarray}
\label{Pimplizitc}
&&P_c^{d-2} \; = - \;  (\varepsilon/A_d)\;{\cal G}_{1,film}( P_c^2).
\end{eqnarray}
For $d>3$, the susceptibility diverges at  $T_{cf}(L)$ and remains infinite for  $T<T_{cf}$  where a finite order parameter exists in analogy to the bulk case below $T_c$ for $d>2$ \cite{cd1998}.
In the following we derive the susceptibility $\hat \chi_f(t,L)=L^2 P_f^{-2}$ and the film part  $\hat F_{film}$ of the free energy density near $T_{cf}(L)$.
For small $P_f^2\ll1$ we have from (\ref{GfilmEinsx}) and from Eq. (4.3a) of \cite{kastening-dohm} for $d>3$
\begin{eqnarray}
\label{calHsmall}
{\cal H}( P_f^2)=(5-d)^{-1}A_{d-1}\; P_f^{d-3} + O(P_f^{d-2}).
\end{eqnarray}
From (\ref{Pimplizitstern}) we then obtain for $P_f^2\ll1$
\be
 \hat \chi_f = \left\{ \begin{array}{r@{\quad\quad}l}
                 \hat\chi_{0\infty} L^2(\hat x - x^*)^{-\gamma^*}\hspace{1.0cm} & \mbox{for} \; T\gtrsim T_{cf}, \\ \infty & \mbox{for} \;
                 T\leq T_{cf},
                \end{array} \right.
\ee
with  $ \hat\chi_{0\infty}= \big\{\varepsilon A_{d-1}/[(5-d)A_d]\big\}^{\gamma^*}$,
and (\ref{3.10}) yields
$ \hat F_{film}  =  (1/2){\cal G}_{0,film}( 0)+ \hat F_{film,s}$
with the singular film part
\begin{eqnarray}
\label{Ffilmunendsing}
 \hat F_{film,s}= \left\{ \begin{array}{r@{\quad\quad}l}
                 \hat F_{f\infty}\;(\hat x - x^*)^{(d-1)\nu^*}\hspace{0.5cm} & \mbox{for} \; T\gtrsim T_{cf}, \\ 0& \mbox{for} \;
                 T\leq T_{cf},
                \end{array} \right.\;\;\;\;
\end{eqnarray}
where $\hat F_{f\infty}=[A_d/(2\varepsilon)]\hat\chi_{0\infty}^{-1}$. The critical exponents
\be
\nu^*=(d-3)^{-1}, \;\;\; \gamma^*=2\nu^*= 2/(d-3)
\ee
at the film transition are indeed identical with the bulk critical exponents of a $(d-1)$-dimensional system for $n=\infty$, as expected on general grounds.
Two-scale-factor universality for {\it isotropic} systems implies that $x^*$, (\ref{xfilmc}), is a universal quantity.
\section*{ACKNOWLEDGMENT}
The author is grateful to M. Hasenbusch for providing the data of Ref. \cite{hasenbusch2011} prior to publication and to D. Dantchev, M. Hasenbusch, A. Hucht, and O. Vasilyev for providing the data of Refs. \cite{dan-krech,vasilyev2009,hasenbusch2010,hucht2011} in numerical form.

\renewcommand{\thesection}{\Roman{section}}
\setcounter{equation}{0} \setcounter{section}{1}
\renewcommand{\theequation}{\Alph{section}.\arabic{equation}}

\section*{Appendix A: Exact results for the Gaussian model and in the large-${\bf n}$ limit}
The following results are valid for systems with the isotropic interaction (\ref{2hh}). They follow from the results for weakly anisotropic systems derived in \cite{dohm2017II}. First we consider the Gaussian model, i.e., (\ref{hamiltonian fourier}) with $u_0=0, {\bf h}={{\bf 0}},r_0=a_0t\geq 0,r_{0c}=0$.
The Gaussian bulk correlation length is
$\xi_G = r_0^{-1/2}= \xi_{0G} t^{-\nu_G}  , \xi_{0G} = a_0^{-\nu_G}$
with $\nu_G=1/2$. The scaling functions of the excess free energy density and  the Casimir force in slab geometry are
\begin{eqnarray}
\label{b23scal}
&&F^{G, {\text ex}}(\tilde x_G, \rho) = (n/2) {\cal
G}_0 (\tilde x_G,\rho),\\
\label{cascalJ}
&&X^G(\tilde x_G, \rho) =-(n/2)\;\rho^{d-1}\nonumber \\&& \times \Big\{2+
\Big[(\tilde x_G/\nu_G)
\partial/\partial \tilde x_G+\rho
\partial/\partial \rho\Big]{\cal J}_0(\tilde x_G, \rho)\Big\},\;\;\;\;\;\;\;\;\;\;\;\;
\end{eqnarray}
with the scaling variable
$\tilde x_G= t (L/\xi_{0G})^{1/\nu_G}$
where ${\cal G}_0$ and ${\cal J}_0$ are defined by (\ref{calG3spec}) and (\ref{calJ3spec}). This yields the finite Gaussian Casimir amplitude in slab geometry
\begin{eqnarray}
\label{cascalJcritslab}
X^G(0, \rho)=-(n/2)\;\rho^{d-1} \big[2+ \rho\;
\partial{\cal J}_0(0, \rho)/\partial \rho\big]. \;\;\;\;\;\;\;\;
\end{eqnarray}
The corresponding results for film geometry are
\begin{eqnarray}
\label{FexfilmscalingGauss}
&&F_{film}^{G,{\text ex}}(\tilde x_{G})=(n/2)  \; {\cal G}_{0,film}\big(\tilde x_{G}\big),\;\;\;\;\;\;\;\;\\
\label{CasimirfilmscalingGauss}
&&X^G_{film}(\tilde x_{G})=(d-1)F^{G,ex}_{film}(\tilde x_{G})\nonumber\\&&-\;
(\tilde x_{G}/\nu_G)\;
\partial F_{film}^{G,ex}(\tilde x_{G})/\partial\tilde x_{G},\;\;\;\;\;\;\\
\label{CasimirGaussAmp}
&&X^G_{\text film}(0)=  (d-1)\;(n/2)\; {\cal G}_{0,film}(0),
\end{eqnarray}
with ${\cal G}_{0,film}$ given by (\ref{GfilmNullx}). $X^G_{\text film}(0)/n$ is identical with the exact low-temperature amplitude (\ref{Xinfinfilmlow}) for $n \to \infty$.

An analytic proof was given in \cite{hucht2011} that the Casimir amplitude at $T_c$ vanishes in cubic systems ($\rho=1,d=3$). This proof, however, is not valid for the Gaussian model which yields the {\it finite} result (\ref{cascalJcritslab}) for general $\rho$ including $\rho=1$ (see Fig. 7(b), dotted line). The reason is a logarithmic divergence of   $F^{G, {\text ex}}(\tilde x_G, 1)$, (\ref{b23scal}),  with
\begin{eqnarray}
\label{calG0decomGaussdiv}
{\cal G}_0(\tilde x_G, 1) \approx  \ln[\tilde x_G/(4\pi^2)]+ {\cal J}_0(0, 1 ),
\end{eqnarray}
for $\tilde x_G \to 0$, see (\ref{calG0decom}). Thus $\tilde x_G \partial F^{G, {\text ex}}(\tilde x_G, 1))/\partial \tilde x_G$ (corresponding to $x \partial \Theta(x,1)/\partial x$ in Eq. (20) of \cite{hucht2011} ) does not vanish for $\tilde x_G \to 0$, and Eq. (22) of \cite{hucht2011} is not valid for the Gaussian model.

Next we consider the scaling functions $\hat F= \lim_{n\to \infty} F/n$  and $\hat X= \lim_{n\to \infty} X/n$ of the free energy density and the Casimir force per component, respectively, of the $\varphi^4$ model in the large-$n$ limit in slab geometry,
\begin{eqnarray}
\label{3.9xx}
&&\hat F(\hat x, \rho)=   \frac{A_d}{2 \varepsilon}
\left[\hat x P^2  -  \frac{2}{d}  P^d \right] + \frac{1}{2}{\cal G}_0( P^2, \rho),\;\;\;\;\;\;\;\;\;\;\;\;\;\;\;\\
\label{X-large-nx-isoslab}
&&\hat X(\hat x,\rho)= (A_d/\varepsilon)\Big\{(1/2)\hat x P^2- [(d-1)/d]P^d\Big\} \nonumber\\&&+\;\hat F_{b\infty}(\hat x) - (\rho^d/2)\;
\partial {\cal J}_0(P^2, \rho)/\partial \rho\;\;\;\;\;\;\;\;\;
\end{eqnarray}
with the scaling variable (\ref{xiunend}).
The function $P(\hat x, \rho)$ is determined by Eqs. (3.7)-(3.9) of \cite{dohm2011}. The bulk part $\hat F_{b\infty}(\hat x)$ is given in (\ref{bulkF}). For $\hat x \to \infty$, $\hat F^{ {\text ex}}(\hat x, \rho)=\hat F(\hat x, \rho)-\hat F_{b\infty}(\hat x)$ vanishes exponentially \cite{dohm2011}.
For $\hat x  \to - \infty$, $\hat F(\hat x, \rho)$ has a logarithmic divergence \cite{dohm2011}
\begin{eqnarray}
\label{4hhx}
&&\hat F(\hat x , \rho) \approx -(1/2) \;\rho^{d-1} \ln|2\hat x |+ \;\;{\cal C}_\infty(\rho) ,\;\;\;\\
\label{constcalC}
&&{\cal C}_\infty(\rho)=\frac{ \rho^{d-1}}{2}\Big\{-1  + \ln \Big[\frac{\varepsilon\rho^{d-1}}{2\pi^2 A_d}\Big]+ {\cal J}_0(0,\rho)\Big\}\;\;\;\;\;\;\;\;\;\;\;
\end{eqnarray}
where we have used (\ref{calG0decom}),
whereas $\hat X$ has a finite limit for $\hat x  \to - \infty$,
\begin{eqnarray}
\label{X-large-nx-low} \hat X(-\infty,\rho) =  - (1/2) \;\rho^{d-1}[ 1 +\rho\;
\partial {\cal J}_0(0, \rho)/\partial \rho].
\end{eqnarray}
Unlike the case of film geometry, the low-temperature amplitude (\ref{X-large-nx-low}) for a slab geometry is not identical with the Gaussian amplitude at $T_c$ in the same geometry, divided by $n$, as given in (\ref{cascalJcritslab}).
\renewcommand{\thesection}{\Roman{section}}
\setcounter{equation}{0} \setcounter{section}{2}
\renewcommand{\theequation}{\Alph{section}.\arabic{equation}}

\section*{Appendix B: 1-loop perturbation theory}
Ordinary perturbation theory for finite systems is based on the decomposition
\begin{eqnarray}
\label{decompfilm}
{\bm \varphi}_j = {\bf M}_{mf} +  {\bm s}_j
\end{eqnarray}
where ${\bf M}_{mf}=M_{mf}\;{\bf e}_h$ is the the bulk mean-field  order parameter.
This approach improves the results of the lowest-mode separation approach well above $T_c$ but suffers from serious shortcomings at and below $T_c$.
We consider only a finite slab geometry.
For general $n$ below $T_c$ it is necessary to work at finite external field ${\bf h}= h{\bf e}_h$.
$ M_{mf}(r_0,h)$ is determined implicitly by
$M_{mf} [r_0+4u_0M_{mf}^2]=h$.
The nonanalyticity of $ M_{mf}(r_0,h)$ at $r_0=0, h=0,$ implies that the ansatz (\ref{decompfilm}) introduces an artificial singularity at $T_c$ where no singularity should exist in a finite volume.
We further decompose $ {\bm s}_j$  into "longitudinal" and "transverse" parts $ {\bm s}_j =  {\bm s}_{{\rm L}j} +  {\bm s}_{{\rm T}j}$ which are parallel and perpendicular with respect to ${\bf h}$ and ${\bf M}_{mf}$. Up to one-loop order the Gibbs free energy density and excess free energy density are
\begin{eqnarray}
\label{free one loop} &&f_{1-loop}(t,h, L,\rho) = \frac{1}{2}r_0M_{mf}^2+u_0 M_{mf}^4-hM_{mf} \nonumber\\&&- n\frac{\ln (2 \pi)} {2 \tilde a^d} + \frac{1}{2V}
{\sum_{\bf k}} \ln\{[r^{mf}_{\rm L}+ \delta \widehat K (\mathbf k)] \tilde a^2\}
\nonumber\\&&+  \frac{n-1}{2V}
{\sum_{\bf k}} \ln
\{[r^{mf}_{\rm T}+ \delta \widehat K (\mathbf k)] \tilde a^2\}+ O(u_0),\;\\
\label{blockexcess}
&&f^{\text ex}_{1-loop}(t, h,  L,\rho)=  \;(1/2) \; \Delta( r^{mf}_{\rm L},  L,\rho) \nonumber\\ &&+ \; (n-1/2) \;  \Delta(r^{mf}_{\rm T},   L,\rho)+ O(u_0),\;\;\;\;\\
\label{LongMF}
&& r^{mf}_{\rm L}(t,h)=r_0+12u_0M_{mf}(r_0,h)^2 ,
\\
\label{TransMF}
&& r^{mf}_{\rm T}(t, h)=r_0+4u_0M_{mf}(r_0,h)^2,
\end{eqnarray}
where  $\Delta$ is defined in (\ref{bb9xy}).
Because of the ${\mathbf k} = {\bf0}$ terms, the sums ${\sum_{\bf k}}$ exist only
for $r^{mf}_{\rm L} > 0$ and $r^{mf}_{\rm T} > 0$. Thus, for $n>1$, $f^{\text ex}_{1-loop}$ diverges for $r_0 < 0$, $h\to 0$ at finite $V$ where $r^{mf}_{\rm T} \to 0$ (whereas the bulk part of (\ref{free one loop}) remains finite for $h\to 0$, see Sec. III). It also diverges for $n\geq1, r_0 = 0,h\to 0$. These singularities of the {\it finite} system are unphysical and are the consequence of the inappropriate perturbation approach based on the decomposition (\ref{decompfilm}). The lowest-mode separation approach avoids these spurious singularities.

First we consider the case $T>T_c$ at $h=0$. The scaling functions
follow from the Gaussian results in App. A combined with the renormalization parallel to that described in Sec. X.A of \cite{dohm2008} and Sec. VI.A of \cite{dohm2011}. The results read for $\tilde x > 0$
\begin{eqnarray}
\label{VIIkoberhalbx}
&&F^{ex, +}_{1-loop}(\tilde x,\rho) = (n/2)\; {\cal G}_0 (\tilde x^{2\nu} ,\rho), \;\\
\label{cascalJplus}
&&X^+_{1-loop}(\tilde x, \rho) =-(n/2)\rho^{d-1}\nonumber \\&& \times \big\{2+\big[
(\tilde x/\nu)
/\partial/\partial \tilde x+ \rho
\partial/\partial \rho\big]{\cal J}_0(\tilde x^{2\nu}, \rho)\big\}\;\;\;\;\;\;\;\;\;\;\;\;
\end{eqnarray}
where ${\cal G}_0$, ${\cal J}_0$, and $\tilde x$ are given by (\ref{calG3spec}), (\ref{calJ3spec}), and (\ref{3jjx}).  We consider this as a reliable result well above $T_c$ at the one-loop level.  It predicts an exponential approach of $f_{1-loop}^{\text ex,+}$ and $X^+_{1-loop}$ to zero for large $\tilde x$.
In the critical limit $\tilde x \to 0_+$,  $F^{ex, +}_{1-loop}$ at $h=0$ has an unphysical logarithmic divergence [see (\ref{calG0decom}) and the dotted line in Fig. 11 (a)] which comes from the ${\mathbf k} = {\bf0}$ terms in (\ref{blockexcess}). This divergence is canceled in the Casimir force which has a finite amplitude at $T=T_c$
\begin{eqnarray}
\label{cascalJcrit1loop}&&X_{c,1-loop}=-(n/2)\rho^{d-1}\big[2+ \rho \partial{\cal J}_0(0, \rho)/\partial \rho \big]\;\;\;\;\;\;\;\;\;\;\;
\end{eqnarray}
as shown in Fig. 7 (b) as dotted line. It is identical with the Gaussian result $X_c^G$, (\ref{cascalJcritslab}), but differs from the result (\ref{CasimiratTc}) of the lowest-mode separation approach for $\rho> 0$.
Below $T_c$ we find from (\ref{blockexcess}),  (\ref{F.155yy}), (\ref{calG0decom}) and (\ref{3nn}) a finite Casimir force  scaling function for $h \to  0$,
\begin{eqnarray}
\label{Scaleinloop} &&X^-_{1-loop}(\tilde{x},\rho) =-\frac{\rho^{d-1}}{2} \Big\{2n+(n-1)\rho
\frac{\partial}{\partial \rho}
{\cal J}_0(0, \rho)\nonumber \\ && +\big[
(\tilde x/\nu)
/\partial/\partial \tilde x+ \rho
\partial/\partial \rho\big]{\cal J}_0(|2\tilde x|^{2\nu}, \rho)\Big\},
\end{eqnarray}
with a finite low-temperature limit $\tilde x \to -\infty$ for $n>1$,
\begin{eqnarray}
\label{lowXoneloop}
\lim_{\tilde x \to -\infty}X^-_{1-loop} =-\frac{n-1}{2}\rho^{d-1}\big[2+\rho
\frac{\partial{\cal J}_0(0, \rho)}{\partial \rho}\big].
\end{eqnarray}
This differs from the result (\ref{lowXn2}) of the lowest-mode separation approach  and also, for $n \to \infty$, from the exact result (\ref{X-large-nx-low})  but agrees with these results for $\rho \to 0$. The result (\ref{lowXoneloop}) is shown in Fig. 5 as dotted line.
Using (\ref{calG0decom}) we find that, in the critical limit $\tilde x \to 0_-$, (\ref{Scaleinloop}) yields the same finite result as (\ref{cascalJcrit1loop}). Thus  $X_{1-loop }$ is finite and continuous at $T_c$ for general $n\geq 1$ but has a singularity at $T_c$ [Fig. 11 (b)].

For $n=1$
below $T_c$ we have
$\lim_{ h\to  0}f_{1-loop}^{\text ex,-} = (1/2) \Delta( -2r_0,  L,\rho)$
with the scaling function
\begin{eqnarray}
\label{block-excess-below-ngleich1}
&&
F_{1-loop}^{\text ex,-}(\tilde x,L,\rho)= (1/2){\cal G}_0((2|\tilde x|)^{2\nu }, \rho)\;\;\;\;\;\;\;\;\;\;\;\;
\end{eqnarray}
which predicts an exponential approach of $f_{1-loop}^{\text ex,-}$ to {\it zero} for $\tilde x \to -\infty$ [Fig. 11 (a)]. This is at variance with the {\it finite} low-temperature result (\ref{lowTemplx}) of the lowest-mode separation approach.
The resolution of this discrepancy is that, for $n=1$, ordinary perturbation theory at finite volume does not capture the configurations of both phases with positive and negative magnetization that exist at $h=0$ below $T_c$, as noted in Sec. VI.B of \cite{dohm2011}.

The results (\ref{VIIkoberhalbx})-(\ref{block-excess-below-ngleich1}) are shown as dotted lines in Fig. 11 for $\rho=1/4,d=3$. The thin line of Fig. 13 (a) of \cite{dohm2011} is part of the dotted line in our Fig. 11 (a) which diverges for $t\to 0$. The thin line of Fig. 14 (a) of  \cite{dohm2011} is part of the dotted line in our Fig. 11 (b). The remark in \cite{dohm2011} that this line diverges for $t\to 0$ is not correct.
\renewcommand{\thesection}{\Roman{section}}
\setcounter{equation}{0} \setcounter{section}{3}
\renewcommand{\theequation}{\Alph{section}.\arabic{equation}}

\section*{Appendix C: Scaling functions well away from $T_c$ }
We rewrite the function ${\cal W}_n(y)$, (\ref{calWx}), as
\begin{eqnarray}
\label{Wn3x}
&&{\cal W}_n\big(y\big)=-(y^2/16) + [(n-2)/2]\ln\Big(4/(-y)\Big) \nonumber \\&&-\ln \Big\{\int_{y/4}^\infty  dt \;\;(1-4t/y)^{(n-2)/2} \exp ( -t^2 )\Big\},\\
\label{Woberhalb}
&&{\cal W}_n(y)=-\ln \Big\{\Big(\frac{2}{y}\Big)^{n/2}\int_0^\infty  dt \;\;t^{(n-2)/2} e^{ -t-4t^2/y }\Big\}\;\;\;\;\;\;\;\;\;
\end{eqnarray}
for $y<0$
and $y>0$, respectively. For $n=2$, ${\cal W}_n (y)$ can be expressed in terms of the error function as
\begin{eqnarray}
\label{calW2x}
{\cal W}_2 (y) =  - \ln \big\{ e^{-y^2/16}\;\text {erfc}(y/4)/(2 \pi^{1/2})\big\}.
\end{eqnarray}

In the following we derive (\ref{excess-scalfreeaniso18}) and (\ref{excess+scalfreeanisoasymx}) from (\ref{scalfreeaniso}).
Eqs. (\ref{6ff}) and (\ref{6gg}) yield the exact implicit relation between $\tilde y$ and $\tilde x$ for general $n$
\begin{eqnarray}
\label{ytildeexact}
A_d^{-1} u^*\bar\rho\;^{d-1}\tilde y^2 = \frac{1}{4}(2|\tilde x| Q^*)^{d\nu}\;\frac{|2\tilde y)|^\alpha}{|\tilde y+12 \vartheta_{2,n}(\tilde y)|^\alpha}\;.
\end{eqnarray}
Employing (\ref{theta2largebelow}) we obtain for large negative $\tilde x$
\begin{eqnarray}
\label{ytildenegative}
\tilde y^2 = &&A_d \bar \rho\;^{1-d}(2|\tilde x| Q^*)^{d\nu}/(4 u^*)\; -6(n-2)\alpha \nonumber\\&& +\; O\big((n-2) |\tilde x|^{-d\nu}\big)\;\;\;\,\;\;\;\,
\end{eqnarray}
for $n\neq 2$. By means of (\ref{6ff}), the first term of (\ref{scalfreeaniso}) should be written as
\begin{eqnarray}
\label{lexplicit} && -A_d \;\tilde l^d/(4d) = -[u^* /(4d)]\big[\tilde y + 12\; \vartheta_{2,n}(\tilde y)\big]^2 \bar\rho\;^{d-1},\;\;\;\;\;\;\;\;\;\;
\end{eqnarray}
and (\ref{theta2largebelow}) should be employed. This yields a contribution $\propto \tilde y^2$ that contains the bulk term $\propto 1/(4d)$ in (\ref{6m}). The logarithmic term $\propto n/2$ in (\ref{scalfreeaniso}) should be decomposed according to $n/2 = 1/2 + (n-1)/2$. From (\ref{Wn3x}) we obtain the asymptotic form for $- \tilde y \gg 1$
\begin{eqnarray}
\label{Wnminus3}
{\cal W}_n(\tilde y)\approx-\frac{\tilde y^2}{16}- \frac{ \ln \pi}{2} - \frac{n-2}{2}\ln\Big(-\frac{ \tilde y}{4}\Big)
\end{eqnarray}
for $n\neq 2$.
For $n=2$ there is only an exponential correction as seen from the asymptotic representation
\begin{eqnarray}
\label{W2asym} {\cal W}_2\big(\tilde y\big)& \approx& -\frac{\tilde y^2}{16}  - \frac{\ln \pi}{2} - \frac{2}{\pi^{1/2}\tilde y}\exp\Big(-\frac{\tilde y^2}{16}\Big)
\end{eqnarray}
which follows from (\ref{calW2x}) using ${\rm erfc}(y/4)= 2-{\rm erfc}(-y/4)$.
For $- \tilde y \gg 1$ we may set $\tilde l_T = 0$ in (\ref{scalfreeaniso}). For large $ \tilde l$ and slab geometry we use (\ref{calG0decom}) in the form
\begin{eqnarray}
\label{calJ0decom}
{\cal J}_0(\tilde l^2, \rho )=-\ln\big[\tilde l^2/(4\pi^2)\big]+ \rho^{1-d}{\cal G}_0(\tilde l^2, \rho)
\end{eqnarray}
with $\tilde l \approx \tilde l_- =  \;  (2 | \tilde x|Q^{*})^\nu$. These steps lead to (\ref{excess-scalfreeaniso18}).

Eqs. (\ref{6ff}) and (\ref{6gg}) yield the exact representation of the flow parameter
$\tilde l = \Big\{|\tilde x| Q^* \big[1+12 \vartheta_{2,n}(\tilde y)/\tilde y\big] \Big\}^\nu$.
For $\tilde y \gg 1$ we employ the asymptotic form of $\vartheta_{2,n}(\tilde y)$, (\ref{theta2largeabove}). This leads to
\begin{eqnarray}
\label{flowex} \tilde l^d = (\tilde x Q^*)^{d\nu} +12n u^* d\nu A_d^{-1}\bar\rho\;^{d-1}+ O(\tilde x^{-d\nu}), \;\;\,
\end{eqnarray}
and the terms in the first curly brackets of (\ref{scalfreeaniso}) become
\begin{eqnarray}
\label{first} && - A_d \big[n/(4d)  +\nu B(u^*)/(2\alpha)\big](\tilde x Q^*)^{d\nu}\nonumber\\&& + n u^* \nu\;\big[-3n + 6B(u^*)\big]\bar\rho\;^{d-1} + O(\tilde x^{-d\nu}).\;\;\,\;\;\,\;\;\,
\end{eqnarray}
From (\ref{Woberhalb}) we obtain
\begin{eqnarray}
&&{\cal W}_n(\tilde y)=(n/2)\ln(\tilde y/2) -\ln \big[\Gamma(n/2)\big]+ O(\tilde y^{-1}),\\
&&\tilde y= ( \tilde x Q^{*})^{d\nu/2}A_d^{1/2} {u^*}^{-1/2}\bar\rho\;^{(1-d)/2}\big[1+ O(\tilde x^{-d\nu})\big].\;\;\;\;\;\;\;\;\;\;\;
%\ee
\end{eqnarray}
For ${\cal J}_0(\tilde l^2, \rho)$  and ${\cal J}_0(\tilde l_T, \rho)$ the same decomposition is used as in (\ref{calJ0decom}), with $\tilde l_T = \tilde l^2[1+ O(\tilde x^{-d\nu})]$ and  $\tilde l \approx \tilde l_+ =  \;  ( \tilde x Q^{*})^\nu$. These steps lead to (\ref{excess+scalfreeanisoasymx}).

\end{document}